\documentclass[defaultstyle,11pt]{thesis}
\newtheorem{definition}{Definition}[section] 
\usepackage{amssymb}		
\usepackage{graphicx}
\usepackage{amsmath}

\usepackage[utf8]{inputenc}
\usepackage{multirow}
\usepackage{booktabs}
\usepackage{todonotes}
\usepackage{float}

\usepackage{amsmath}
\usepackage{amssymb}
\usepackage{amsfonts}
\usepackage{bbm}
\usepackage{blindtext}
\usepackage{graphicx}

\DeclareMathOperator*{\argmax}{arg\,max}

\setkeys{Gin}{draft=false}
\usepackage{url}
\usepackage{csquotes}

\usepackage[utf8]{inputenc}
\usepackage{palatino}
\usepackage{amsmath}
\usepackage{amssymb} 
\usepackage{graphicx}

\usepackage{multirow}
\usepackage{figsize}

\usepackage[labelfont=bf]{caption}
\usepackage{cite}
\usepackage{xcolor}
\usepackage{overpic}

\title{Popularity Bias in Recommendation:\\
A Multi-stakeholder Perspective}

\author{Himan}{Abdollahpouri}

\otherdegrees{B.S., Bu Ali Sina University, 2009 \\
	      M.S., Iran University of Science and Technology, 2012}

\degree{Doctor of Philosophy}		
	{Ph.D., Information Science}		

\dept{Department of}			
	{Information Science}		

\advisor{Prof.}				
	{Prof. Robin Burke}			

\reader{Prof.~Christine Lv} 		
\readerThree{Prof. ~Edward Malthouse }
\readerFour{Prof. ~Bamshad Mobasher}		
\readerFive{Prof.~Stephen Voida}
\date{29th July}

\abstract{  \OnePageChapter	
Recommendation and ranking systems are ubiquitous. They are used in a variety of different domains where they can influence what news people read, what products they buy, who they date and marry, and what cultural artifacts they consume. Traditionally, especially in academic research in recommender systems, the focus has been solely on the satisfaction of the end-user. While user satisfaction has, indeed, been associated with the success of the business, it is not the only factor. In many recommendation domains, there are other stakeholders whose needs and interests should be taken into account in the recommendation generation and evaluation. In this dissertation I describe the notion of multi-stakeholder recommendation and discuss its main building blocks, challenges, and connections to prior research. In particular, I study one of the most important challenges in recommendation research, popularity bias, from a multi-stakeholder perspective since, as I show later in this dissertation, it impacts different stakeholders in a recommender system. Popularity bias is a well-known phenomenon in recommender systems where popular items are recommended even more frequently than their popularity would warrant, amplifying long-tail effects already present in many recommendation domains. Prior research has examined various approaches for mitigating popularity bias and enhancing the recommendation of long-tail items overall. The effectiveness of these approaches, however, has not been assessed in multi-stakeholder environments. In this dissertation, I study the impact of popularity bias in recommender systems from a multi-stakeholder perspective. In particular, I investigate how this bias impacts users and suppliers of the recommendations. That is, how this bias causes the recommendations to deviate from the users’ preferences towards popular items and how it leads to unfair exposure for different supplier groups. Finally, I propose several algorithms each approaching the popularity bias mitigation from a different angle and compare their performances using several metrics with some other state-of-the-art approaches in the literature. I show that, often, the standard evaluation measures of popularity bias mitigation in the literature do not reflect the real picture of an algorithm’s performance when it is evaluated from a multi-stakeholder point of view.

}

\dedication[Dedication]{	
	This dissertation is a small gift to my parents. To my hard working father who started working when he was a teenager after his father passed away. He never went to school so he could support his mom and siblings. A construction worker who spent his entire life supporting his family. Every evening when he was coming back from work, the sound of his motorcycle was the most pleasant thing we could hear coming from the distance and slowly approaching us. He has been always my hero and the symbol of honesty, authenticity, and integrity. Also to my mother: the selfless woman who never asked for anything but always was the provider to her husband and children. She can't read and use smartphones so she used to go to the neighbors and relatives asking them if she could call her son in the U.S. I still remember the excitement and brightness in their eyes when I first told them I was accepted as a PhD student. It has been five years since the last time I saw them and I hope I can see them soon. Every time I talk to them over the phone they keep saying "We are sorry that we can't financially support you living in a foreign country and all we can offer you is our prayers" little do they know this is the best thing I could ever ask for. Thanks for everything you have giving me till today.
	}

\acknowledgements{	\OnePageChapter	
	First and foremost, I would like to express my sincere gratitude to my supervisor Prof. Robin Burke for all the effort and patience he has put in mentoring my PhD and guiding my research. I thoroughly enjoyed our brainstorming sessions and learned a lot during this five years both personally and professionally.

Also, special thanks to my PhD committee members Prof. Christine Lv, Prof. Edward Malthouse, Prof. Bamshad Mobasher, and Prof. Stephen Voida for their great and invaluable comments both on my proposal and dissertation defense. 
I am also thankful to the RecSys community for supporting me and my research. 
Finally, I would also like to thank the National Science Foundation for supporting my research during my PhD. 
	}

\begin{document}

\newtheorem{theorem}{Theorem}

\newcommand{\diff}[2]{\frac{\partial #1}{\partial #2}}
\newcommand{\diffr}[1]{\diff{#1}{r}}
\newcommand{\diffth}[1]{\diff{#1}{\theta}}
\newcommand{\diffz}[1]{\diff{#1}{z}}

\newcommand{\vth}{V_{\theta}}

\newcommand{\twochoices}[2]{\left\{ \begin{array}{lcc}
        \displaystyle #1 \\ \vspace{-10pt} \\
        \displaystyle #2 \end{array} \right. } 

\newcommand{\threechoices}[3]{\left\{ \begin{array}{lcc}
        #1 \\ #2 \\ #3 \end{array} \right. }    

\newcommand{\fourchoices}[4]{\left\{ \begin{array}{lcc}
        #1 \\ #2 \\ #3 \\ #4 \end{array} \right. }      

\newcommand{\twovec}[2]{\left(\begin{array}{c} #1 \\ #2 \end{array}\right)}
\newcommand{\threevec}[3]{\left(\begin{array}{c} #1 \\ #2 \\ #3 \end{array}\right)}
\newcommand{\twomatrix}[4]{\left(\begin{array}{cc} #1 & #2 \\ #3 & #4 \end{array}\right)}

\chapter{Introduction}\label{Introduction_chapter}
\section{Research Goals and Contributions}

This research work focuses on popularity bias problem in recommender systems. Generally speaking, popularity bias refers to the problem where the recommendation algorithm favors a few popular items while not giving deserved attention to the majority of other items. In particular, the goal of this thesis is to evaluate the impact of popularity bias in recommendation on different stakeholders on a given recommender system platform such as users and item suppliers. In addition, several algorithms for mitigating this bias are proposed. With that being said, the following are the research goals and contributions of this thesis:
\begin{itemize}
    \item \textbf{RG-1}: Proposing a comprehensive paradigm called multi-stakeholder recommendation for evaluation and development of recommender systems. Especially in academia, recommender systems are typically evaluated based on the satisfaction of one stakeholder: the end user. \\
    \textbf{Contributions:} In this dissertation, I introduce the notion of multi-stakeholder recommendation where the needs and interests of multiple stakeholders are taken into account in the recommendation generation and evaluation, in addition to that of users. In many real-world applications of recommender systems users are not the only stakeholder so the multi-stakeholder paradigm is crucial in such systems. Take a music streaming service such as Spotify as an example. On the one hand, there are listeners who use the service to find and listen to the music they would enjoy. On the other hand, the creators of those songs (i.e. the artists) also benefit from the platform by their songs being played for / by different users and hence gaining utility. The popularity bias in recommendation can negatively impact both sides of the platform.  
    \item \textbf{RG-2}: Investigating the impact of popularity bias in recommendation on different users. Popularity bias plays an important role in the unfairness of the recommendations towards different stakeholders. There has been research on the popularity bias and long-tail recommendation which focuses on the consequences of such bias from the system's perspective such as item coverage, aggregate diversity etc or how this bias can negatively impact the accuracy of the recommendations. \\
    \textbf{Contributions:} In this thesis, I investigate how the popularity bias problem impacts different user groups with a varying degree of interest in item popularity. For instance, how users with niche tastes are impacted by this bias versus those with a more inclination towards mainstream products. 
    \item \textbf{RG-3}: Investigating the impact of popularity bias in recommendation on different item providers (aka suppliers). As mentioned earlier, suppliers of the items on a recommendation platform are also stakeholders whose needs and preferences may need to be addressed. Typically, in many recommendation domains, the platform may not necessarily own any of the recommended items itself but rather acts as a bridge that connects the suppliers of the items to the users. For instance, in music recommendation, each song or album is produced by an artist; on Airbnb, each listing is provided by a host; on eBay, each product is provided by a seller. Therefore the dynamics of how recommendations impact these item suppliers is important. \\
    \textbf{Contributions:} In this thesis, I investigate how popularity bias in recommendation impacts different item suppliers with a varying degree of popularity. For instance, I investigate whether this bias causes over-promotion of items from certain item suppliers while under-recommending items from some others. 
    
    \item \textbf{RG-4}: Proposing algorithms to mitigate popularity bias in recommendation and evaluating their performance from a multi-stakeholder perspective. \\
    \textbf{Contributions:} Most of the existing work on long-tail in recommendation concentrates on either the accuracy of the recommendation in the presence of popularity bias and how to address that or they are mainly focusing on the holistic system-level picture of this bias and its impact on overall recommendation coverage. In this thesis, I propose several algorithms for controlling popularity bias and see how each of these perform from the perspective of different stakeholders. By "control" I mean reducing the over-concentration of the recommendations on few popular items and to give more chance to other items to be seen by the users. In particular, the algorithm based on calibrating recommendations to the users' historical interaction with respect to item popularity shows benefits both to the users and the suppliers of the recommendations.   
\end{itemize}

\section{Structure of the Thesis}
This dissertation is structured as follows:
\begin{itemize}
    \item Chapter 2 gives a brief summary of the field of recommender systems. In particular, different types of recommendation methods are explained and the general evaluation methodologies for evaluating the recommendations are described. Then, I move on to give a short background of several non-accuracy concepts in recommender systems such as diversity, novelty, and long-tail recommendation. 
    \item Chapter 3 gives a thorough description of the characteristics of
multi-stakeholder recommendation and particular considerations involved in creating and evaluating multi-stakeholder recommender systems. The connection between prior work in non-accuracy measures and the proposed multi-stakeholder paradigm is also studied. 
\item Chapter 4 describes the methodology and data used for experiments in this dissertation. Details regarding the datasets, the methodology for running the experiments and also the evaluation metrics that are used in other chapters of this dissertation are all fully described. 

\item Chapter 5 describes the popularity bias problem and how it can impact different stakeholders in recommender systems. Two sources of bias including data and algorithms are explained using two datasets. 

\item Chapter 6 gives the details of the proposed algorithms in this dissertation to mitigate the popularity bias problem. All algorithms are evaluated using different evaluation metrics described in Chapter 4 and two state-of-the-art baselines for popularity bias mitigation are also included for comparison.
\item Chapter 7 gives a short summary of the findings in this dissertation and proposes several interesting open questions and possible future work. 
\end{itemize}

Table \ref{tab:publications} contains the list of main publications that are the result of these dissertations and are appeared in peer-reviewed journals, conferences, and workshops. 
\begin{table}[]
\centering
\captionof{table}{List of Publications that are the result of this dissertation and their corresponding chapter.} 
\label{tab:publications}
\begin{tabular}{|l|l|}
\hline
\textbf{Publiction}                                                                                                                                                                                                                                                                                                                               & \textbf{Chapter} \\ \hline
\begin{tabular}[c]{@{}l@{}}{[}1{]} Abdollahpouri, Himan, Gediminas Adomavicius, Robin Burke, \\ Ido Guy,  Dietmar Jannach, Toshihiro Kamishima, Jan \\ Krasnodebski, and Luiz Pizzato."Multistakeholder recommendation:\\  Survey and research directions." User Modeling \\ and User-Adapted Interaction 30, no. 1 (2020): 127-158.\end{tabular} & 3                \\ \hline
\begin{tabular}[c]{@{}l@{}}{[}2{]} Abdollahpouri, Himan, and Robin Burke. "Multi-stakeholder\\  recommendation and its connection to multi-sided fairness.\\ " Workshop on Recommendation in Multi-stakeholder Environments \\ RMSE in Conjunction with RecSys (2019).\end{tabular}                                                               & 3                \\ \hline
\begin{tabular}[c]{@{}l@{}}{[}3{]} Burke, Robin D., Himan Abdollahpouri, Bamshad Mobasher, \\ and Trinadh Gupta. "Towards Multi-Stakeholder  Utility \\ Evaluation of Recommender Systems." In UMAP \\ (Extended Proceedings). 2016.\end{tabular}                                                                                                 & 3                \\ \hline
\begin{tabular}[c]{@{}l@{}}{[}4{]} Abdollahpouri, Himan, Robin Burke, and Bamshad \\ Mobasher. "Recommender systems as multistakeholder environments.\\ " In Proceedings of the 25th Conference on User Modeling, Adaptation\\  and Personalization, pp. 347-348. 2017.\end{tabular}                                                              & 3                \\ \hline
\begin{tabular}[c]{@{}l@{}}{[}5{]} Burke, Robin, and Himan Abdollahpouri. "Educational \\ recommendation with multiple stakeholders." In 2016\\  IEEE/WIC/ACM International Conference\\  on Web Intelligence Workshops (WIW), pp. 62-63. IEEE, 2016.\end{tabular}                                                                                & 3                \\ \hline
\begin{tabular}[c]{@{}l@{}}{[}6{]} Burke, Robin, and Himan Abdollahpouri. "Patterns of \\ Multistakeholder Recommendation." Workshop on \\ Value-aware and Multi-stakeholder Recommendation \\ VAMS in Conjunction with RecSys (2017).\end{tabular}                                                                                               & 3                \\ \hline
\begin{tabular}[c]{@{}l@{}}{[}7{]} Abdollahpouri, Himan, Masoud Mansoury, Robin Burke, \\ and Bamshad Mobasher. "The unfairness of popularity\\  bias in recommendation." arXiv preprint arXiv:1907.13286 (2019).\end{tabular}                                                                                                                    & 5                \\ \hline
\begin{tabular}[c]{@{}l@{}}{[}8{]} Abdollahpouri, Himan, and Masoud Mansoury.\\ "Multi-sided Exposure Bias in Recommendation."  \\ Workshop on Industrial Recommendation  Systems (IRS), \\  in Conjunction with KDD 2020.\end{tabular}                                                                                                           & 5                \\ \hline
\begin{tabular}[c]{@{}l@{}}{[}9{]} Abdollahpouri, Himan, Robin Burke, and Bamshad \\ Mobasher. "Controlling popularity bias in learning-to-rank\\  recommendation." In Proceedings of the Eleventh ACM Conference \\ on Recommender Systems, pp. 42-46. 2017.\end{tabular}                                                                        & 6                \\ \hline
\begin{tabular}[c]{@{}l@{}}{[}10{]} Abdollahpouri, Himan, Robin Burke, and Bamshad \\ Mobasher. "Managing popularity bias in recommender\\  systems with personalized re-ranking."\\  In The Thirty-Second International Flairs Conference. 2019.\end{tabular}                                                                                    & 6                \\ \hline
\begin{tabular}[c]{@{}l@{}}{[}11{]} Abdollahpouri, Himan, Masoud Mansoury, Robin Burke, \\ and Bamshad Mobasher. "Addressing the Multistakeholder Impact \\ of Popularity Bias in Recommendation \\ Through Calibration." arXiv preprint arXiv:2007.12230 (2020).\end{tabular}                                                                    & 6                \\ \hline
\end{tabular}
\end{table}
\section{Definitions and Notations}

I summarise here the main definitions and notations that I use throughout this dissertation. If needed, additional notation will be introduced in chapters where it applies.

A recommendation problem is defined as suggesting items from a collection of items $I$ to a group of users $U$. I also show an individual user by $u$ and an individual item by $i$. I show the interaction data as a matrix $R$ where it contains the ratings that users have given to particular items. In certain cases, instead of ratings, we have the information about whether or not a user has interacted with a certain item (liked, clicked, watched, played etc). The rating a user $u$ has given to an item $i$ is denoted as $r_{ui}$ and the predicted rating for that user and item pair returned by the recommendation algorithm is denoted by $\hat{r}_{uj}$. For any user $u$, we typically recommend a list of items denoted by $\ell_u$. The user for whom we generate the recommendations is often referred to as the \textit{target user}. The list of all the recommendation lists given to all users is denoted by $L$. Additionally I show the profile (i.e. the interaction history) of user $u$ by $\rho_u$ which is $\{i \in I, (u,i)\in R\}$. Each item $i$ is provided by a supplier $s_i$. $S$ denotes a set of supplier groups where each supplier group $j$ is shown by $S_j$. The set of all individual suppliers in a given dataset is shown by $\mathcal{P}$.

\chapter{Background}\label{background_chapter}
\section{Recommender Systems}

It is always good to have a choice. However with the massive amount of information,
data, products that are created everyday, it is constantly becoming more difficult to handle this overwhelming number of choices and make a final decision. Information retrieval systems such as search engines can help a user seeking information by filtering from a large collection of documents (or items or products), those that are most relevant to the user’s needs. Users express their needs in terms the type of information they are looking for as a query, which for example could be a list of keywords, or
some other explicit description of the requirement. In recommender systems, however, often there
is no explicit query. Instead, the system uses its knowledge about the user to reduce the
cognitive load of choice by predicting which items are most relevant to her while still leaving the final decision making for the user.

Recommender systems are intelligent tools and algorithms that help users find relevant items according to their preferences \cite{ricci2011introduction}.
Especially in areas where the number of possible choices is overwhelming, recommender systems been successfully used in finding the ones that are most relevant to the user. By offering personalized recommendations, these systems lower the decision effort for the users, making the item selection process much
easier. Recommender systems can be seen as personal shoppers who preselect the
best items based on the knowledge they have about the customer, such that a she has to only decide which to pick from a small set of items.

\begin{figure}
    \centering
    \includegraphics[width=5in]{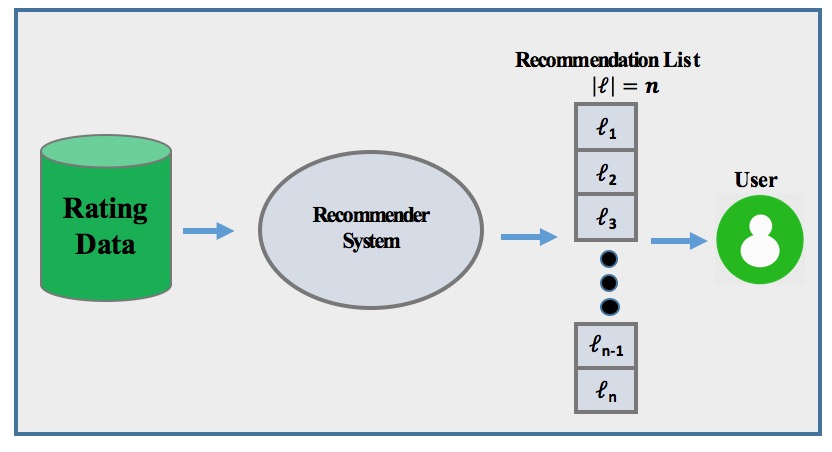}
    \caption{A general schema for recommender systems}
    \label{fig:recsys}
\end{figure}

The general schema for a given recommender system is shown in Figure \ref{fig:recsys}. The algorithm uses the information about the users and items stored in the rating data to give a list of recommendations $\ell$ of size $n$ to a user. We can think of this schema as an abstract view that often can be modified to serve different purposes as we will see in Chapter \ref{tackle_pop_bias}. 

Recommender systems have become popular in the last decades, however
the recommendation process is much older than that. Even in our daily life, people often make decisions
based on the suggestions made by others. For example, we may get recommendations from our friends about movies, music, restaurants or even a job. We trust these recommendations if we know that the
person who gave it has some knowledge of the topic, has similar taste to ours or knows our taste, or
that the past shows we can trust that person's suggestions; this is known as "word of mouth" \cite{anderson1998customer}. What recommender systems do is to mimic the above behaviour on a larger scale, by utilizing viewpoints of many users that do not have to know each other and have similar tastes. This is known as the collaborative filtering.

Recommender systems are not solely tools that were developed to help users in the item selection process. In fact, there are a number of other
reasons why different service providers adapt recommender systems in their services
\cite{ricci2011introduction}. From the business’ perspective, probably the most important is
the ability to increase sales through recommendations \cite{schafer1999recommender}. Another example of business-oriented goals is to diversify sales, such that they do not center around a small set of very popular items. It might be hard to achieve this goal without recommender systems as there is a risk and cost associated with every item promotion. Promoting less popular items is more risky, however recommender
systems can help to recommend less popular items to the right users. It is shown that a
well-designed recommender system can also improve users’ satisfaction with the
service \cite{nguyen2018user}, and their loyalty to the service \cite{schafer1999recommender}. A better understanding of what the users really like can be utilized not only by a recommendation engine, but for other operational or marketing purposes.

Not only are there different reasons we deploy recommender systems, but
also different scenarios which may depend on the domain that we are working in. Historically, the most commonly considered scenario is the recommendation task of movies or music (based on datasets such as MovieLens \cite{harper2015movielens}, Netflix \cite{bennett2007netflix}, or Yahoo! Music \cite{dror2011yahoo}), since these are the type of data that are publicly available. However recommender systems can be used in more complex settings and in different domains such as e-commerce \cite{oestreicher2012recommendation}, job recommendation \cite{paparrizos2011machine}, and even in online dating platforms \cite{pizzato2010recon}.

In recommender systems, the users interaction with the system is often referred to as a transaction. For example, a transaction could be a user listening to a song, or rating a recently booked accommodation. Commonly, with a transaction we may have an associated feedback, in this case the
act of playing a song, or rating the accommodation, represent, respectively, implicit
and explicit feedback \cite{jawaheer2014modeling}. The main difference between these
two is that for the explicit feedback, user consciously expresses her opinion or preference, where in case of the implicit feedback the preference
is inferred from other actions recorded by the system \cite{oard1998implicit} —
if a user plays a song and skips it after a few seconds, this could represent a dislike
rather than a like. In the case of the implicit feedback, it is the recommender system's job to figure out the
strength of the signals and what each action could indicate for different users. On the other hand, in the
case of explicit feedback, the type of the feedback may vary. Feedback on contents
like movies and books have been commonly expressed through numerical ratings
(like 1-5 stars, or 1-10 points), written reviews, or simple like/dislike indicators. In modern systems, often multiple sources of feedback are available all of which can impact the design of
a recommender system.

Another factor affecting the design is the consumption pattern of items, how
often we interact with an item, and how long the item is ‘fresh’. Depending on the
domain, a transaction may reoccur, e.g. an item is repurchased by the user \cite{chen2016recommendation}, and there are other scenarios where interactions are
rather unique, or the reoccurrence is unlikely to happen. For example, in a music recommendation system it is natural that the same songs and artists
are recommended to users multiple times. However, recommending the same movie to
watch a second time may make sense to some but not to all. Similarly, while it may
be desirable to recommend coffee beans every few weeks, recommending another
coffee machine does not sound reasonable. In the news domain, recommending the
same news article to a user is rather useless, but this is not the only problem. News are very dynamic and often the most recent news are more important to the users than old news. The
lifetime of news stories is typically very short. All these factors have to be taken into
account when designing systems like a news recommendation system \cite{burke2011matching,garcin2013personalized,liu2010personalized}.

Lastly, an important aspect that has to be considered is the type of recommendations
that the system should produce, and the use cases in which these are used. In some cases, we might
be interested in accurately predicting how the user will rate an item when presented
to the user. On the other hand, we might want to generate a list of best items to the user without focusing on the accuracy of the predicted ratings per se. These represent the two main types
of problems that have been distinguished in the literature \cite{steck2013evaluation}: the rating
prediction and the ranking problem. Historically, the rating prediction was mostly
considered and it was the main evaluation metric used in Netflix Challenge in 2006 \cite{bennett2007netflix}. However, more recently the ranking task has been found to be more
relevant to the users and business since, in the end, it is how the users would like the recommendations in a list that matters not how close our predicted ratings are to the original ratings given by the user. The choice of the recommendation task significantly affects
the evaluation methodology of recommendations.

For all these different reasons and scenarios, a number of different algorithms have been proposed that generate recommendations to the users, which I discuss next.

\subsection{Types of Recommendation Algorithms}

Recommender system algorithms can be classified according to a number of criteria.
Burke \cite{burke2007hybrid} distinguished six different classes of recommender systems: content-based,
collaborative filtering, demographic, utility-based, knowledge-based, and
hybrid recommender systems. I focus my review on the three, most common of
them: collaborative filtering, content-based, and hybrid.

\textit{Content-based Recommender Systems}

In the content-based recommender systems \cite{lops2011content,pazzani2007content}, the main task of the system is to analyze items with which users interacted and build users’ interests profiles which shows the overall taste of that user in terms of different types of items. This is typically done based on item features that are used
to describe those items. In the recommendation process, the system matches the features
of the user’s profile against the features of items to find the best items for the user.
For example, if a user has shown an interest in drama movies, this will be
indicated in his or her interests profile, and the system will recommend movies that are also related to drama movies.

In Lops et al. \cite{lops2011content}, a number of advantages and disadvantages of content-based recommendation have been identified, especially if compared with the collaborative filtering approach (which
I will discuss next). Content-based systems depend solely on the item features
and items that the current user interacted with—we do not need any information
about other users. In cases of
new users without an established interest profiles, users may not receive useful
recommendations. This does not happen when a new item is introduced, as long
as the item is properly described in terms of its associated features, and there are other items sharing those features. The
process of describing items requires domain knowledge, and may be tedious and
costly. However, there are real-life examples where that has been done (like the Music
Genome Project \footnote{https://www.pandora.com/about/mgp}) and is leveraged for generating the recommendations. Assuming the set of features associated with different items are well-defined, it is easier to explain recommendations to the users, which
positively impacts the users' experience with the system. On the other hand, by
design, this approach favours items that are highly similar to those that the user has
already seen, making the recommendations of low novelty (and diversity as well), or
of little surprise and engagement to the user, which we consider as an undesirable
situation we would like to avoid.

\textit{Collaborative Filtering Recommender Systems}

Collaborative filtering methods, by contrast,
make recommendations to the current user based on the similarity of rating behavior between the user and other users, on the assumption that if two users have similar taste, they may be interested in
what the other user likes \cite{goldberg1992using}. The similarity in taste is based on
the similarity in users’ feedback on different items, e.g. how similarly the users have rated different items. While, depending on the type of interaction data (explicit vs implicit) different
approaches may be needed, collaborative
filtering methods are often known to be domain-independent, which is their advantage
compared with content-based methods where they have high dependency on the contents of the items. Also, in many cases collaborative filtering approaches have been more effective.

Most of the collaborative filtering approaches fall into two main categories: model-based and memory-based. Model-based recommender
systems are similar to parametric models \cite{rasmussen2003gaussian} in machine learning. In a model-based
algorithm, a recommendation model, depending on a number of parameters is
proposed. These parameters are then learned using the user-item interactions in a given dataset.
Thereafter, recommendations can be made, without further recourse to the data. However, unlike typical classification methods in machine learning, the model-based recommendation approaches are personalized meaning the result for each prediction varies depending on the user for whom we make prediction. In a memory-based algorithm, unlike model-based techniques, the dataset is explicitly queried for each new recommendation.

Memory-based methods (also called neighborhood-based) have been originally
tasked with the rating prediction for items that are unseen to users. This is typically
done in two ways, known as user-based
\cite{hill1995recommending,konstan1997grouplens,ning2015comprehensive} or item-based recommendations \cite{sarwar2001item,deshpande2004item}.

User-based methods predict the interest of the current user for a given item using the
interaction of other users, called neighbors, who also were interested in that
item. Typically we do not consider all users interested in the item, but the $k$ most
similar neighbors to the current user. The similarity of different users is calculated using some metrics such as cosine similarity, adjusted cosine similarity, or Pearson correlation.
We can then, with the use of the similarity function, calculate the score reflecting
the predicted relevance of an unseen item for a given user. In the simplest form, the score can be
expressed as a sum over feedback of the $k$ closest neighbors to the target user on
the target item, weighted by the similarity of those neighbors to the target user:
\begin{equation}
    \hat{r}_{ui}= Z \sum_{v \in N^k(u)} sim(u,v)r_{vi}
\end{equation}

\noindent where $u$ is the target user, $sim(u,v)$ is the similarity between users $u$ and $v$, and $N^k(u)$ denotes the neighborhood of user $u$ of size $k$, and $Z$ is a normalization constant. This
normalization constant can be used to customize the equation to fit different recommendation
tasks. Originally, constant $Z=\frac{1}{\sum_{v \in N^k(u)}|sim(u,v)|}$ has been used for the rating
prediction task \cite{adomavicius2005toward} — it is required
to scale the rating prediction to the correct range of values, same as the original ratings
— however it has been pointed that for the ranking task the constant of $Z = 1$ is
generally more effective \cite{aiolli2013efficient,canamares2017probabilistic,cremonesi2010performance}.

Several variations of neighborhood-based methods have been considered. Here we presented a neighborhood of a fixed size but another
option is to select neighbors by thresholding the similarity value \cite{bellogin2013comparative}. In other words, considering a user to be a neighbor for another user only if the similarity between them is above a certain threshold. An interesting approach has been proposed by Said et al. \cite{said2012increasing},
where instead of taking the $k$ closest neighbors, the $k$ furthest neighbors are
chosen instead. This is done based on the assumption that items disliked by users that
are dissimilar to the target user may be of interest to that user. This approach also
improves diversity of the recommendations. Some probabilistic versions of the method
have been also proposed. Adamopoulos and Tuzhilin \cite{adamopoulos2014over} proposed a probabilistic neighborhood selection in the context of over-specialisation and concentration bias, where Cañamares and Castells \cite{canamares2017probabilistic} provided a formal probabilistic formulation and analysed the problem of popularity bias in recommendations.

In the item-based methods \cite{itemRec}, instead of similarity between users, the similarity
between items is used to generate the recommendations. The basic intuition is that items similar to those
that the user already liked, are good choices to be recommended to the user. In the simplest
form, the predicted score can be calculated as a sum over item similarities between the target
item and previously interacted items weighted by user’s preference, given the target
item belongs to the $k$ closes neighbors of each item in the user’s profile:
 
\begin{equation}
    \hat{r}_{ui}= Z \sum_{j \in \rho_u} \mathbbm{1}_{N^k(j)}(i)sim(i,j)r_{uj}
\end{equation}

\noindent where $u$ is the target and the indicator function $\mathbbm{1}_{N^k(j)}(i)$ filters out items if they do not belong to the
neighborhood $N^k(j)$, $sim(i, j)$ is the similarity between items $i$ and $j$, $\rho_u$ is the collection of items that user u interacted with (i.e. the profile of user $u$), $r_{uj}$ denotes the preference of user $u$ for item $j$. Similarly as with the user-based methods, a number of extensions have been proposed — many of them were actually proposed for both techniques as they are similar in principle.

One of the advantages of the neighborhood-based methods \cite{ning2015comprehensive} is that
they happen to be simple, requiring little parameterization. They are also intuitive in the way they
work which makes it easier to justify the recommendations to users. From the
computational point of view, since they require no training, they can efficiently generate
recommendations. They may require pre-computation of the neighborhoods,
however this is generally much cheaper than training a model. On the other hand,
neighborhood-based methods are known to suffer from limited item coverage. In other words, they often concentrate on few popular items. They are also sensitive to sparseness of data (datasets generally contain only a fraction of all
possible user-item interactions) and may not perform well on sparse datasets.

In the model-based methods, generally there are two phases that can be distinguished:
a learning phase and a prediction phase. In the learning phase a model is
learned from the interaction data, and in the prediction phase the model is utilized
to generate recommendations. This approach is similar to many machine learning
methods, such as linear models, (deep) neural networks, clustering, latent factor
models and others. In the field of recommender systems, latent factor models have
attracted the most attention among model-based methods, especially to perform
dimensionality reduction of a ratings matrix. These models aim to uncover latent
features that would explain user preferences hidden in observed ratings. Examples
include Probabilistic Latent Semantic Analysis (PLSA) \cite{hofmann2001unsupervised}, Latent
Dirichlet Allocation \cite{blei2003latent}, or matrix factorization \cite{koren2009matrix}.

In the context of collaborative filtering, the most commonly used model-based
method is matrix factorization. Its goal is to learn two low-rank matrices, $P \in ^{R|U|\times k}$
and $Q \in ^{R|I|\times k}$, which represent, respectively, all users and items in k-dimensional
latent space. Typically, $k$ is  much smaller than the number of users or items, and
it is a model parameter that needs to be tuned. $P$ and $Q$ matrices are learned through
the optimization of an objective function $L(P,Q)$. There are different optimization
techniques that can be used, such as alternating least squares \cite{pilaszy2010fast,takacs2012alternating}, stochastic gradient descent \cite{jahrer2011collaborative,rendle2014improving,shi2012climf}, or maximum
margin matrix factorization \cite{rennie2005fast,weimer2008cofi}. After learning $P$ and $Q$, recommendations can be made based on the
score
\begin{equation}
    \hat{r}_{ui}=p_u^Tq_i
\end{equation}

where $p_u$ is the row vector of $P$ for the user $u$, and $q_i$ is the row vector of $Q$
corresponding to the item $i$. $T$ indicates the transpose of a given matrix.

\textit{Hybrid Recommender Systems}

In the hybrid methods \cite{adomavicius2005toward,burke2002hybrid}, we combine
two or more recommendation algorithms to generate recommendations that overcome
limitations of the individual methods. This is similar to fusion strategies in information retrieval which are usually adopted to improve
performance of individual retrieval methods \cite{wu2013weighted}, or in machine
learning where ensemble methods have been used to combine several machine
learning models \cite{dietterich2000ensemble} for improving the performance. In Burke \cite{burke2007hybrid}, seven classes of hybrid methods have been proposed: weighted, switching, mixed, feature combination, cascade,
feature augmentation, and meta-level.

First three techniques — weighted, switching, and mixed — assume that several
recommendation algorithms simultaneously generate recommendations and a
hybridization strategy is applied to combine these recommendations. The simplest strategy is the mixed strategy, in
which recommendations from all recommender systems are presented, either in a single list or in multiple recommendation lists. In the weighted strategy, each
recommender system returns recommendations along with scores for each recommendation. These recommendations from each system are
weighted by the weight assigned to that system. Such weighting systems generally
favors items that are considered relevant by many systems (the maajority vote), and recommendations
coming from trustful systems — typically a high weight is assigned to systems
that have proven to be successful in producing useful recommendations in the past. In the
switching strategy, instead of combining recommendations, at a given time, one algorithm is selected
based on some conditions to provide recommendations. Both weighted
and switching strategies require some meta-information on how to select/combine systems.
For example, these can be selected arbitrarily, or learned in similar fashion as stacking in
machine learning \cite{dietterich2000ensemble}, or using a reinforcement learning
technique \cite{sutton1998introduction}.

The feature combination methods can be seen as a data engineering approach (and not necessarily and algorithm)
where multiple different data sources are combined together, and the
enhanced final dataset is used to train a single recommendation algorithm. In other words, the hybridization is done on data and not on the algorithms.

The last group of strategies—cascade, meta-level—
require staged processing, and feature augmentation, where results of the first stage is used by the second
stage recommender system. The methods differ in what is produced in the first
stage, and how it is used in the next stage. For example, in cascade method,
recommendations produced by a recommender system are refined by the second
stage system. In meta-level recommender system, model learned by one technique is used as input to another. Finally, in the feature augmentation strategy, output coming from one technique is added to the feature sets used by the next stage algorithm.

\section{Evaluation Methodologies}
So far I have discussed a number of different recommendation algorithms. The question is how to assess the quality of the recommendations generated by any of these algorithms. In order to select one that performs the best for a given scenario or which parameters to use for each algorithm, typically a set of experiments is run where performance
of the algorithms are compared. Following Shani and Gunawardana \cite{shani2011evaluating}, three main types of evaluation are typically considered — offline experiments, user studies, and online
experiments.

\subsection{Offline Evaluation}

Offline evaluation is the most common approach used in the literature to evaluate the
effectiveness of recommender systems. It is performed by using a pre-collected
dataset of users and items rated/chosen by them. The idea behind offline evaluation is based on the assumption
that user characteristics at the time the data was collected will be similar to when a system
trained on that data is used. The main advantage of this type of evaluation is its efficiency in testing many algorithms at a low cost. However this type of evaluation cannot answer many questions related to the quality of the recommendations and the success of the recommender system in general.  

Datasets used in offline evaluation should match as closely as possible the data in
the online system. In other words, a great amount of care must be taken in order to make sure no additional biases are introduced,
as bias correction is often difficult. These biased may be introduced in the filtering and
splitting process that some recommender systems require — favouring algorithms
that work better with e.g. denser data — but also at the data collection stage. For example, in the
case of explicit feedback, people tend to rate items that they like which means that data is not missed at random.

In order to evaluate the system, offline evaluation requires a way to simulate
users interacting with the system. A common approach — popular in machine
learning — is to set aside a subset of the recorded interactions and treat it
as the ground truth. This splits the dataset into the training set—made of ‘known’
interactions — and the test set — made of ‘hidden’ interactions. There are different
possibilities for performing the split. One simple approach is to randomly
split all the recorded past interactions. This, however, assumes that the time aspect
of the system is unimportant and users preferences does not change over time. Other techniques include a time-aware split of the
dataset, or time-aware split performed separately for each user, or to select a fixed
number of known items per user, or a fixed number of hidden items per user. As
there are many options, it is necessary to understand which technique matches the
target recommendation scenario better.
\\
\subsection{Online Evaluation}

In online evaluation, users are interacting with a live recommender system and react to the recommendations generated by that system. In many real-world
recommender systems, online testing mechanisms are implemented through which experiments are
run and algorithms are compared. A common approach is to select a subgroup of users and redirect them to the recommendation algorithm that is being tested
and record its performance. It is important to select a subgroup randomly so there
is no bias involved, and fair comparisons can be made. Although, in some scenarios users should be systematically chosen if the algorithm is particularly needs to be tested on a certain group of users. Also, conditions of the
experiments should be controlled — this means if one aspect of the system is under
experimentation, then other components ideally should be fixed.

Performing an online evaluation of a new recommender system comes with risk. For example the algorithm may not work as expected and may cause harm to user experience and hence harming the business. Therefore, in practice, before conducting an online experiment, an offline evaluation is performed and if the results are stable and reasonable, online testing can be performed next.

\subsection{User Studies}

If access to a live system with real users is not possible or too risky, user studies can be conducted as a mean to
gather real feedback from users of the system. Generally a number of test subjects (i.e. users) is
asked to interact with the system, or a simulation of such interaction is shown and
subjects are asked to judge their recommendations. User studies can go even beyond evaluating users' interaction with a recommender system. For example, they can be used to survey users to directly
gather more insights on the system and learn users’ opinion on certain aspects of
the system that could not be collected during offline or online evaluation. Running
user studies is however expensive as it requires finding subjects that would be willing to spare their time and to participate in the study. Also the process
of selecting subjects is complex as to perform a fair user study we should have a
representative sample of the population of users in the system. In addition, in user studies users often have no fear of the costs of their choice. In reality, when a user buys an item, rents an apartment, or even watches a recommended movie, s/he should accept some kind of cost either financial or simply the time s/he spends on using that item. This is usually not the case in user studies. Finally, we should be aware of different biases, coming from different sources. For example, users might act differently than normally if they know that they are being surveyed or observed; others might be more interested in or aware of the topic of the survey, which may have an impact on the results. All these aspects (and many more) can directly affect the results.
\\

\section{Evaluation Metrics}

Besides the type of experiment used to perform the evaluation, another important
aspect is the type of the problem that the recommender system is used for. It affects
the choice of algorithms, form of output recommendations, and methods used to
evaluate the performance. In Steck \cite{steck2013evaluation}, two types of problems have been identified:
the rating prediction task, and the ranking task. In the realy days of research on recommender systems, rating prediction was more
common. However, later on researchers realized that
ranking is more closely related to real recommendation
scenarios (\cite{cremonesi2010performance}).

\subsection{Rating Prediction}

In rating prediction, the goal of the system is to
accurately predict how a user $u$ would rate an item $i$ in the test set. Also it is worth noting that explicit ratings are required in order to perform rating prediction. With respect to evaluation, rating prediction task is
commonly evaluated through one of the error metrics such as the Mean-Absolute
Error (MAE), or the Root-Mean-Square Error (RMSE):

\begin{equation}
    MAE=\frac{1}{|\tau|}\sum_{(u,i) \in \tau} |r_{ui}-\hat{r}_{ui}|
\end{equation}

\begin{equation}
    RMSE=\sqrt{\frac{1}{|\tau|}\sum_{(u,i) \in \tau} (r_{ui}-\hat{r}_{ui})^2}
\end{equation}

where $\tau$ represents the test set, $r_{ui}$ the original rating that user $u$ has given to item $i$, and $\hat{r}_{ui}$ is the predicted rating. Values closer to $0$ are considered better, however obtaining low error values does not necessary indicate that users will be satisfied with the results of recommendations. This is due to the fact that ratings
are not missing at random which is commonly assumed by systems. Missing at random in the context of recommender systems means the probability of a rating to be
missing does not depend on its value. If data are missing at random, using only the observed data for statistical analysis yields ’correct’ results (e.g., unbiased maximum-likelihood
parameter estimates, and predictions) \cite{little2019statistical,rubin1976inference}. In contrast, if the data are missing not at random (MNAR), the missing data mechanism cannot be ignored in general, and has to be modeled precisely as to obtain correct results \cite{little2019statistical,rubin1976inference}. In other words, if the missing data mechanism is ignored, the resulting recommendation accuracy may be degraded significantly. Also, in many systems, explicit ratings are not available.

\subsection{Ranking }

\begin{figure*}
    \centering
    \includegraphics[width=2.5in]{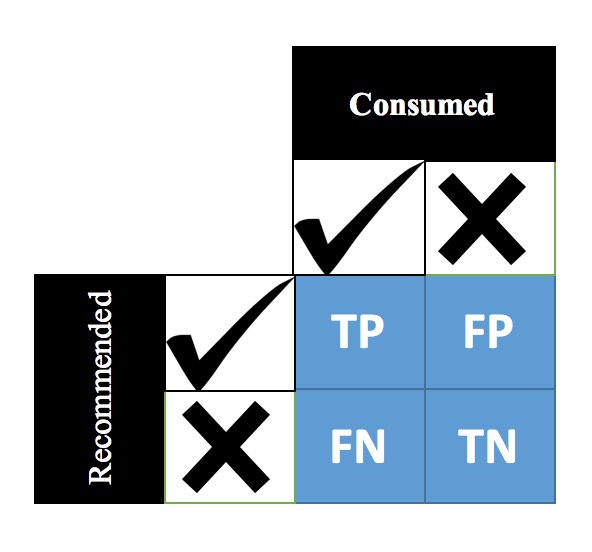}
    \caption{Contingency Table}
    \label{fig:contingency}
\end{figure*}

In the ranking task, instead of predicting how a user would rate an item, the
recommender system is asked to select items that it thinks are relevant to the user and recommend them to the user.
Commonly the selection of top-N relevant recommendations is considered. Formally
the task can be expressed as follows: given a set of items, the system is tasked to
create a ranked list $\ell_u$ for each user $u$. The ranking is determined by scores $\hat{r}_{ui}$
calculated by a recommendation system. Then scores are sorted in a descending
order, and the top-N items are kept.

Figure ~\ref{fig:contingency} shows a typical contingency table for a recommendation task. Given a list of recommendation $\ell$ given to the user, each element in the contingency table are interpreted as follows:

\begin{itemize}
    \item \textbf{True Positive (TP)}: When a recommended item is consumed by the user.
    \item \textbf{False Positive (FP)}: When a recommended item is not consumed by the user.
    \item \textbf{True Negative (TN)}: When an item is not recommended to the user and it is not in user's profile either. 
    \item \textbf{False Negative (FN)}: When an item is not recommended to the user but it is in the user's profile. 
\end{itemize}

Common metrics used to evaluate the ranking task are precision, recall, and Normalised
Discount Cumulative Gain (NDCG) \cite{jarvelin2017ir} of recommendations,
which can be defined as:

\begin{equation}
    Precision=\frac{TP}{TP+FP}
\end{equation}

Which can be calculated as:
\begin{equation}
    Precision(\ell_u)=\frac{\sum_{i \in \ell_u} \mathbbm{1}(i \in \rho_u)}{|\ell_u|}
\end{equation}
where $\mathbbm{1}(.)$ is an indicator function which returns 1 if its argument is true and 0 otherwise. 
Similarly,
\begin{equation}
    Recall=\frac{TP}{TP+FN}
\end{equation}
Which can be calculated as:
\begin{equation}
    Recall(\ell_u)=\frac{\sum_{i \in \ell_u} \mathbbm{1}(i \in \rho_u)}{|\rho_u|}
\end{equation}

If the order of the correct recommendations matters in the list (in contrast to precision and recall where the order is not evaluated), then we can use NDCG which is defined as follows:

\begin{equation}
    NDCG(\ell_u)=\frac{1}{iDCG}\sum_{i \in \ell_u} \frac{2^{\mathbbm{1}(i \in \rho_u)}-1}{\log_2(K_i+1)}
\end{equation}

\noindent where $k_i$ is the position of the item $i$ in the list and $iDCG$ is the normalisation factor set as the maximum possible value of $NDCG$ for an
ideal ranking and it can be measured as:
\begin{equation}
    iDCG(\ell_u)=\sum_{k=1}^{min(|\ell_u|,|\rho_u|)}\frac{1}{\log_2(k+1)}
\end{equation}

Typically the mean of these values over all test users is reported as the overall system performance.

Although a ranked list of recommendations can be generated via predicting a score for each item and pick the ones with the highest scores, it has been found
that better ranking performance can be obtained with algorithms that address the ranking
task directly. Such methods are known as learning-to-rank techniques \cite{karatzoglou2013learning}, which originated from the machine learning and information retrieval fields, as an alternative to other tasks such as regression and classification that have been commonly used to create rankings. Learning-to-rank methods can be split into
three main categories, depending on the way the ranking model is learned. These
are: pointwise \cite{koren2008factorization}, pairwise \cite{jahrer2012collaborative,takacs2012alternating}, and listwise \cite{shi2010list}. The pointwise ranking models are simple models where item ranks are
independent of other items and the item’s feedback—like ratings—indicates the
position in the ranked list—e.g. items rated as 5 should be higher in the list than
items rated as 4. In contrast, pairwise and listwise methods consider relationships
between items to learn the ranking model. In the pairwise ranking, pairs of
items are compared, and the relationship between them defines their position. For
example, if most of the times, item $A$ has higher utility than item $B$, then item $A$
should be recommended higher in the list. Finally, the listwise ranking model tries
to learn relationships between the entire set of items for which a user expressed her
or his preferences.

\subsection{Non-user Evaluation}

User-oriented evaluation of recommender systems is the main type that is commonly
considered. While it is important to measure user satisfaction, it is not the only
perspective on a system evaluation. For example, businesses do not only deploy recommender
systems to improve users’ satisfaction, but also to optimise sales, e.g. by promoting
less popular items, or items on which the business profits more. These are all important system
performance metrics that should also be considered while evaluating a recommender
system.

One of the common metrics for measuring how the recommender system performs with respect to covering different items from the catalog is \textit{Aggregate Diversity} (aka catalog coverage) \cite{adomavicius2014optimization}. 
Aggregate diversity measures the ratio of unique items out of all the available items that the recommender system was able to recommend to different users. This is an important metric as many recommendation algorithms often are biased towards popular items \cite{celma2008hits} and therefore they have low coverage. Businesses want to make sure the recommendation algorithm can recommend different items to their desired audience. Aggregate diversity of a recommender system is defined as follows:

\begin{equation}\label{agg_diver}
    Agg\mbox{-}{Div}=\frac{\left| \bigcup_{u \in U}\ell_u\right|}{|I|}
\end{equation}

Aggregate diversity counts a recommended item even if it is only recommended once. Therefore it is possible for an algorithm to have a high aggregate diversity while some of the items are recommended only few times. Another metric that gives a better sense of the distributional properties of the recommendations is \textit{Gini Index} which measures how evenly the recommendations are distributed across different items. In an ideal situation where every items are recommended equally, the Gini index is zero. In contrast, an algorithm with extreme inequality of recommendation frequency for different items will have a Gini index of 1. More formally, the Gini index is defined as follows:
    
     \begin{equation}
        Gini(L)=1-\frac{1}{|I|-1} \sum_{j=1}^{|I|}(2j-|I|-1)p(i_j|L)
    \end{equation}
    \noindent where $p(i|L)$ is the ratio of occurrence of item $i$ is $L$.
    
In addition to the metrics I described in this section, several additional metrics for evaluating other aspects of recommendation related to multi-stakeholder paradigm will be presented and fully discussed in Chapter \ref{data_method}.

\section{Challenges in Recommendation}

While many proposed recommendation solutions have been proven to be
successful on the task of recommending items to users, there are still many open
issues that worsen the experience to some users, or limits the usefulness of such
systems. I am going to depict and discuss a few of them now.

In any machine learning system it is generally true that the more data we have,
the more accurate models we can learn. The same applies to recommender systems
—if we have more user-item interactions, the more we know about what users like
and who likes certain items, the more accurate recommendations we can make.
However when we have a new user joining the platform, or a new item is added, we
know very little about them which makes the recommendation task much harder or technically impossible and of higher risk. This type of a problem is commonly known as the cold start
problem \cite{kluver2014evaluating}. Generally, content-based models are less prone to suffer from the cold start problem since the only information needed is some content information about the new item or some features about the new user. However, collaborative filtering methods cannot generate personalized recommendations for a completely new user or recommend a new item to the users in a personalized way since no previous interaction data exist for those type of users or items. Nevertheless, they can still generate recommendations that are non-personalized such as randomly recommending items or recommending the most popular items. 

In addition, certain types of recommender systems — mostly content-based methods, are known to be affected by the filter bubble, or over-specialisation,
where recommendations are similar to what the user already knows and this causes the user to be stuck in a bubble. While this might be seen as a good indicator of the ability of the recommender system to identify user interests and provide personalized recommendations, such recommendations may not be engaging and would not help the user discover new content. Similarly,
recommending lists of similar items, even if relevant, may not provide the value that
the users are looking for, and more diverse recommendations are rather expected \cite{zhang2008avoiding,ziegler2005improving}.

Finally, many online systems have been known to suffer from the so-called the long-tail
effect, or the popularity bias \cite{anderson2006long}. Most of the interactions that are recorded by a system happen to refer to a relatively small set of items, commonly known as
the short-head, and the rest is called the long-tail. Due to such bias in the interactions towards
the head, many collaborative filtering techniques are prone to also recommend these
popular items, making them even more popular \cite{zhao2013opinion}. This situation is
not desirable from both user and business perspectives—popular items are likely to
be already known by the users but also of little utility as they are less surprising or
engaging; where for businesses, recommendations focusing on most popular items
lead to poor utilization of the whole catalogue of items.

\subsection{Diversity, Novelty and Long-tail}

Traditionally, the focus of recommender systems has been put on recommending as
many relevant items as possible and maximising the overall accuracy of a system.
While high accuracy is desirable, Herlocker et al. \cite{herlocker2004evaluating} stated that high accuracy on its own may not be enough to provide useful and engaging recommendations, and that other properties should be simultaneously considered. In work of Kaminskas and Bridge \cite{kaminskas2016diversity} and McNee et al. \cite{mcnee2006being}, number of properties have been identified, such as novelty, coverage, and diversity.

Starting from novelty, looking at the Merriam Webster dictionary, it is defined as
‘the quality or state of being new, different, and interesting’\footnote{http://www.merriam-webster.com/dictionary/novelty}. In the context of recommendations, novelty typically refers to two aspects — an item being unknown
to the user, and an item being different from what the user has seen before \cite{kaminskas2016diversity}. The former of the above two reflects the global, system utility of recommendations, and the latter user-level novelty. Simply put, the global aspect mainly describes whether a recommended item is known to the user or not. Typically, popular items are already known to the users and therefore recommending them would result in low novelty. 

Diversity refers to the fact that the items in a given recommendation list is better to be different from each other so the user has enough choice \cite{kaminskas2016diversity,noveltydiversityvargas}. For instance, in a movie recommendation, a list containing Scream1, Scream2 and Scream3 is not a good recommendation even if the user is interested in horror movies since there is not enough choice in the list for the user and it would have been better if the list contained a diverse set of horror movies.

The main property that I focus on in this dissertation is the novelty of the recommendations. Popular items have low novelty to the users while less popular, long-tail items are considered to be more novel as the users are less likely to know about them. In Chapter \ref{pop_bias} I will describe the long-tail and popularity bias in detail.

\chapter{Multi-stakeholder Recommendation}\label{ms_rs}
\section{Introduction}
\label{sec:introduction}
\noindent  Recommender systems provide personalized information access, supporting e-commerce, social media, news, and other applications where the volume of content would otherwise be overwhelming. They have become indispensable features of the Internet age, found in systems of many kinds. One of the defining characteristics of recommender systems is personalization. In research contexts, recommender systems are typically evaluated on their ability to provide items that satisfy the needs and interests of the end user. Such focus is entirely appropriate. Users would not make use of recommender systems if they believed such systems were not providing items that matched their interests. Still, it is also clear that, in many recommendation domains, the user for whom recommendations are generated is not the only stakeholder in the recommendation outcome. Other users, the providers of products, and even the system's own objectives may need to be considered when these perspectives differ from those of end users. 

In many practical settings, such as e-commerce, recommendation is viewed as a form of marketing and, as such, the economic considerations of the retailer will also enter into the recommendation function \cite{leavitt2006recommendation,pathak2010empirical}. A business may wish to highlight products that are more profitable or that are currently on sale, for example. More recently, system-level objectives such as fairness and balance have been considered by researchers, and these social-welfare-oriented goals may at times run counter to individual preferences. Sole focus on the end user hampers researchers' ability to incorporate such objectives into recommendation algorithms and system designs. 

I believe that, far from being special ``edge cases", these examples illustrate a more general point about recommendation, namely, that recommender systems often serve multiple goals and that the purely user-centered approach found in most academic research does not allow all such goals to enter into their design and evaluation. What is needed is an inclusive approach that expands outward from the user to include the perspectives and utilities of multiple stakeholders.

It is relevant to note a shift that occurred in microeconomics in the early part of the 21st century with the development of the theory of multi-sided platforms~\cite{rochet2003platform,evans_platform_2011}. Prior to that time, the traditional business model focused on a firm's ability to produce products and deliver them to customers at a price that could ensure profitability. By contrast, multi-sided platforms create value by bringing buyers and sellers together, reducing search and transaction costs. Many online systems are exactly such multi-sided platforms ~\cite{evans_matchmakers:_2016}. 

As noted above, when it comes to the study of personalized information access in the form of recommender systems, academic research has, with few exceptions, examined only a single side of these interactions. The stage was set historically by the first recommender systems implementations, which either operated on objects with no associated price (newsgroup posts~\cite{konstan1997grouplens}) or were external to any commerce associated with their recommendations (such as non-commercial music, movie, and restaurant recommenders~\cite{shardanand1995social,breese1998empirical,burke1997findme}). These systems brought users and products together, but they were not themselves party to any transactions. While academic research has largely concentrated on the user, commercial systems have regularly taken a broader view of recommendation objectives \cite{rodriguez2012multiple,nguyen2017multi}. There is, therefore, a gap between the complexity of real-world applications of recommender systems and those on which academic research has focused. 

The integration of the perspectives of multiple parties into the design of recommender systems is the goal underlying the sub-field of \textit{multi-stakeholder recommendation}~\cite{abdollahpourimultistakeholder2020,abdollahpouri_recommender_2017,soappaper,nguyen2017multi} which I first introduced in \cite{soappaper} and I describe in detail in this dissertation. This chapter is intended to describe the current state of the art in multi-stakeholder recommendation research and to outline research questions that should be addressed to support the demands of recommendation applications in environments where the perspectives of multiple parties are important.

\section{Recommendation Stakeholders and Multi-stakeholder Paradigm }
\label{sec:MS_recommendation}
The concept of a stakeholder appears in business management literature as a way to discuss the complexities of corporate governance. According to \cite{goodpaster1991business}, the term `stakeholder' appears to have been invented in the early '60s as a deliberate play on the word `stockholder' to signify that there are other parties having a `stake' in the decision-making of the modern, publicly-held corporation in addition to those holding equity positions.

In his classic work, \textit{Strategic Management: A Stakeholder Perspective}, Freeman extends older definitions that emphasize a ``stake'' as a kind of investment, and instead defines stakeholders as ``any groups or individuals that can affect, or are affected by, the firm's objectives'' \cite{freeman2010strategic}, pg. 25. I adopt this definition for our aims, focusing specifically on recommender systems.

\begin{definition}{}
A \textit{recommendation stakeholder} is any group or individual that can affect, or is affected by, the delivery of recommendations to users.
\end{definition}

As recommender systems are elements of an organization's operations, they will necessarily inherit the large and wide-ranging set of stakeholders considered in the management literature. However, only some of these stakeholders will be particularly salient in the generation of recommendations. In this dissertation, I will consider three key groups of stakeholders who are particularly close to the recommendation interaction: 

\begin{description}
    \item[\textbf{Consumers (aka users)}:] The consumers are the end-users who receive / consume recommendations. They are the individuals whose choice or search problems bring them to the platform, and who expect recommendations to satisfy those needs.  
    \item [\textbf{Providers (aka suppliers)}:] The (item) suppliers are those entities that supply or otherwise stand behind the recommended objects. 
    \item [\textbf{System}:] The final category is the organization itself, which has created a platform and associated recommender system in order to match consumers with suppliers. The platform may be a retailer, e-commerce site, broker, or other venue where consumers seek recommendations.
\end{description}

\begin{figure}
    \centering
    \SetFigLayout{3}{2}
 \subfigure[Traditional View of Recommendation]{\includegraphics[width=5in]{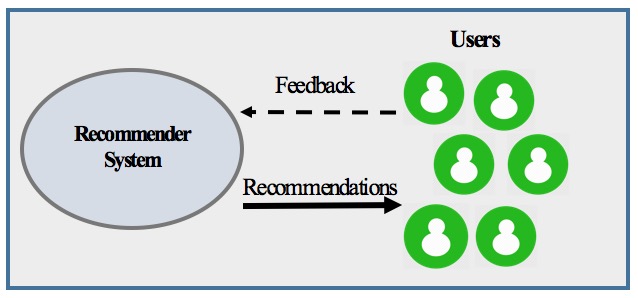}\label{fig:recs1}}
 \subfigure[Multi-stakeholder View of Recommendation]{\includegraphics[width=5in]{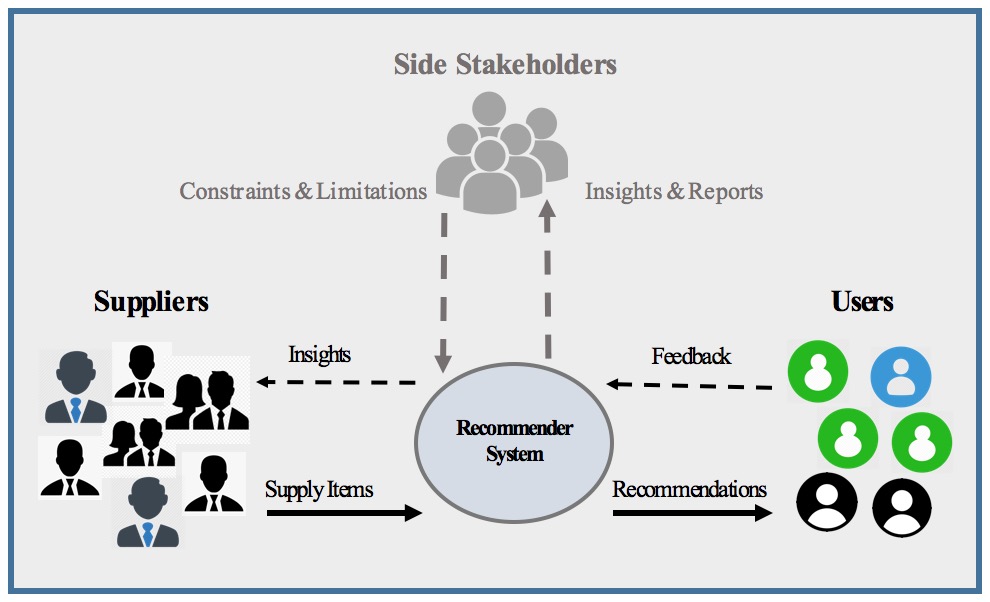}\label{fig:recs3}}

 \caption{Traditional schema for recommendation versus multi-stakeholder schema.}
\label{schema}
\end{figure}

Of course, none of these are stakeholder groups are unitary entities --- not even the system, which stands in for various internal groups within an organization who may have different, possibly competing demands on a recommender system. Differentiation among stakeholder groups may be necessary, depending on the application. This taxonomy does, however, highlight an important consequence of a multi-stakeholder perspective, namely the foregrounding of the multi-sided nature of recommendation, which has been slow to emerge in the research literature.

Figure \ref{schema} contrasts the traditional view of recommendations versus the multi-stakeholder one. Traditionally, the system is assumed to generate a set of recommendations to the users and its performance is measured from the users' perspective with an average value across all users. In other words the differences between users and groups of users are often omitted from the evaluation. However, even if we only focus on users as the sole stakeholders, they are not a single entity but rather different individuals with certain characteristics who might belong to different groups often protected by law (e.g. gender, race etc. ). In addition, the multi-stakeholder schema considers the suppliers of the items as another important party in the recommendation process whose needs and satisfaction should be taken into account. The recommender system might even give insights to the suppliers based on how the users interact with their items and the feedback they receive from the users. On top of that, some \textit{side stakeholders} may also impose certain constraints and limitations on the recommendations. For example, on a music recommendation system, the society or law regulations may also be a considered as stakeholders to enforce certain human values in the recommendations such as recommending local versus international music in a fair way or making sure female artists are fairly represented in the recommendations given to different users. Of course, the term \textit{fair} could be interpreted in many different ways \cite{narayanan2018translation} but often there are certain law regulations that businesses need to follow \cite{hacker2018teaching}.

The multi-stakeholder perspective on recommendation can manifest itself in various aspects of recommender systems research and development. We may adopt a multi-stakeholder view of evaluation: asking the question of how different groups of stakeholders are impacted by the recommendations a system computes. A developer may employ the multi-stakeholder perspective in designing and optimizing an algorithm, incorporating the objectives of multiple parties and balancing among them. Finally, the creation of the recommender itself may be a multi-stakeholder endeavor, in which different stakeholders participate in developing a design.

\subsection{Multi-stakeholder Evaluation}

In research settings, recommender systems traditionally are evaluated from the users' perspective. Metrics such as precision, recall, NDCG, diversity, novelty, etc. all capture different aspects of recommendation quality as experienced by the end user. In on-line testing, click-through rate, dwell time, and other interaction metrics capture similar types of outcomes. These metrics are typically averaged over all users to produce a single outcome. This methodology in entirely reasonable and logical as, in the end, users are one of the most important stakeholders of any recommender system. We will take it as a given that any recommender system evaluation will include an outcome of this type, the \textit{user summary} evaluation.

A multi-stakeholder perspective on evaluation, however, brings to light additional aspects of system performance that may be quite important. As mentioned above, multi-sided platforms such as eBay, Etsy, or AirBnB, have a key business requirement of attracting and satisfying the needs of providers as well as users. A system of this type will need to evaluate its performance from the provider perspective.

As mentioned above, even when only a single group of stakeholders is under consideration, a methodology that relies on a single point estimate of system performance under some metric may miss differences among stakeholder groups, even among users, who are typically the target of evaluation. Stakeholder theory recognizes that subgroups within the stakeholder categories may experience a range of different impacts from a firm's decisions~\cite{freeman2010strategic}. In recommendation, recent work has shown that, depending on the algorithm, male and female users may experience different outcomes in movie recommendations~\cite{ekstrand2018all}. Therefore, a multi-stakeholder evaluation is one that includes different stakeholder groups in its assessment, in addition to the user summary evaluation.

\begin{definition}{}
A \textit{multi-stakeholder evaluation} is one in which the quality of recommendations is assessed across multiple groups of stakeholders, in addition to a point estimate over the full user population.
\end{definition}

A multi-stakeholder evaluation may entail the use of different kinds of metrics and different evaluation methodologies than typically encountered in recommender systems research. For example, typical cross-validation methodologies that distribute user profiles between test and training sets may not yield reliable results when assessing outcome for other stakeholders, especially providers. We will take up the issue of provider-side metrics in Section~\ref{sec:provider-metric}.

\subsection{Multi-stakeholder Algorithms}
In implementing a given recommender system, a developer may choose to use the metrics associated with multi-stakeholder evaluation in algorithm design, implementation, and optimization. In general, this will entail balancing the objectives of different stakeholders, as it is unlikely that the optimal solution for one will be the best for all. Some solutions may combine all such objectives into a single optimization, a  challenging approach given the methodological complexities noted above; others use a multi-stage approach incorporating different stakeholder concerns throughout a pipeline. Both approaches are discussed in the examples in Chapter \ref{tackle_pop_bias}.

Multi-stakeholder algorithms are particularly differentiated from typical recommendation approaches when the stakeholder concerns lie on different sides of the recommendation platform. For example, it is not difficult to change the loss function associated with a factorization algorithm to prefer balanced outcomes over multiple subgroups of users, rather than a simple overall mean. However, system or provider objectives may be orthogonal to user concerns and form a separate optimization problem that cannot be simply combined with the users'. I therefore define a  multi-stakeholder recommendation algorithm with special attention to this subclass.

\begin{definition}{}
A \textit{multi-stakeholder recommendation algorithm} takes into account the preferences of multiple parties when generating recommendations, especially when these parties are on different sides of the recommendation interaction.
\end{definition}

\subsection{Multi-stakeholder Design}
Beyond implementing metrics and tuning algorithms, any fielded recommender system will also go through phases of design in which the system's role within a particular platform and its specific requirements are formulated. System designers may choose to engage in a design process such as participatory design~\cite{kensing1998participatory} in which external stakeholders are incorporated into decision making. Although these techniques are well-established in the HCI community, they have not seen much discussion in recommender systems research. Anecdotal information suggests that commercial platforms with multiple stakeholders do engage in stakeholder consultation with item providers in particular~\cite{semerci2019homepage}.

\begin{definition}{}
A \textit{multi-stakeholder design process} is one in which different recommendation stakeholder groups are identified and consulted in the process of system design.
\end{definition}

\subsection{The Landscape of Multi-stakeholder Recommendation}

The above list of the key recommendation stakeholders provides an outline for understanding and surveying the different types of multi-stakeholder recommendation. We can conceptualize a recommender system as designed to maximize some combined utility associated with the different stakeholders, and consider how different types of applications yield different stakeholder concerns.

\subsubsection{Consumer-side issues}
If we concentrate only on the individuals consuming recommendations, multi-stakeholder issues arise when there are tradeoffs or disparities between groups of users in the provision of recommendations. For example, \cite{ekstrand2018all} explored the performance of recommender system algorithms on users belonging to different demographic groups (gender, age) and observed that some algorithms perform better for certain groups than others. Other researchers have found that users with niche or unusual tastes can be poorly served by particular recommendation algorithms, lapses that are not de\-tect\-able from point estimates of accuracy measures~\cite{abdollahpouriRMSE2019,abdollahpouri2020multi,ghazanfar2014leveraging}.

In some settings, such differences in system performance for different users may be considered examples of unfair treatment. For example, in Chapter \ref{tackle_pop_bias} we will see that the recommendation algorithms perform different for users with different levels of tendency towards popular items. In particular, we will see that the recommendations for users who have lesser tendency towards popular items have a higher deviation from what they expected to receive from the recommender system.

\subsubsection{Supplier-side issues}
As noted above, multi-sided platforms need to satisfy both the consumers of recommendations and the suppliers of items that are being recommended. The health of such a platform depends both on a user community and a catalog of items of interest. Suppliers whose items are not recommended may experience poor engagement from users and lose interest in participating in a given platform. Platforms that facilitate peer-to-peer interactions, such as the online craft marketplace Etsy, may be particularly sensitive to the need to attract and retain sellers. 

Depending on the application, suppliers may have particular audiences in mind as appropriate targets for their items. A well-known example is computational advertising in which advertisers seek particular target markets and ad platforms try to personalize ad presentation to match user interests~\cite{internetadvertisingyuan}. In this application, market forces, expressed through auction mechanisms, serve to mediate between supplier interests. In other cases, such as online dating, preferences may be expressed on both sides of the interaction but it is up to the recommender system to perform appropriate allocation.

\subsubsection{System/platform issues}
In many real-world contexts, the system may gain some utility when recommending items, and is therefore a stakeholder in its own right. Presumably, an organization creates and operates a recommender system in order to fulfill some business function and generally that is to enhance user experience, lower search costs, increase convenience, etc. This would suggest a consumer stakeholder point of view is sufficient.

However, there are cases in which additional system considerations are relevant, as noted above, and internal stakeholders have an impact on how a recommender system is designed. For example, in an e-commerce platform, the profit of each recommended item may be a factor in ordering and presenting recommendation results. This marketing function of recommender systems was apparent from the start in commercial applications, but rarely included as an element of research systems. 

Alternatively, the system may seek to tailor outcomes specifically to achieve particular objectives that are separate from either provider or consumer concerns. For example, an educational site may view the recommendation of learning activities as a curricular decision and seek to have its recommendations fit a model of student growth and development. Its utility may, therefore, be more complex than a simple aggregation of those of the other stakeholders. 

\subsubsection{Other stakeholders}

Complex online ecosystems may involve a number of stakeholders directly impacted by recommendation delivery beyond item suppliers and recommendation consumers. For example, an online food delivery service, such as UberEats, depends on delivery drivers to transport meals from restaurants to diners. Drivers are affected by recommendations delivered to users as the efficiency of routing and the distribution of order load will be a function of which restaurants receive such orders. 

\section{Related Research}

\noindent Multi-stakeholder recommendation brings together research in a number of areas within the recommender systems community and beyond: (1) in economics, the areas of multi-sided platforms and fair division; (2) the growing interest in multiple objectives for recommender systems, including such concerns as fairness, diversity, and novelty; and, (3) the application of personalization to matching problems.

\subsection{Economic foundations}

\noindent The study of the multi-sided business model was crystallized in the work of \cite{rochet2003platform} on what they termed ``two-sided markets.'' Economists now recognize that such contexts are often multi-sided, rather than two-sided, and that ``multi-sidedness'' is a property of particular business platforms, rather than a market as a whole~\cite{evans_matchmakers:_2016}. 

Many of today's recommender systems are embedded in multi-sided platforms and hence require a multi-sided approach. The business models of multi-sided platforms are quite diverse, which means it is difficult to generalize about multi-stakeholder recommendation as well. A key element of the success of a multi-sided platform is the ability to attract and retain participants from all sides of the business, and therefore developers of such platforms must model and evaluate the utility of the system for all such stakeholders. 

The theory of just division of resources has a long intellectual tradition going back to Aristotle's well-known dictum that ``Equals should be treated equally.'' Economists have invested significant effort into understanding and operationalizing this concept and other related ideas. See~\cite{moulin2004fair} for a survey. In recommendation and personalization, we find ourselves on the other side of Aristotle's formulation: all users are assumed unequal and unequal treatment is the goal, but we expect this treatment to be consistent with diverse individual preferences. Some aspects of this problem have been studied under the subject of social choice theory~\cite{arrow2010handbook}. However, there is not a straightforward adaptation of these classical economic ideas to recommendation applications as the preferences of users may interact only indirectly and in subtle ways. For example, if a music player recommends a hit song to user A, this will not in any way impact its desirability or availability to user B. On the other hand, if a job recommender system recommends an appealing job to user A, it may well have an impact on the utility of the same recommendation to user B who could potentially face an increased competitive environment if she seeks the same position.

\subsection{Multi-objective recommendation}
Multi-stakeholder recommendation is an extension of recent efforts to expand the considerations involved in recommender system design and evaluation beyond simple measurements of accuracy. There is a large body of recent work on incorporating diversity, novelty, long tail promotion and other considerations as additional objectives for recommendation generation and evaluation. See, for example, \cite{abdollahpouri2017controlling,diversitySmyth,diversityziegler,noveltydiversityvargas,jannach2016recommendations,abdollahpouri2019managing}. There is also a growing body of work on combining multiple objectives using constraint optimization techniques, including linear programming. See, for example, \cite{jambor_optimizing_2010,agarwal_click_2011,svore_learning_2011,rodriguez_multiple_2012,jiang_optimization_2012,agarwal_personalized_2012}. These techniques provide a way to limit the expected loss on one metric (typically accuracy) while optimizing for another, such as diversity. The complexity of these approaches increases exponentially as more constraints are considered, making them a poor fit for the general multi-stakeholder case. Also, for the most part, multi-objective recommendation research concentrates on maximizing multiple objectives for a single stakeholder, the end user. 

Another area of recommendation that explicitly takes a multi-objective perspective is the area of health and lifestyle recommendation. Multiple objectives arise in this area because users' short-term preferences and their long-term well-being may be in conflict~\cite{lin2011motivate,ponce2015quefaire}. In such systems, it is important not to recommend items that are too distant from the user's preferences -- even if they would maximize health. The goal to be persuasive requires that the user's immediate context and preferences be honored.

Fairness is an example of a system consideration that lies outside the strict optimization of an individual user's personalized results. Therefore, recent research efforts on fairness in recommendation are also relevant to this work~\cite{lee2014fairness,burke_multisided_2017,DBLP:conf/recsys/KamishimaAAS14,kamisha-akaho-fatrec-2017,yao_huang_fatml-2017,mehrotra2018towards}. The multi-stakeholder framework provides a natural ``home'' for such system-level considerations, which are otherwise difficult to integrate into recommendation. See \cite{abdollahpourimultistakeholder2020} for a more in-depth discussion.

\subsection{Personalization for matching} 

\noindent The concept of multiple stakeholders in recommender systems is suggested in a number of prior research works that combine personalization and matching. The earliest work on two-sided matching problems~\cite{roth1992two} assumes two sets of individuals, each of which has a preference ordering over possible matches with the other set. The task to make a stable assignment has been shown to have an $O(n^2)$ solution. This formulation has some similarities to reciprocal recommendation. However, it assumes that all assignments are made at the same time, and that all matchings are exclusive. These conditions are rarely met in recommendation contexts, although extensions to this problem formulation have been developed that relax some of these assumptions in online advertising contexts~\cite{bateni2016fair}.

Researchers on reciprocal recommendation have looked at bi-lateral considerations to ensure that a recommendation is acceptable to both parties in the transaction. A classical example is on-line dating in which both parties must be interested in order for a match to be successful~\cite{reciprocal,reciprocaldating}. Other reciprocal recommendation domains include job seeking~\cite{rodriguez_multiple_2012}, peer-to-peer ``sharing economy'' recommendation (such as AirBnB, Uber and others), on-line advertising \cite{targetadvertisingbiding}, and scientific collaboration~\cite{lopes2010collaboration,tang2012cross}. See \cite{abdollahpourimultistakeholder2020} for a detailed discussion. 

The field of computational advertising has given considerable attention to balancing personalization with multi-stakeholder concerns. Auctions, a well-es\-tab\-lished technique for balancing the concerns of multiple agents in a competitive environment, are widely used both for search and display advertising \cite{internetadvertisingyuan,mehta2007adwords}. However, the real-time nature of these applications and the huge potential user base makes recommender-style personalization computationally intractable in most cases.

\section{Evaluation}

At this point in the development of multi-stakeholder recommendation research, there is a diversity of  methodological approaches and little agreement on basic questions of evaluation. In part, this is a reflection of the diversity of problems that fall under the multi-stakeholder umbrella. Multi-stakeholder evaluation and algorithm development do not always use the same methodologies. 

A key difficulty is the limited availability of real-world data with multi-stakeholder characteristics. The reason is these type of data are highly business-critical and often including private information such as margins associated with each supplier and the commissions negotiated by the platform. Close collaboration is required to obtain such sensitive proprietary data. Some researchers have built such collaborations for multi-stakeholder research, but progress in the field requires replicable experiments that proprietary data does not support. Areas of multi-stakeholder research that involve public, rather than private, benefit may offer advantages in terms of the availability of data: see, for example, the data sets available from the crowd-funded educational charity site DonorsChoose.org\footnote{https://data.donorschoose.org/explore-our-impact/}.

\subsection{Simulation}

In the absence of real-world data with associated valuations, researchers have typically turned to simulations. Simulated or inferred supplier data is useful for transforming publicly-available recommendation data sets in standard user, item, rating format into ones that can be used for multi-stakeholder experimentation. The experiments in \cite{surer2018multistakeholder} provide an example of this methodology: each item in the MovieLens 1M data set was assigned to a random supplier, and the distribution of utilities calculated. To capture different market conditions, the experimenters use two different probability distributions: normal and power-law. There is no accepted standard for producing such simulations and what are reasonable assumptions regarding the distribution of provider utilities or the formulation of system utilities, except in special cases.  

Researchers have also used objective aspects of data sets to infer proxy attributes for multi-stakeholder evaluation. In \cite{soappaper}, the first organization listed in the production credits for each movie was treated as the provider -- a significant over-simplification of what is a very complex system of revenue distribution in the movie industry. In other work, global metrics such as network centrality \cite{ValuePick2010} have been used to represent system utility for the purposes of multi-stakeholder evaluation. \cite{burke2018synthetic} demonstrated an alternate approach to generate synthetic attribute data based on behavioral characteristics that can be used to evaluate system-level fairness properties.

More sophisticated treatments of profitability and recommendation are to be found in the management and e-commerce literature, some using public data as seen in  \cite{Oestreicher-Singer:2012:VHD:2398302.2398303,ChenWuYoon:04a,DBLP:conf/icis/AdamopoulosT15}, but these techniques and associated data sets have not yet achieved wide usage in the recommender system community. 

\subsection{Models of utility}
A multi-stakeholder framework inherently involves the comparison of outcomes across different groups of individuals that receive different kinds of benefits from the system. In economic terms, this entails utility calculation and comparison. As with data, different researchers have made different assumptions about what types of utilities accrue from a recommender system and how they are to be measured. A standard assumption is that the output of a recommendation algorithm can be treated as an approximation of user utility. Yet, research has confirmed that users prefer diverse recommendation lists \cite{pu2011user}, a factor in tension with accuracy-based estimates of user utility. 

These are, of course, more concentrated on the short-term definitions of utility and things that can be measured from the users feedback in a single run of recommendation generation. More research is therefore required to understand the potential positive and negative long-term effects of profit-aware recommendation and other strategies that are not strictly user-focused. Future models could also consider the price sensitivity and willingness-to-pay of individual consumers in the recommendation process. 

\subsection{Off-line experiment design}
A standard off-line experimental design which I described in Chapter \ref{background_chapter} makes a bit less sense in a multi-stakeholder context, and this is where the essential asymmetry of the stakeholders comes into play. Suppliers are, in a key sense, passive -- they have to wait until users arrive at the system in order to have an opportunity to be recommended. The randomized cross-fold methodology measures what the system can do for each user, given a portion of their profile data, the potential benefit to be realized if the user visits and a recommendation list is generated. Evaluating the supplier side under the same conditions, while a commonly-used methodology, lacks a similar justification.

A more realistic methodology from the suppliers' point of view is a temporal one, that takes into account the history of the system up to a certain time point and examines how supplier utilities are realized in subsequent time periods. See \cite{Campos2014} for a comprehensive discussion of time-aware recommender systems evaluation. However, time-aware methods have their own difficulties, forcing the system to cope with cold-start issues possibly outside of the scope of a given project's aims. Therefore, for certain models of utility where the dynamics of suppliers (such as price, profit margin etc) may be of less important, offline evaluation can be still valuable. For example, our evaluation of the popularity bias in recommendation from the perspectives of different stakeholders is done in an offline setting as we will see in Chapter \ref{data_method} and \ref{pop_bias}.

\subsection{User studies}
User studies are another instrument available to researchers that has not been extensively applied to multi-stakeholder recommender systems. As usual for such studies, the development of reliable experimental designs is challenging as the participants' decision situation typically remains artificial. Furthermore, as in the study by Azaria et al. \cite{Azaria:2013:MRS:2507157.2507162}, familiarity biases might exist -- in their study participants were willing to pay more for movies that they already knew -- which have been observed for other types of user studies in the recommender systems domain \cite{JannachLercheEtAl2015a}. Ultimately, more field tests -- even though they are typically tied to a specific domain and specific business model -- are needed that give us more insights into the effects of multi-stakeholder recommendations in the real world.

\subsection{Metrics}
The building block of multi-stakeholder evaluation is the measurement of the utility each of the stakeholders gets within a recommendation platform. Common evaluation metrics such as RMSE, precision, NDCG, diversity, etc. are all different ways to evaluate the performance of a recommender system from the user's perspective. As noted above, these measures are implicitly a form of system utility measure as well: system designers optimize for such measures under the assumptions that (1) higher accuracy metrics correspond to higher user satisfaction and (2) higher user satisfaction contributes to higher system utility through customer retention, trust in the recommendation provided, etc. However, the formulation of multi-stakeholder recommendation makes it possible to characterize and evaluate system utility explicitly. 

Typically, evaluation metrics are averaged over all users to generate a point score indicating the central tendency over all users. However, it is also the case that in a multi-stakeholder environment additional aspects of the utility distribution may be of interest. For example, in an e-commerce context, suppliers who receive low utility may leave the eco-system, suggesting that the distribution of supplier utility may be important as well as the mean. One suggested practice would be to report results for significant user groups within their populations as I will discuss in Chapter \ref{data_method} and \ref{tackle_pop_bias}.

\subsubsection{Supplier metrics}
\label{sec:provider-metric}
When evaluating the utility of a recommender system for a particular supplier, we may take several different stances. One views the recommender as a way to garner user attention. In this case, the relevance of an item to a user may be a secondary consideration. Another perspective views the recommender as a source of potential leads. In this view, recommending an item to uninterested users is of little benefit to the supplier. In the first situation, simply counting (with or without a rank-based discount) the number of times a supplier's products appears in recommendation lists would be sufficient. In the second situation, the metric should count only those recommendations that were considered ``hits'', those that appear positively rated in the corresponding test data.

Another supplier consideration may be the reach of its recommended products across the user population. A metric could count the number of unique users to whom the supplier's items are recommended. Multi-stakeholder applications may differ in their ability to target specific audiences for their items. In a targeted system, it would make sense to consider reach relative to the target population. For example, in an online dating application where the user can specify desired properties in a match, an evaluation metric might be the fraction out of the target audience receiving the recommendation. 

Finally, where the consideration is the accuracy of system's predictions, we can create a supplier-specific summary statistic of a measure like RMSE. \cite{ekstrand2018exploring} uses this method to examine differences in error when recommending books by male and female authors. Since the statistic by itself is not that useful for a single provider, a better metric would indicate the provider's position relative to other providers in the overall distribution

\subsubsection{System metrics}
System utility may, in many cases, be a simple aggregation of the utilities of other parties. For example, in a simple commission-oriented arrangement, the profit to the system might be some weighted aggregate of the \textit{Hits} metric, taking item price and commission rate into account. However, other cases arise where the system has its own targeted utility framework.

An important such context is algorithmic fairness \cite{yao2017beyond,ekstrand2018exploring}. In general, we should not expect that suppliers will care if the system is fair to others as long as it provides them with good outcomes. Any fairness considerations and related metrics will therefore be ones defined by system considerations.

As we will see in Chapter \ref{pop_bias}, popularity bias in recommendation is a multi-stakeholder problem that impacts different sides of the recommendation platform including the users and the suppliers. It also might create challenges for the system itself if, for example, the system designer cares about giving certain amount of exposure to different items with different levels of popularity. In next chapter, I discuss our methodology for measuring the popularity bias from the perspective of different stakeholders: users, suppliers, and the system. 

\chapter{Data and Methodology}\label{data_method}
In this chapter we describe the details of the datasets we used for the experiments in this dissertation and several evaluation metrics that we introduce for measuring the multi-stakeholder aspect of the popularity bias.

\section{Datasets}
In Chapter \ref{ms_rs}, we highlighted the main stakeholders of a given recommender system: users, suppliers and the system itself.
To incorporate our analysis for the suppliers, we needed datasets where the item suppliers could be identified. We found two publicly available datasets for our experiments: the first one is a sample of the Last.fm (LFM-1b) dataset \cite{schedl2016lfm} used in \cite{dominik2019unfairness}. The dataset contains user interactions with songs (and the corresponding albums). We used the same methodology in \cite{dominik2019unfairness} to turn the interaction data into rating data using the frequency of the interactions with each item (more interactions with an item will result in higher rating). In addition, we used albums as the items to reduce the size and sparsity of the item dimension, therefore the recommendation task is to recommend albums to users. We considered the artists associated with each album as the supplier. We removed users with less than 20 ratings so only consider users for which we have enough data. The resulting dataset contains 274,707 ratings by 2,697 users to 6,006 albums. Total number of artists is 1,998. 
 
 The second dataset is the MovieLens 1M dataset \cite{harper2015movielens} \footnote{Our experiments showed similar results on MovieLens 20M, and so we continue to use MovieLens 1M for efficiency reasons.}. This dataset does not have the information about the suppliers. We considered the director of each movie as the supplier of that movie and we extracted that information from the IMDB API. Total number of ratings in the MovieLens 1M data is 1,000,209 given by 6,040 users to 3,706 movies. Overall, we were able to extract the director information for 3,043 movies reducing the ratings to 995,487. The total number of directors is 831. 
 
 For convenience, the characteristics of both datasets is shown in Table ~\ref{datasets}.
 \begin{table*}[t]
\centering
\large
\captionof{table}{Number of users $|U|$, number of items $|I|$, number of ratings $|R|$, and number of suppliers $|\mathcal{P}|$ in two datasets MovieLens and Last.fm} 
\label{datasets}
\begin{tabular}{lllll}
          & $|U|$   & $|I|$   & $|R|$     & $|\mathcal{P}|$   \\\midrule
MovieLens & 6,040 & 3,043 & 995,487 & 831   \\
Last.fm   & 2,697 & 6,006 & 274,707 & 1,998
\end{tabular}
\end{table*}

\section{Recommendation Algorithms}
As we mentioned in Chapter \ref{background_chapter}, there are different types of recommendation algorithms each of which have their own cons and pros. Some are neighborhood-based and some are based on matrix factorization methods. We also mentioned content-based models where the content information of the items (or users) is used to generate the recommendations. In this dissertation, our main focus is on collaborative filtering models where no content information is used in the algorithm. Both neighborhood-based and factorization-based models are considered collaborative filtering techniques since they use the rating (or interaction data) information given by the users to the items. We have used several well-known recommendation algorithms in the following chapters that cover both neighborhood-based models and also factorization-based approaches (either rating prediction-based or ranking-based).

\begin{itemize}
    \item \textbf{Alternative Least Square Ranking (RankALS) \cite{takacs2012alternating}}: This method is a pair-wise learning-to-rank technique that directly optimizes the order of the items in the recommendation lists without the need for calculating the predicted ratings for all items and sorting them. The idea is, if one item is ranked higher than another item in rating data, this order should be kept in the recommended list. The algorithm tries to minimize the pairwise ranking discrepancies between every pair of items in rating data. 
    \item \textbf{Biased Matrix Factorization (Biased-MF)} \cite{koren2009matrix}: This is another factorization-based technique but, unlike RankALS, it does not optimize for ranking but rather seeks to minimize the distance between the predicted rating and the original rating for any given <user,item> pair. Therefore, the optimization is done in the rating prediction step and once the predicted ratings for <user,item> pairs are calculated, they are sorted and the top $n$ items will be returned as the recommendations. 
    
    \item \textbf{User-based Collaborative Filtering (User-CF)} \cite{aggarwal2016neighborhood}: This algorithm is a neighborhood-based technique meaning the rating prediction is directly using the similarities between users. For any given user, the algorithm finds top most similar users and uses their rating information for items that the target user has not rated. In order to find the similarity between users, distance metrics such as Pearson Correlation or Cosine similarity are used.  
    
    \item \textbf{Item-based Collaborative Filtering (Item-CF)} \cite{sarwar2001item}: Similar to \textit{User-CF}, this algorithm is also neighborhood-based but, unlike \textit{User-CF} that leverages similarity between users, it uses the similarity between items that the user has rated before and other items. Again, the similarity calculation is done using distance metrics such as Pearson Correlation or Cosine similarity. The algorithm then recommends most similar items. 
\end{itemize}

\section{Stakeholder Grouping}\label{grouping}
Throughout this dissertation, for better illustration of certain patterns and behavior of different algorithms, I use grouping (either items, users, or suppliers) so I can discuss how each algorithm performs for items, users, or suppliers in different groups. Many researchers have used demographic information such as age, gender, race etc to define groups of users and to measure demographic parity between those groups \cite{ekstrand2018exploring,ekstrand2018all,alvarez2019promoting}. However, these works assume there is no intrinsic differences among individuals in different groups which is not supported by prior work where significant differences have been found \cite{egli2011influence,barnett2006boredom}. In this dissertation, our grouping strategy is consistent with the problem I am addressing. The problem I am concentrating on is popularity bias and its impact of different stakeholders and hence a natural way of grouping would have the popularity component in its consideration. 

For all stakeholder types (items, users, and suppliers), the grouping is based on the popularity of the individuals or their interest in popular items in each stakeholder type. In other words, items are grouped based on their popularity, users are grouped based on their interest towards popular items, and suppliers are grouped according to the average popularity of their items.  
Before going any further, it is important to define what exactly I mean by popularity.

\begin{definition}
\textbf{Item Popularity:} The popularity of an item is defined as the ratio of users who have interacted (or rated) that particular item. More formally:
\begin{equation}
    Pop(i)=\frac{\sum_{u \in U}\mathbbm{1}(i \in \rho_u)}{|U|}
\end{equation}
\noindent where $\rho_u$ is the list of rated (or interacted with) items by user $u$. 
\end{definition}
Therefore, the more users rate or interact with a certain item, the higher its popularity will be. In a sense, popularity of an item measures the amount of attention the items receives. 

Here is how the grouping is done for each stakeholder type:
\begin{figure*}
\centering
\SetFigLayout{3}{2}
 \subfigure[MovieLens]{\includegraphics[width=5.9in]{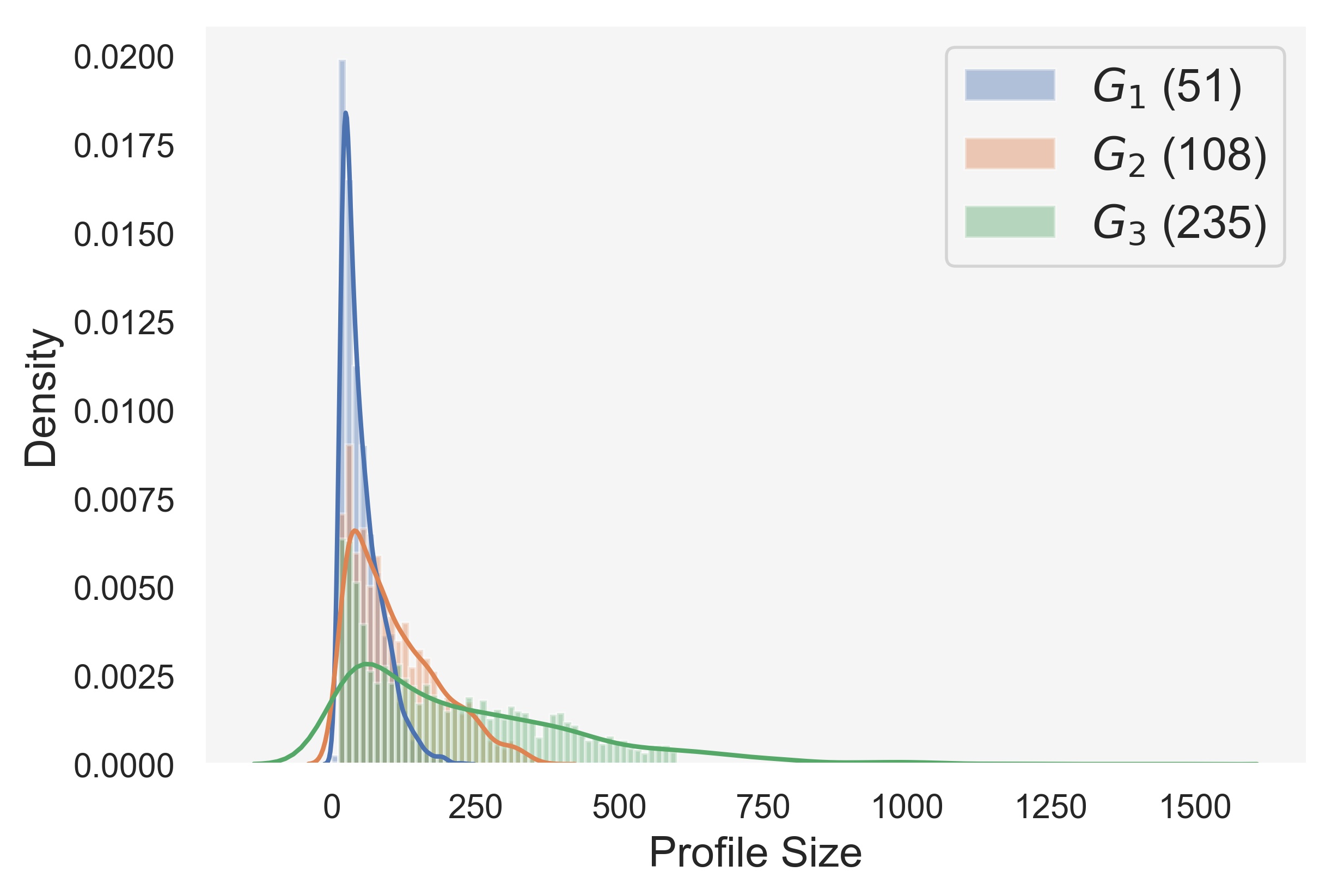}}
\subfigure[Last.fm]{\includegraphics[width=5.9in]{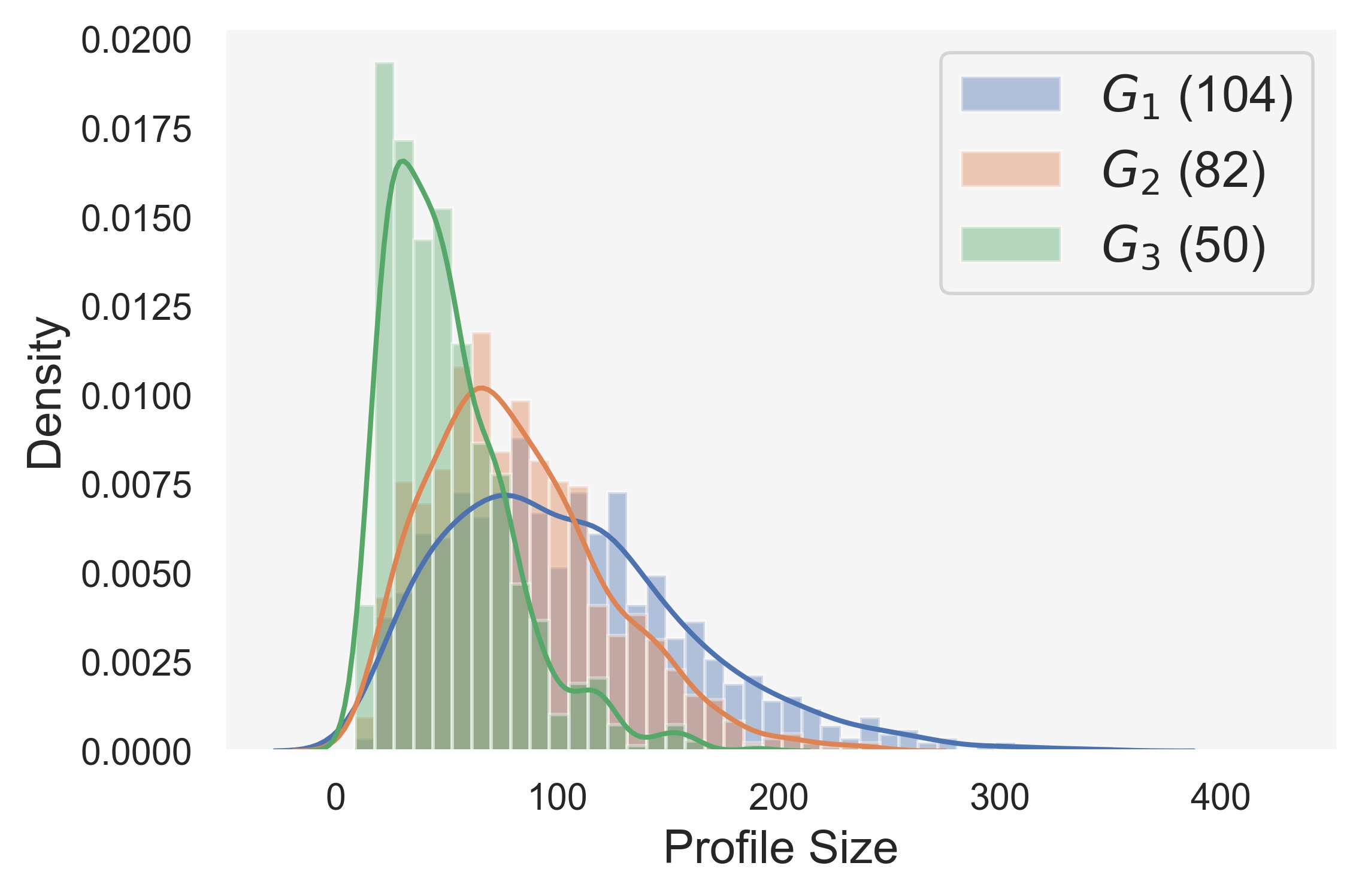}}

\caption{Profile size of users in different user groups. Numbers within the parentheses show the average profile size of the users in each group} \label{user_groups_psize}
\end{figure*}

\begin{figure*}
\centering
\SetFigLayout{3}{2}
 \subfigure[MovieLens]{\includegraphics[width=5.9in]{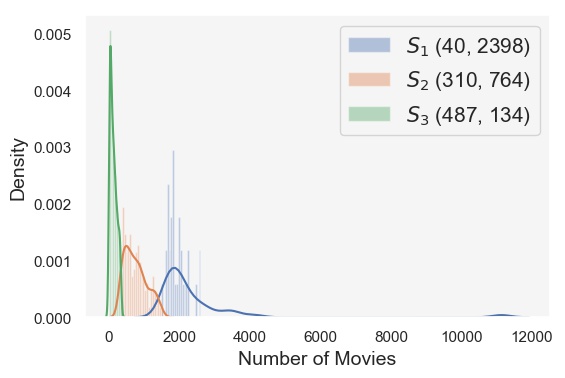}}
\subfigure[Last.fm]{\includegraphics[width=5.9in]{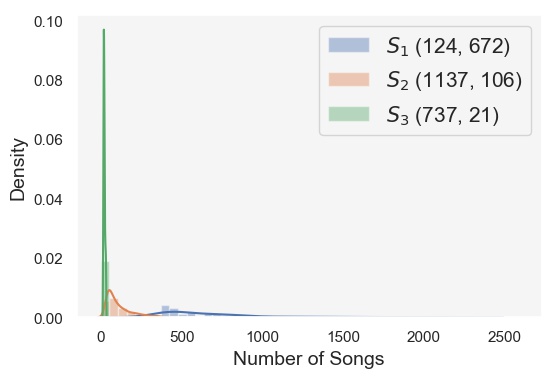}}

\caption{Number of songs / movies for artists / directors in each supplier group. The first number in the parentheses shows the size of each group and the second number shows the average number of songs / movies for the suppliers in each group.} \label{supplier_groups_psize}
\end{figure*}
\begin{itemize}
    \item \textbf{Item Grouping}: Generally speaking, in recommendation domains there are a small number of very popular items and many other items have either medium popularity or are extremely non-popular. These few popular items are referred to as the \textit{short-head} in the literature which take up roughly 20\% of the ratings according to the Pareto Principle \cite{sanders1987pareto}. The rest of the items in a long-tail distribution are usually divided into two other parts \cite{celma2008hits} as I also do in this dissertation: Tail items ($T$) are the larger number of less popular items which collectively take up roughly 20\% of the ratings\footnote{Tail items may also be new items that have yet to find their audience and will eventually become popular. In this way, popularity bias is related to the well-known item cold-start problem in recommendation.}, and \textit{Mid} items ($M$) include a larger number of items in between that receive around 60\% of the ratings, collectively. 
    
    \item \textbf{User Grouping}: Users have different interest in popular items. Some are mainly focused on popular and blockbuster products and some are more inclined towards less popular, or niche, items. There are also some who fall in between and they have a more diverse taste in terms of the popularity of the items they tend to consume. In this dissertation, I sort the users from the highest interest towards popular items to the lowest based on the ratio of different item groups ($H$, $M$, and $T$) in their profiles. Users are lexicographically sorted according to the ratio of $H$, $M$, and $T$ items in their profiles with priority given to $H$ followed by $M$ and $T$. Sorted users are divided into three equal-sized bins $G=\{G_1, G_2, G_3\}$ from most popularity-focused ($G_1$) to least ($G_3$).

    Figure \ref{user_groups_psize} shows the distribution of profile size of the users within each group for both MovieLens and Last.fm datasets. We can see that, on MovieLens, groups that have higher interests towards popular movies have smaller average profile size compared to the groups with lesser interest in popular movies. In fact, the average profile size for the users in $G_3$ (group with the least interest in popular movies) is almost five time larger than the highly-popularity-focused group ($G_1$). Interestingly, this trend on Last.fm seems to be the exact opposite where the average profile size of the user groups who have higher focus on popular songs is larger. This is indeed an interesting distinction between the movie and music domain. 
    
    \item \textbf{Supplier Grouping}: Suppliers also have different levels of popularity. In movies, some directors are making more popular movies and therefore they have a higher popularity according to our definition of popularity which is the average popularity of the items for that supplier. Similar to items, supplier popularity has also an extreme long-tail shape and therefore I need to group them in a way that the resulting groups vary in terms of popularity. Therefore, I, again, use the Pareto Principle for the suppliers to determine the appropriate cut-off points to create the groups. I have defined three supplier groups based on their popularity $S=\{S_1, S_2, S_3\}$: $S_1$ represents few popular suppliers whose items take up 20\% of the ratings, $S_2$ are larger number of suppliers with medium popularity whose items take up around 60\% of the ratings, and $S_3$ are the less popular suppliers whose items get 20\% of the ratings.

    Figure \ref{supplier_groups_psize} shows the distribution of the number of items owned by the suppliers in each supplier group and also the number of suppliers in each group. In both datasets, the supplier groups with higher popularity own larger number of items compared to the groups whose suppliers do not have high popularity.

\end{itemize}

\section{Evaluation Metrics}
A multi-stakeholder evaluation entails multiple metrics representing the perspective of different stakeholders on system performance. In this dissertation, in order to evaluate different recommendation algorithms with respect to their performance in terms of handling popularity bias, I use several well-established metrics in the literature. I also introduce several other metrics that are able to better differentiate the performance of different algorithms when they are assessed from the perspective of multiple stakeholders. In particular, I propose metrics that are sensitive to the performance across different groups of stakeholders: suppliers in different popularity categories, items in different popularity categories, and users with different levels of interests in item popularity.

\subsubsection{Overall}
To capture overall long-tail performance, a variety of metrics have been used in prior research.  I compute the following metrics:
\begin{itemize}
    \item \textbf{Aggregate Diversity:} The ratio of unique recommended items across all users which I defined in Chapter \ref{background_chapter}.
    
    
    
     \item \textbf{Long-tail Coverage (LC):} The concept of long-tail coverage is used in many prior works in popularity bias \cite{adomavicius2011improving,abdollahpouri2019managing} and it measures the ratio of long-tail items (in our case it is $M \cup T$ since I have divided the long-tail into two separate groups) that appear in the recommendation lists of different users. Effectively, this is \textit{Agg-Div} applied only to the $M \cup T$ portion of the catalog.  
    \begin{equation}
         LC=\frac{\left| \bigcup_{u \in U}(\ell_u \cap (M \cup T))\right|}{|M \cup T|}
    \end{equation} 
    $I$ be the set of all items in the catalog and $U$ is the set of all users.
    \item \textbf{Gini Index:} Measures the inequality across the frequency distribution of the recommended items which was also defined in Chapter \ref{background_chapter}.

    

\end{itemize}

 \begin{figure}
    \centering
    \includegraphics[width=5in]{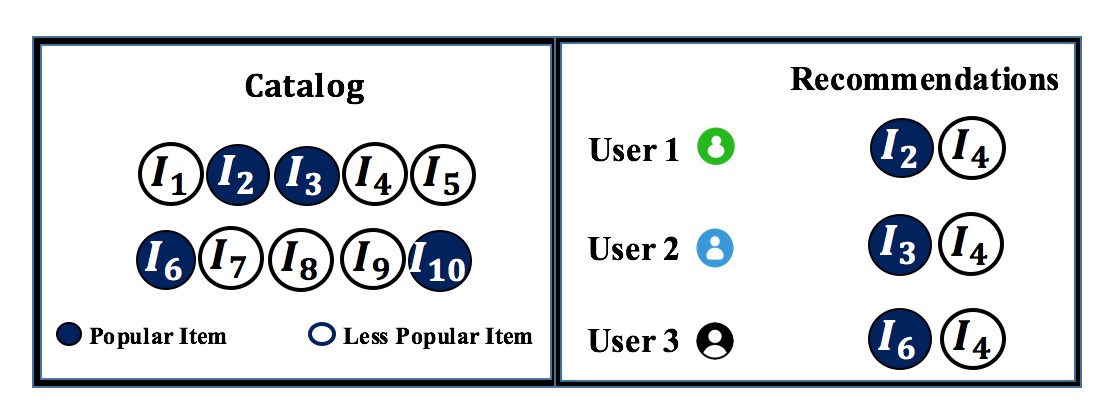}
    \caption{List of two Recommendations given to three users. Dark circles are popular items and light ones are long-tail items. There are ten items overall in the catalog 6 of which are long-tail and the other 4 are popular. \textit{Agg-Div}=$\frac{4}{10}=0.4$, $APL=\frac{1}{2}=0.5$ and $LC=\frac{1}{6}=0.166$.}
    \label{fig:recs_example}
\end{figure}

Another metric for measuring long-tail performance of a given algorithm which I introduce in this dissertation is \textit{Average Percentage of Long-tail Items} ($APL$) which I define as follows: 

 \begin{equation}
      APL=\frac{1}{|U|}\sum_{u \in U} \frac{|\{i, i \in (\ell_u \cap (M \cup T)) \} |}{|\ell_u|}
 \end{equation}

 This metric tells us what percentage of items in users' recommendation lists belongs to the long-tail of the item distribution ($M \cup T$). This is purely a user-centric metric as it does not look at the overall percentage of long-tail items but rather only in a user-level. With that said, it is possible for an algorithm to have high $APL$ meaning every user has received a high percentage of long-tail items but, overall, it may not have covered many items from the long-tail when it is looked at across all users. Therefore, $LC$ and $APL$, together, can give a better picture of how an algorithm is performing in terms of long-tail promotion. Look at Figure \ref{fig:recs_example} for example. There are ten items in the catalog $I_1...I_{10}$. Popular items are shown by dark circles and less popular (long-tail) one shows by light circles. There are six long-tail items and four popular items in the catalog. Three users in this figure have each received two items as the recommendations. We can see that, overall, four unique items are recommended across all users so the aggregate-diversity will be 0.4. Each user has received 1 long-tail item out of a list with size 2 so the $APL$ will be $\frac{1}{2}=0.5$. However, as we can see, all users have received the same long-tail item ($I_4$) so, overall, only one long-tail item is recommended and $LC$ will be $\frac{1}{6}=0.166$. The $APL$ of 0.5 indicates 50\% of the recommendations given to the users are long-tail items. This might create the sense that the recommendation algorithm has done a good job in terms of covering long-tail items. However, looking at $LC$ we can see that out of 6 long-tail items only 1 has been recommended to different users. So, depending on the purpose of the recommender system, each of the $APL$ and $LC$ metrics could show a different side of the recommendation performance and, therefore, both can be measured to capture a better picture of the recommendation performance in terms of long-tail.

\subsubsection{Supplier Groups}
Similar to \cite{mehrotra2018towards}, I operationalize the concept of fairness for the suppliers (i.e. artists and directors in Last.fm and MovieLens datasets, respectively) using their popularity. In \cite{mehrotra2018towards}, the authors grouped the artists in Spotify data into 10 equal-sized bins. As I noted in Section~\ref{grouping}, I group the suppliers into three different bins based on their position in the popularity spectrum $S=\{S_1,S_2,S_3\}$ with the $S_1$ being the most popular suppliers followed by $S_2$ and $S_3$ that have lower popularity.

Following \cite{mehrotra2018towards}, I use \textit{Equality of Attention Supplier Fairness} (ESF), which operationalizes the idea that a supplier's chance of being recommended should be independent of the popularity bin to which they belong. The \textit{ESF} of a list of recommendations ($L$) given to all users from the suppliers' perspective is defined as:


\begin{equation}
     ESF=\sum_{i=1}^{|S|}\sqrt {\sum_{j \in L} \mathbbm{1}(A(j) \in S_i)}
 \end{equation}

\noindent where $S_i$ is the list of suppliers belonging to the popularity bin $i$, and $A(j)$ is a mapping function that returns the supplier of item $j$. $|S|$ is the number of supplier groups which in this dissertation it is 3.

\textit{ESF} rewards sets that are diverse in terms of the different supplier bins represented, thus providing a fairer representation of different supplier bins. Given the nature of the function, there is more benefit to selecting suppliers from a bin not yet having one of its suppliers already chosen. When a supplier from a particular bin is represented in the recommendations, other suppliers from the same bin will be penalized due to the concavity of the square root function (e.g. $\sqrt{1}+\sqrt{1}>\sqrt{2}$). 

The \textit{ESF} measure does not, however, consider the inherent popularity of suppliers in different bins. I introduce an alternative metric that does so: \textit{Supplier Popularity Deviation} (SPD). 

For any supplier group $s$, $SPD(s)=q(s)-p(s)$ where $q(s)$ is the ratio of recommendations that come from items of supplier group $s$ (i.e. $q(s)=\frac{\sum_{u \in U}\sum_{j \in \ell_u}\mathbbm{1}(A(j) \in s)}{n\times|U|}$), and $p(s)$ is the ratio of ratings that come from items of supplier group $s$ (i.e. $p(s)=\frac{\sum_{u \in U}\sum_{j \in \rho_u}\mathbbm{1}(A(j) \in s)}{\sum_{u \in U} |\rho_u|}$). The average \textit{SPD} across different groups of suppliers can be calculated as:

\begin{equation}
SPD=\frac{\sum_{i=1}^{|S|}|SPD(S_i)|}{|S|}
\end{equation}

Lower values for $SPD$ indicate a better match between the distribution of items from different suppliers in rating data and in the recommendations. The inverse $1-SPD$ can be considered a type of proportional fairness metric since it measures how the items from different supplier groups are exposed to different users \textit{proportional} to their popularity in rating data. 

\subsubsection{Item Groups}
We can use the same formalization as \textit{SPD} to define a metric \textit{Item Popularity Deviation} (IPD) that looks at groups of items across the popularity distribution. For any item group $c$, $IPD(c)=q(c)-p(c)$ where $q(c)$ is the ratio of recommendations that come from item group $c$ (i.e. $ q(c)=\frac{\sum_{u \in U}\sum_{j \in \ell_u}\mathbbm{1}(j \in c)}{n\times|U|}$). $p(c)$ is the ratio of ratings that come from item group $c$ (i.e. $p(c)=\frac{\sum_{u \in U}\sum_{j \in \rho_u}\mathbbm{1}(j \in c)}{\sum_{u \in U} |\rho_u|}$). The average \textit{IPD} across different groups of items can be measured as:

\begin{equation}
IPD=\frac{\sum_{c \in C}|IPD(c)|}{|C|},
\end{equation}


\subsubsection{User Groups}
Finally, based on our finding above that users with different interests in popular items may receive results with different degrees of calibration, we can examine the popularity calibration of recommendation results across groups of users with different popularity propensity. This metric is \textit{User Popularity Deviation} (UPD). For any user group $\textsl{g}$, $UPD(\textsl{g})=\frac{\sum_{u \in \textsl{g}}\mathfrak{J}(P(\rho_u),Q(\ell_u))}{|g|}$. The average \textit{UPD} across different groups of users is:
\begin{equation}
    UPD=\frac{\sum_{g \in G}UPD(\textsl{g})}{|G|}
\end{equation}
$UPD$ can be also seen as the average popularity miscalibration of the recommendations from the perspective of users in different groups.

For all three metrics \textit{IPD}, \textit{SPD} and \textit{UPD}, lower values are desirable. 
\begin{table}[]
\footnotesize
\centering
\captionof{table}{The summary of the evaluation metrics used in this dissertation, their abbreviations and their purpose.} 
\label{tab:eval}
\begin{tabular}{|l|l|l|}
\hline
\textbf{Metric Name}                          & \textbf{Abbreviation}                      & \textbf{Purpose}                                                                                                                                                                                                 \\ \hline
Aggregate Diversity                   & \textit{Agg-Div} & \begin{tabular}[c]{@{}l@{}}The ratio of unique \\ recommended items compared to all \\ the available items.\end{tabular}                                                                                    \\ \hline
Average Percentage of Long-tail Items & APL                               & \begin{tabular}[c]{@{}l@{}}The average percentage of \\ long-tail items  in individual \\ recommendation lists.\end{tabular}                                                                                 \\ \hline
Long-tail Coverage                    & LC                                & \begin{tabular}[c]{@{}l@{}}The percentage of long-tail \\ items that are covered by \\ the recommender system.\end{tabular}                                                                                 \\ \hline
Gini Index                            & Gini                              & \begin{tabular}[c]{@{}l@{}}Measures how evenly \\ the recommendations are distributed \\ across all items.\end{tabular}                                                                                     \\ \hline
Equality of Attention Supplier Fairness & ESF                               & \begin{tabular}[c]{@{}l@{}}Measures the fairness \\ of exposure for different supplier \\ groups based on the equality of \\ appearance  for different supplier groups.\end{tabular}                                    \\ \hline
Supplier Popularity Deviation         & SPD                               & \begin{tabular}[c]{@{}l@{}}Measures the deviation of \\ exposure for items from different \\ supplier groups compared to their \\ original popularity in data.\end{tabular}                                 \\ \hline
Users Popularity Deviation            & UPD                               & \begin{tabular}[c]{@{}l@{}}Measures the average amount  \\ of deviation the algorithms impose on \\ users in different  groups \\ compared to what their \\ original interest in popularity was.\end{tabular} \\ \hline
Items Popularity Deviation            & IPD                               & \begin{tabular}[c]{@{}l@{}}Measures the deviation of\\  exposure for items from different \\ item groups compared to their original \\ popularity in data.\end{tabular}                                     \\ \hline
\end{tabular}
\end{table}

Table \ref{tab:eval} summarizes the evaluation metrics I described in this chapter. I will use these in the following chapters. 

\chapter{Popularity Bias in Recommendation}\label{pop_bias}

Skew in wealth distribution is well-known: less than 1\% of the human population has more wealth than more than 40\% of the world combined \footnote{https://inequality.org/facts/global-inequality/}; in recommender systems a similar problem exists: a small number of popular items appear frequently in user profiles and a much larger number of niche items appear rarely. In many consumer taste domains, where recommender systems are commonly deployed, there is a highly-skewed distribution of user interest in items. A music catalog might contain artists whose songs have been played millions of times (Beyonc\'{e}, Ed Sheeran) and others whose songs have a much smaller audience (Iranian musician Kayhan Kalhor, for example). Recommender systems often recommend popular items at a much higher frequency than even their high popularity would call for, amplifying the bias already present in the data.

Recommending serendipitous items from the long-tail (items with low popularity) are generally
considered to be valuable to the users \cite{anderson2008long,shani2011evaluating}, as these are items that
users are less likely to know about. Brynjolfsson and his colleagues
showed that 30-40\% of Amazon book sales are represented by titles that would not normally be found in brick-and-mortar stores at the time of their writing \cite{brynjolfsson2006niches}. Access to long-tail items is a strong driver for e-commerce growth: the consumer surplus created by providing access to these less-known book titles is estimated at more than seven to ten times the value consumers receive from access to lower prices online.

Long-tail items are also important for generating a fuller understanding of users’ preferences. Systems that use active learning to explore each user’s profile will typically need to present more long-tail items because these are the ones that the user is less likely to have rated, and where user’s preferences are more likely to be diverse \cite{nguyen2014exploring,resnick2013bursting}.

Finally, long-tail recommendation can also be understood as a social good. A market that suffers from popularity bias will lack opportunities to discover more obscure products and will be, by definition, dominated by a few large brands or well-known artists \cite{celma2008hits}. Such a market will be more homogeneous and offer fewer opportunities for innovation and creativity.

As Anderson discusses in his book \cite{anderson2008long}, in an era without the constraints of physical shelf space and other bottlenecks of distribution, narrowly targeted products and services can be as economically attractive as mainstream fare. When you can dramatically lower the costs of connecting supply and demand, it changes not just the numbers, but the entire nature of the market. This is not just a quantitative change, but a qualitative one, too. Bringing niches within reach reveals latent demand for noncommercial content. Then, as demand shifts towards the niches, the economics of providing them improve further, and so on, creating a positive feedback loop that will
transform entire industries – and the culture. "We are turning from a mass market into a niche nation, defined now not by our geography but by our interests" Anderson says in his book. 

The constraints of physical shelf space was particularly a barrier to the idea of long-tail and niche market in the past \cite{bentley2009physical} as it was not economically effective for the businesses to store a large number of products some of which had a limited number of customers interested in them. In the era of digitalization, however, the physical storage of the goods is no longer a necessity and hence there is more opportunity for the niche market. For instance, movie and music platforms such as Netflix, Hulu, Spotify, and Pandora are connecting millions of consumers to a large number of movies and songs completely on a digital platform. In such a large marketplace, the popular items are still more attractive and getting more attention but the niche products also can reach a specific range of audience. That being said, the recommendations generated on such platforms are often still biased towards hits and popular items and a mechanism for alleviating this bias is needed to give a fair exposure to different items from different popularity levels in the popularity spectrum.

The popularity bias in recommender systems can originate from different sources: the underlying data for user interactions and the inherent bias of the recommendation algorithms. 

\section{Bias in Data}
Rating data in many recommendation domains is generally skewed towards more popular items--there are a few popular items which take up the majority of rating interactions while the majority of the items receive small attention from the users. For example, few songs and movies are internationally popular, few people on social media have millions of followers, a small percentage of research papers have thousands of citations, and few tourism destinations attract millions of people.   

\begin{figure*}
\centering
\SetFigLayout{3}{2}
 \subfigure[MovieLens]{\includegraphics[width=2.5in]{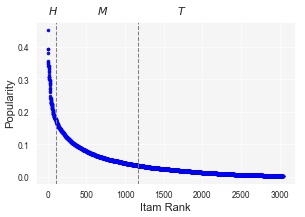}}
\subfigure[Last.fm]{\includegraphics[width=2.5in]{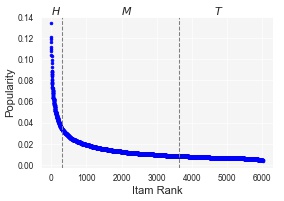}}
  \hfill
\caption{The Long-tail Distribution of Rating data} \label{longtail}
\end{figure*}

Figure ~\ref{longtail} shows the percentage of users rated different items in MovieLens and Last.fm datasets: the popularity of each item. Items are ranked from the most popular to the least with the most popular item being on the far left. The curve has a long-tail shape \cite{anderson2006long,celma2008hits} indicating few popular items are taking up the majority of the ratings while many other items on the far right of the curve have not received much attention. This type of distribution can be found in many other domains such as e-commerce where a few products are best-sellers, online dating where a few profiles attract the majority of the attention, and social networking sites where a few users have millions of followers. 

The bias in rating data could be due to two different reasons:
\begin{itemize}
    \item \textbf{External Bias:} Some items and products are inherently more popular than others even outside of the recommender systems in the real world. For instance, even before music streaming services emerged, there were always a small number of artists that were nationally or internationally popular such as \textit{Shakira}, \textit{Jennifer Lopez}, or \textit{Enrique Iglesias}. As a result of this external bias (or tendency) towards popular artists, users also often listen to those artists more on streaming services no matter if they are recommended to them or not, and hence they get more user interactions. Another external motive for some items being more popular could be due to social influence \cite{kim2007impact,kwahk2012effects}, word of mouth \cite{fan2012effect}, and other external factors that impact the popularity of an item and the users' intention to buy those products.  
    \item \textbf{Feedback Loop}: As we will see later in this chapter, recommendation algorithms have a higher tendency towards recommending popular items. That means, the popular items have a higher chance to be recommended to the users and hence garnering a larger number of interactions from the users. Therefore, when these interactions are logged and stored in the system as the new data, the popularity of those items increases since they get more and more interactions over time \cite{mansoury2020feedback,jiang2019degenerate}.  
\end{itemize}

\section{Algorithmic Bias}

Due to this imbalance property of the rating data, often algorithms inherit this bias and, in many cases, intensify it by over-recommending the popular items and, therefore, giving them a higher opportunity of being rated by more users \cite{park2008long,steck2011item,jannach2015recommenders}: \textit{the rich get richer and the poor get poorer.} 

\begin{figure*}
    \centering
    \includegraphics[width=5in]{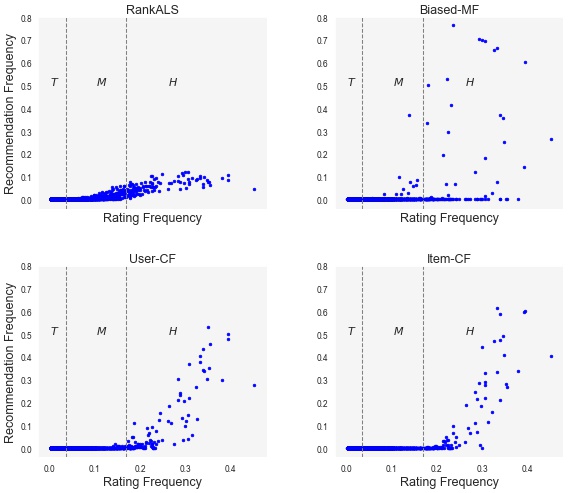}
    \caption{Item popularity versus recommendation popularity (MovieLens)}
    \label{corr_scatter_movielens}
\end{figure*}

Figure ~\ref{corr_scatter_movielens} shows the percentage of users who have rated an item on the x-axis and the percentage of users who received that item in their recommendation lists using four different recommendation algorithms \textit{RankALS} \cite{takacs2012alternating}, \textit{Biased Matrix Factorization (Biased-MF)} \cite{koren2009matrix},\textit{User-based Collaborative Filtering (User-CF)} \cite{aggarwal2016neighborhood}, and \textit{Item-based Collaborative Filtering (Item-CF)} \cite{sarwar2001item}. In other words, the plot aims to show the correlation between the popularity of an item in the rating data versus how often it is recommended to different users using different algorithms. It is clear that in all four algorithms, many items are either never recommended or just rarely recommended.

Among the four algorithms, \textit{Item-CF} and \textit{User-CF} show the strongest evidence that popular items are recommended much more frequently than the others. In fact, they are recommended to a much greater degree than even what they initial popularity warrants. For instance, the popularity of some items have been amplified from roughly 0.4 to 0.7 indicating a 75\% increase. Both \textit{Item-CF} and \textit{User-CF} are over-promoting popular items (items on the right size of the x-axis) while significantly hurting other items by not recommending proportionate to what their popularity in data warrants. In fact, the vast majority of the items on the left are never recommended indicating an extreme bias of the algorithms towards popular items and against less popular ones. On \textit{RankALS} the positive correlation between popularity in rating data and how frequently an item is recommended can still be observed but it is nowhere as extreme as \textit{User-CF} and \textit{Item-CF}. \textit{Biased-MF} does not show a positive correlation between popularity in data and in recommendations although some items are still over-recommended (have much higher popularity in recommendations versus what they had in rating data). However, this over-recommendation is not concentrated on only popular items and some items from lower popularity values are also over-recommended.

\begin{figure*}
    \centering
    \includegraphics[width=5in]{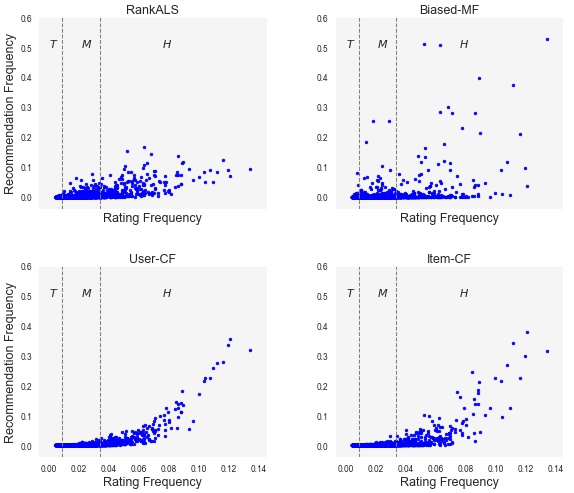}
    \caption{Item popularity versus recommendation popularity (Last.fm)}
    \label{corr_scatter_lastfm}
\end{figure*}

Figure ~\ref{corr_scatter_lastfm} shows the same information on Last.fm data. The magnitude of the popularity bias of different recommendation algorithms seems to be lower on Last.fm dataset. The reason is the original rating data in Last.fm is less biased towards popular items as we can see in y-axis of Figure \ref{longtail} where the most popular item in MovieLens has a popularity close to 0.45 where in Last.fm this value is less than 0.14.

\section{Multi-stakeholder Impact of Popularity Bias}

As I discussed in Chapter \ref{ms_rs}, recommender systems are multi-stakeholder environments where, in addition to the users, some other stakeholders such as the supplier of the items also benefit from the recommendation of their items and gaining a larger audience. On top of that, the recommender system designer might also have certain goals such as recommending a particular group of items more frequently in certain times depending on their business goals such as sales promotions, items that need to be sold before a certain time etc. 

Algorithmic popularity bias can negatively impact different parts of a recommender system platform. In this section I intend to demonstrate how this bias can impact different groups of items, users, and the suppliers of the items.  

\subsection{Impact on Item Groups}

\begin{figure*}[tb]
\centering
\SetFigLayout{2}{1}
 \subfigure[MovieLens]{\includegraphics[width=5.8in]{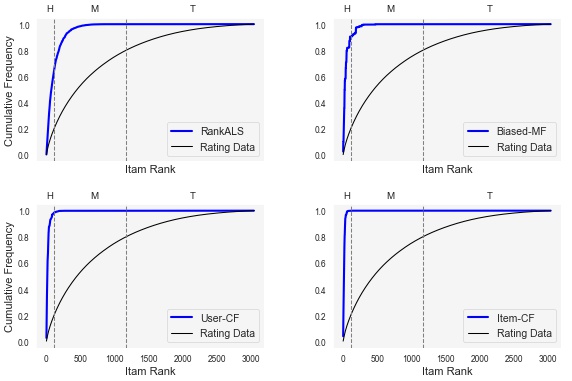}}
  \hfill
 \subfigure[Last.fm]{\includegraphics[width=5.8in]{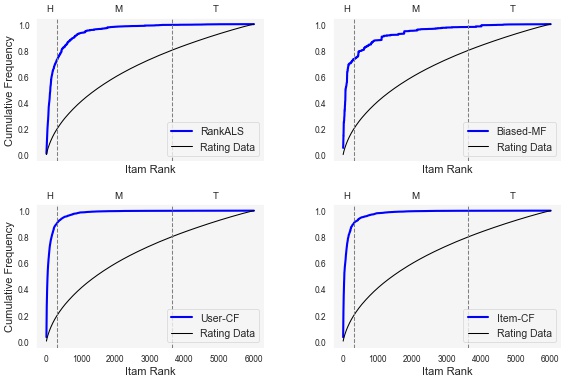}}
  \hfill
\caption{Popularity bias amplification (Item Groups)} \label{ms_impact-Items}
\end{figure*} 

Figure ~\ref{ms_impact-Items} contrasts item popularity and recommendation popularity for four well-known recommendation algorithms \textit{RankALS}, \textit{Biased Matrix Factorization (Biased-MF)},\textit{User-based Collaborative Filtering (User-CF)}, and \textit{Item-based Collaborative Filtering (Item-CF)}. The black curve shows the cumulative frequency of ratings for different items at a given rank in the MovieLens 1M \cite{movielens} data set. The x-axis indicates the rank of each item when sorted from most popular to least popular. The y-axis indicates the cumulative prevalence of that item in the data. As we can see, a few highly ranked items dominate the entire rating history. For instance, only 111 items (less than \%3 of the items) take up more than \%20 of the ratings. A similar type of distribution can be seen in many other domains. These few popular items are referred to as short-head ($H$) in the literature, followed by the middle-tail ($M$), which are items that are not blockbusters but have significant popularity, and tail items ($T$), which are more obscure and appeal to niche groups of users~\cite{celma2008hits}.\footnote{Tail items may also be new items that have yet to find their audience and will eventually become popular. In this way, popularity bias is related to the well-known item cold-start problem in recommendation.}

The blue curves in each figure show popularity bias at work across the four algorithms. In \textit{Item-CF}, no item beyond rank 111 is recommended. The head of the distribution constituting less than 3\% of the items are actually taking up 100\% of the recommendations -- a ``lion's share'' indeed. In \textit{User-CF} this number is 99\%. The other algorithms are only slightly improved in this regard, where the head items take up more than 64\% and 74\% of the recommendations in \textit{RankALS} and \textit{Biased-MF}, respectively. 

\subsection{Impact on User Groups}

\begin{figure*}[tb]
\centering
\SetFigLayout{2}{1}
 \subfigure[MovieLens]{\includegraphics[width=4in]{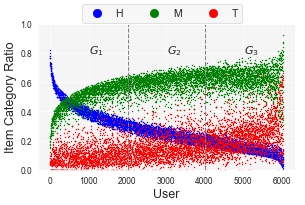}}
  \hfill
 \subfigure[Last.fm]{\includegraphics[width=4in]{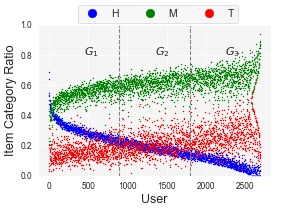}}
  \hfill
\caption{Users' Propensity towards item popularity} \label{user_propensity}
\end{figure*}

We can also take a consumer- / user-centered view of popularity bias. Not every user is equally interested in popular items. In cinema, for instance, some might be interested in movies from Yasujiro Ozu, Abbas Kiarostami, or John Cassavetes, and others may enjoy more mainstream directors such as James Cameron or Steven Spielberg. Figure ~\ref{user_propensity} shows the ratio of rated items for three item categories $H$, $M$, and $T$ in the profiles of different users in the MovieLens 1M and Last.fm datasets. Users are sorted from the highest interest towards popular items to the least and divided into three equal-sized bins $G_{1..3}$ from most popularity-focused to least, as I described in Chapter \ref{data_method}. The y-axis shows the proportion of each user's profile devoted to different popularity categories. The narrow blue band shows the proportion of each users profile that consists of popular items, and its smooth decrease reflects the way the users are ranked. Note, however, that all groups have rated many items from the middle (green) and tail (red) parts of the distribution, and this makes sense: there are only a small number of really popular movies and even the most blockbuster-focused viewer will eventually run out of them. 

\begin{figure*}[tb]
\centering
\SetFigLayout{2}{1}
 \subfigure[MovieLens]{\includegraphics[width=4in]{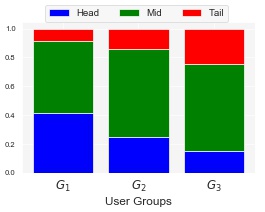}}
  \hfill
 \subfigure[Last.fm]{\includegraphics[width=4in]{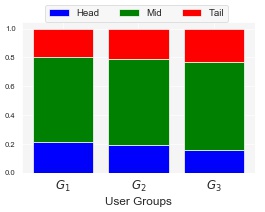}}
  \hfill
\caption{User Groups' Propensity towards item popularity} \label{user_group_propensity}
\end{figure*}

Figure \ref{user_group_propensity} shows the same information as Figure \ref{user_propensity} but all ratios are calculated as an average value across users in each user group. The differences between different user groups in this figure are more visible. For example, looking at the head ratio (blue color), we can clearly see that users in group $G_1$ has the highest ratio of head items followed by $G_2$ and $G_3$.

\begin{figure*}[tb]
\centering
\SetFigLayout{2}{1}
 \subfigure[MovieLens]{\includegraphics[width=4in]{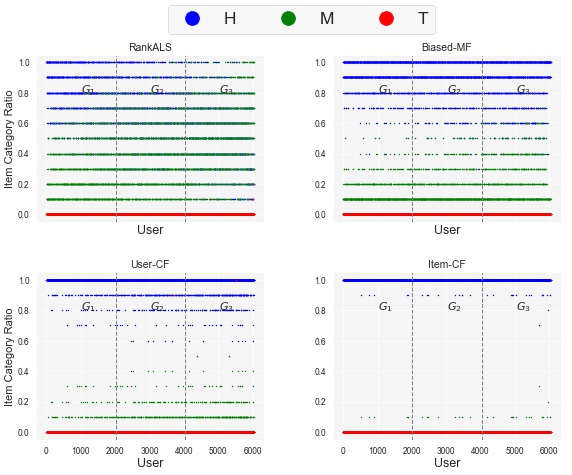}}
  \hfill
 \subfigure[Last.fm]{\includegraphics[width=4in]{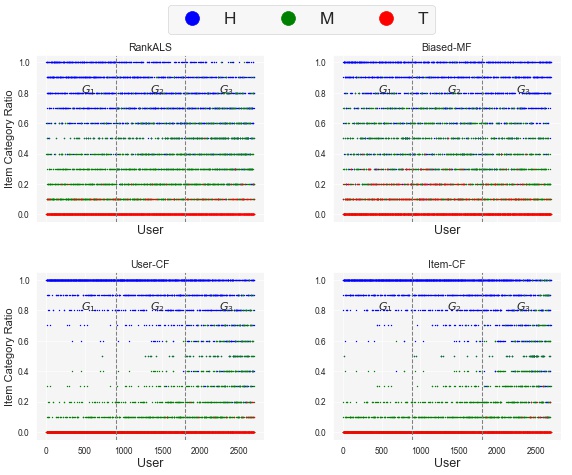}}
  \hfill
\caption{Users' centric view of popularity bias} \label{ms_impact-users}
\end{figure*}

The plots in Figure ~\ref{ms_impact-users} are parallel to Figure ~\ref{user_propensity}, with the users ordered by their popularity interest, but now the y-axis shows the proportion of recommended items using different algorithms from the different item popularity categories. The difference with the original user profiles in rating data especially in the case of $Item-CF$ and $User-CF$ is stark. Where the users' profiles are rich in diverse popularity categories, the generated recommendations are nowhere close that to what the user has shown interest at. In fact, in $Item-CF$ almost 100\% of the recommendations are from the head category, even for the users with the most niche-oriented profiles. Tail items do not appear at all. I demonstrated here that popularity bias in the algorithm is not just a problem from a global, system, perspective. It is also a problem from the user perspective: users are not getting recommendations that reflect the diversity of their profiles and the users with the most niche tastes ($G_3$) are the most poorly served. 

\begin{figure*}[tb]
\centering
\SetFigLayout{2}{1}
 \subfigure[MovieLens]{\includegraphics[width=4in]{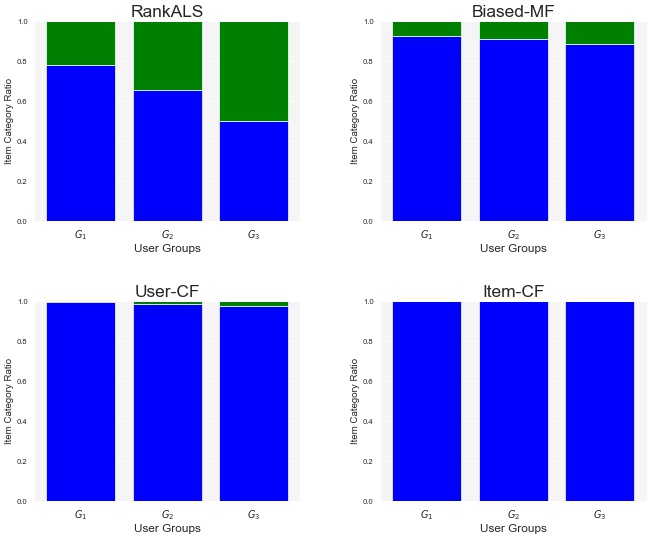}}
  \hfill
 \subfigure[Last.fm]{\includegraphics[width=4in]{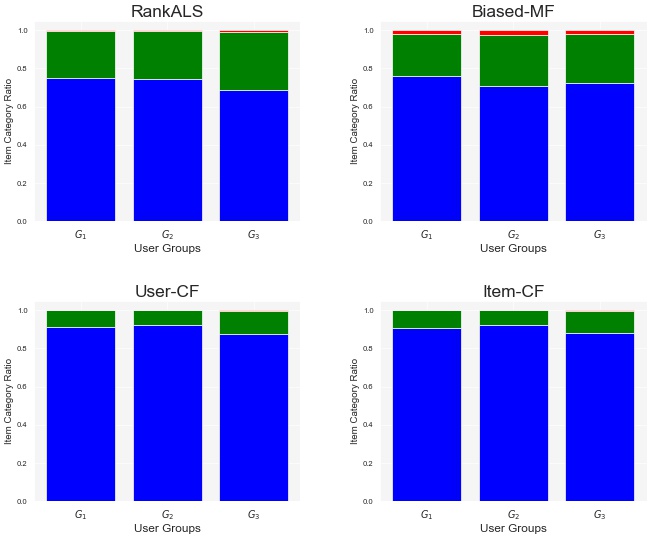}}
  \hfill
\caption{Users' centric view of popularity bias (Groups)} \label{ms_impact-user_groups}
\end{figure*}

Similar to Figure \ref{user_group_propensity}, Figure \ref{ms_impact-user_groups} shows the ratio of different item groups in the profiles of users in different user groups as an average value. Here, we can clearly see the differences between algorithms in term of how the have performed for different user groups. \textbf{Item-CF} clearly consists of only head items for all user groups.

\subsection{Impact on Supplier Groups}
As noted above, multi-stakeholder analysis in recommendation also includes providers or as termed here \textit{suppliers}, ``those entities that supply or otherwise stand behind the recommended objects''~\cite{abdollahpourimultistakeholder2020}. We can think of many different kinds of contributors standing behind a particular movie. As I mentioned in Chapter \ref{data_method}, I consider the director of each movie and the artist for each song as the suppliers in MovieLens and Last.fm respectively.

\begin{figure*}[tb]
\centering
\SetFigLayout{2}{1}
 \subfigure[MovieLens]{\includegraphics[width=5in]{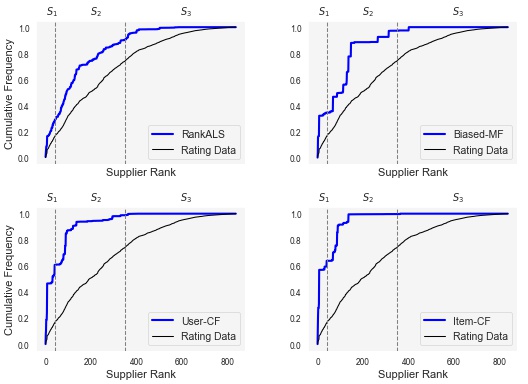}}
  \hfill
 \subfigure[Last.fm]{\includegraphics[width=5in]{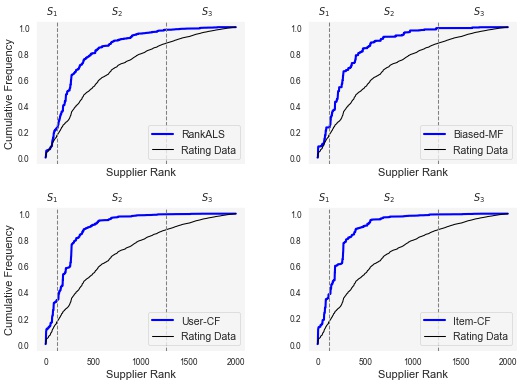}}
  \hfill
\caption{Suppliers' centric view of popularity bias} \label{supplier_centric}
\end{figure*}

Figure~\ref{supplier_centric} is similar to Figure~\ref{ms_impact-Items} but shows the rank of different directors by popularity and the corresponding recommendation results from the \textit{Item-CF} algorithm. A similar pattern of popularity bias is evident. The recommendations have amplified the popularity of the popular suppliers (the ones on the extreme left) while suppressing the less popular ones dramatically. Strikingly, movies from just 3 directors (less than 0.4\% of the suppliers here) take up 50\% of recommendations produced, while the tail of the distribution is seeing essentially zero. 

What this analysis demonstrates is that popularity bias is a multi-stakeholder phenomenon, that entities on different sides of the recommender system are impacted by it (sometimes quite severely) and that any evaluation of it must capture the perspective of different stakeholders. 

In next chapter, I will propose several recommendation algorithms for mitigating the popularity bias and evaluate their performance using the multi-stakeholder paradigm to see how they perform from the perspective of different stakeholders.

\chapter{Tackling Popularity Bias}\label{tackle_pop_bias}

As I mentioned in Chapter \ref{background_chapter} and \ref{ms_rs}, traditionally recommendation algorithms have been developed in a way that could maximize the accuracy of the delivered recommendations to the users. However, this over-emphasis on being accurate could hurt other important factors of a successful recommendation such as diversity, serendipity and novelty~\cite{Vargas:2011:RRN:2043932.2043955,ge2010beyond,castells2011novelty} and impacts across system stakeholders \cite{abdollahpourimultistakeholder2020}. 

One of the problems that is beyond being accurate is the popularity bias phenomena I discussed in Chapter \ref{pop_bias}. We saw that this bias can negatively impact different stakeholders in a given recommender system and hence the algorithms for mitigating such bias need to be evaluated from the perspective of different stakeholders.

The solutions for tackling popularity bias in the literature can be categorized into two groups:\footnote{A third option, preprocessing, is generally not useful for popularity bias mitigation because undersampling the popular items greatly increases the sparsity of the data.}

\begin{itemize}
    \item \textbf{Model-based:} In this group of solutions, the rating prediction step is modified, so that the popularity of items are taken into account in the rating prediction \cite{vargas2014improving,sun2019debiasing}. 
     \item \textbf{Re-ranking:} Most of the solutions for tackling popularity bias fall into this category \cite{adomavicius2011maximizing,adomavicius2011improving}. A re-ranking algorithm takes an output recommendation list extracts a re-ordered list with improved long-tail properties.
\end{itemize}

\section{Model-based solutions}
\begin{figure*}
    \centering
    \includegraphics[width=4in]{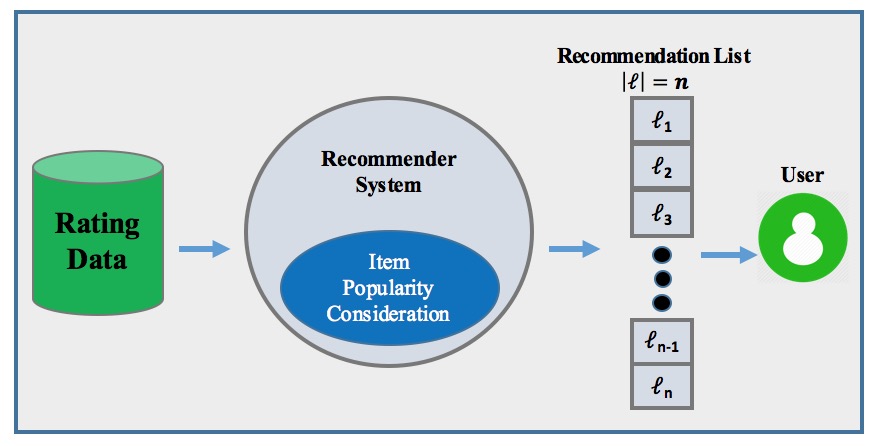}
    \caption{The Schema for the model-based popularity mitigation}
    \label{fig:model-based}
\end{figure*}

The general schema for the model-based approach to tackle popularity bias is shown in Figure ~\ref{fig:model-based}. This is similar to Figure \ref{fig:recsys} in Chapter \ref{background_chapter} with one distinction: an extra component which I call \textit{Item Popularity Consideration} is incorporated in the recommendation algorithm before the final recommendation list is generated to the user.

This consideration of item popularity can be incorporated in different ways such as in the rating prediction step whether it is in neighborhood-based models \cite{desrosiers2011comprehensive} or in matrix factorization techniques based on rating prediction \cite{koren2009matrix} or in the calculation of the ranking order of preferred items from the user's perspective in learning-to-rank approaches \cite{karatzoglou2013learning,steck2013evaluation}. Note that the calculation of the ranking order of preferred items in learning-to-rank models is still within the recommendation algorithm.   

Controlling popularity bias can be incorporated into this step before the recommendation list is generated to the user. For example, Vargas et al. in \cite{vargas2014improving} propose a creative way to reduce the popularity bias and to increase the sales diversity by reversing the role of items and users. Generally, as discussed in Chapter \ref{background_chapter}, the goal of the recommender system is to recommend a list of items to the users. However, due to popularity bias, some items are not recommended to anyone or are not recommended enough. By reversing the role of items and users, the goal would be, for each item, find a set of users who might be interested in that item. This way, less popular items are no longer ignored and they will reach to a set of users which leads to reducing the bias on popular items. In another work, Adamopoulos and Tuzhilin in \cite{adamopoulos2014over} propose a modification to the standard neighborhood-based recommender systems to handle the over-specialization and concentration bias of the recommendations. The most important step in the neighborhood-based algorithms is the selection of the most similar users to the target user in order to use their rating pattern to find interesting and relevant items to be recommended to the target user. However, selecting the most similar users often lead to dictating the preferences of some users (those that appear in many neighborhood lists of other users) on other users. Therefore, authors in this paper use a probabilistic approach to select the neighbors for each user instead of selecting the most similar neighbors. Authors assign an initial weight to each candidate neighbor according to its distance from the target user and then the candidates are sampled, without replacement, proportionally to their assigned weights. By having a more diverse set of neighbors, as opposed to top most similar ones, the recommendations will be less prone to concentrate only on popular items. 

Incorporating the popularity bias control in the rating prediction step in factorization models often is done by modifying the objective function used to minimize the rating prediction error. For instance, Kamishima et al. in \cite{DBLP:conf/recsys/KamishimaAAS14} introduced the notion of neutrality in recommendation algorithms with respect to a certain feature or characteristic. In particular, the authors explore this concept in correcting the popularity bias in recommendations. Their method is a variant of the probabilistic matrix factorization model that adopts a constraint term for enhancing neutrality. The neutrality of a user towards item popularity (or any other aspect) is calculated using users' profiles based on how tolerant they are with respect to popular items. In another work, \cite{coba2018novelty} Coba et al. propose a regularization-based technique for counteracting popularity bias in a matrix factorization-based technique by adding an extra constraint to the general matrix factorization objective function that minimizes the distance between the popularity of a given item and the average popularity of the items in the users' profile. Therefore we can say this method is a personalized approach for mitigating popularity bias. 

The model-based techniques can be also used for improving the diversity of the recommendations. For instance, Wasilewski and Hurley in \cite{wasilewski2016incorporating} proposed a modification of a ranking-based recommendation algorithm ($RankALS$) by adding a regularization term to its objective function that minimizes the similarity (or increases the diversity) of the items within each list. Generally speaking, many other types of recommendation properties such as novelty, diversity and even fairness can be incorporated into the underlying recommendation algorithm as a model-based technique \cite{wasilewski2019bayesian,zhu2018fairness,lee2014fairness}.

\subsection{Proposed Model-based Technique Using Regularization} \label{model_based_RG}
\noindent

In this section, I present a flexible framework for a model-based popularity mitigation built on the top of a learning-to-rank algorithm. When applied to long-tail items, this approach enables the system designer to tune the application to achieve a particular trade-off between ranking accuracy and the inclusion of long-tail items. However, the framework is sufficiently general that it can be used to control recommendation bias for or against any group of items. 



\textit{Learning to Rank}
\noindent

I focus on matrix factorization approaches to recommendation
in which the training phase involves learning a low rank
$k \times |U|$ latent user matrix $P$ and a low-rank $k \times |I|$ latent item matrix $Q$, such that the estimated rating $\hat r_{ui}$ can be expressed as
$\hat{r}_{ui}=p^{T}_{u} q_{i}$ where $p^{T}_{u}$ is the $u^{th}$ row of $P$, $q^{T}_{i}$ is the $i^{th}$ row of $Q$, and $k$ is the chosen dimension of the latent space. $P$ and $Q$ are learned through the minimization of an accuracy-based objective~\cite{koren2015advances}. 

In this method, I focus on the learning-to-rank problem and I add a regularization term to its objective function that adjusts the scoring and gives a higher chance to less popular items to be included in the recommendation lists. Generally, the objective function of any factorization-based technique is based on minimizing a loss function that measures the error of a recommendation algorithm. In rating prediction models, this error is the squared difference between the original rating a user has given to an item and the predicted rating returned by the algorithm ($(r_{ui}-\hat{r}_{uj})^2$). In a learning-to-rank method, however, the error is the difference between the original rating and predicted rating between a pair of items. In other words, in learning-to-rank, what matters is the correct order of any pair of items in the user's profile and in the recommendations ($((\hat{r}_{ui}-\hat{r}_{uj})-(r_{ui}-r_{uj}))^{2}$). That is, if item $i$ is above item $j$ in a user's profile based the rating the user has given to these two items, in the recommendations this order should be also the same. 

For this work, I use the objective function proposed in \cite{jahrer2012collaborative} for a learning-to-rank recommendation system and further developed in \cite{takacs2012alternating}:\footnote{A number of such objectives have been proposed in the literature and the regularization method I propose here could be incorporated with any such pair-wise objective.}

\begin{equation}\label{learning_to_rank}
    \sum_{u \in U}\sum_{i \in I} c_{ui} \sum_{j \in I}s_{j} [(\hat{r}_{ui}-\hat{r}_{uj})-(r_{ui}-r_{uj})]^{2}+\beta (\left\|Q\right\|^2+\left\|P\right\|^2)
\end{equation}

The role of $c_{ui}$ is to select user-item pairs corresponding to positive feedback from all possible pairs. I consider the implicit feedback case in which $c_{ui}$ = 0 if $r_{ui}$ = 0, and 1 otherwise. $s_{j}$ is an importance weighting for item $j$. In this work, because I am trying to reduce the influence of popular items, I use uniform importance weighting $s_j=1$, for all $j$. $\beta$ is the standard regularisation parameter of norm-based regularization to avoid over-fitting.

\textit{Fairness-aware Regularization}
\noindent

Regularization is one of the main techniques to control the behavior of an objective function. Generally speaking, it is mainly used to avoid overfitting in machine learning models \cite{alpaydin2020introduction} by adding an additional penalty term in the error function. This additional term controls the excessively fluctuating function such that the coefficients don't take extreme values. For example, the terms $\beta (\left\|Q\right\|^2+\left\|P\right\|^2)$ in equation \ref{learning_to_rank} is a form of regularization that controls the magnitude of the entries in $P$ and $P$ to discourage the complexity of the model and avoid overfitting.

In my work, I use regularization as a tool to add extra constraint to the original objective function such that, in addition to ranking accuracy, the fairness of representation between popular and less popular items is taken into account in the optimization process. This work was inspired by the work of Wasilewski and Hurley~\cite{wasilewski2016incorporating} in which regularization is used to enhance recommendation diversity in a learning-to-rank setting, specifically using the RankALS algorithm~\cite{takacs2012alternating}. The authors calculate a pair-wise dissimilarity matrix $D$ for all pairs of items, demonstrate that intra-list diversity (ILD) can be represented in terms of this matrix, and then experiment with different ways that increased ILD can be made into an optimization objective.

Following a similar approach, I explore the use of regularization to control the popularity bias of a recommender system. I start with an optimization objective of the form:

\begin{equation}
     \min\limits_{P,Q} acc(P,Q)+ \lambda  reg(P,Q)
\end{equation}
where $acc(.)$ is the accuracy objective similar to Equation \ref{learning_to_rank}, $reg(.)$ is our regularization term, and $\lambda$ is a coefficient for controlling the effect of regularizer.

Our goal therefore is to identify a regularization component of the objective that will be minimized when the distribution of recommendations is fair in terms of popular versus less popular (long-tail) items. I define a \textit{fair recommendation list} as one that achieves a 50/50 balance between long-tail ($M \cup T$) and short-head ($H$) items which is referred to as equality of attention (or opportunity) in the literature \cite{hardt2016equality}. 

Wasileski and Hurley start from a dissimilarity matrix $D$, which contains all pair-wise dissimilarities between items. The regularizer is computed from $D$ in such a way that it pushes the optimization solution towards recommendation lists that balance ranking accuracy with intra-list diversity (ILD). 

In my case, I do not aim for diversifying the recommendation lists in terms of content (movie genre, music genre etc) but rather based on different item groups according to their popularity. This framework is a binary method meaning it needs two groups of items so it can balances the recommendations between them. I define the first group as popular items (short-head) $H$. For the second group which represents long-tail items, I combine the $M$ and $T$ groups defined in Chapter \ref{data_method} and consider them as long-tail items $\gamma$ ($\gamma=M \cup T$). I also define a co-membership matrix $D$, over these sets such that, for any pair of items $i$ and $j$, $d(i,j)=1$ if $i$ and $j$ are in the same set and $0$ otherwise.

Our equivalent for intra-list distance is a measure of the lack of fairness for the two sets of items in a given recommendation list to user $u$, $\ell_u$. I define intra-list binary unfairness (ILBU) as the average value of $d(i,j)$ across all pairs of items $i,j$. 

\begin{equation}
     ILBU(\ell_{u})=\frac{1}{n(n-1)}\sum_{i,j \in \ell_{u}} d(i,j)
\end{equation}

\noindent where $n$ is the number of items in the recommendation list. The fairest list is one that contains equal number of items from each set, which can be easily seen: if we start with a list containing only $H$ items and replace an item from $H$ with one from $\gamma$, the highest number of non-zero pairs occur when there is an equal number of items from $H$ and $\gamma$. If every item in the set are from the same group (either $H$ or $\gamma$), $ILBU$ will be high since all d(i,j) pairs are 1 which means higher unfairness.

Note that I do not in my definition of fairness require that each item have an equal chance of being recommended. There are many fewer short-head items and we are, essentially, allocating half of the recommendation list to them.\footnote{Note that it would be possible to adjust the balance between long-tail and short-head items by creating a smaller penalty for larger numbers of short-head items. These possibilities I leave for future work.} The 50/50 heuristic is a starting point for exploring the effectiveness of our regularization-based technique.

As the ILBU value has the same form as the ILD measure used in \cite{wasilewski2016incorporating}, I can take advantage of their experience in deriving various regularization terms from it. I chose the LapDQ regularizer, which performed well in Wasilewski and Hurley's experiments. LapDQ is based on the Laplacian $L_D$ of the $D$ matrix and has the form $tr(Q^{T}L_{D}Q)$, where $tr$ is the trace of the matrix. This version of the regularizer only influences the item factors and therefore the $p$ step of the RankALS algorithm, which computes and modifies the user factors, is unchanged. 

The details of the derivation of the regularizer is not the focus of this dissertation and readers are referred to \cite{wasilewski2016incorporating} for further information. One way to understand the function of $LapDQ$ is to note that our $D$ matrix is block-structured, and has the same structure as the adjacency matrix for a graph consisting of two separate components of fully interconnected items. For such a graph, the partition is represented by the eigenvector of the Laplacian corresponding to the largest eigenvalue. In our case, we already know what this partition is -- it is the division of the nodes into short-head and long-tail items. The trace of $Q^{T}L_{D}Q$ is the product of each latent factor with the Laplacian of the membership matrix. If the item factors most resemble the partitioning eigenvector -- that is, they contain elements from both partitions in opposite signs, then this vector product, and therefore the matrix trace, is minimized. The regularizer therefore pushes each item factor in $Q$ to span both partitions as evenly as possible.



\textit{Evaluation}

A multi-stakeholder evaluation entails multiple metrics representing the perspective of different stakeholders on system performance. I examine the behavior of this proposed algorithm relative to different metrics of long-tail performance that I described in Chapter \ref{data_method}. Conventional metrics are those that look at a system's overall recommendation of long-tail items. I show these results, but in addition, I also use metrics that are sensitive to the performance across different groups of stakeholders as I discussed in Chapter \ref{data_method}: suppliers in different popularity categories, items in different popularity categories, and users with different levels of interests in item popularity. 

\begin{figure}
    \centering
    \SetFigLayout{3}{2}
 \subfigure[Precision, NDCG, Gini, APL,  Aggregate Diversity, and Long-tail coverage for our model-based approach on MovieLens data ]{\includegraphics[width=4in]{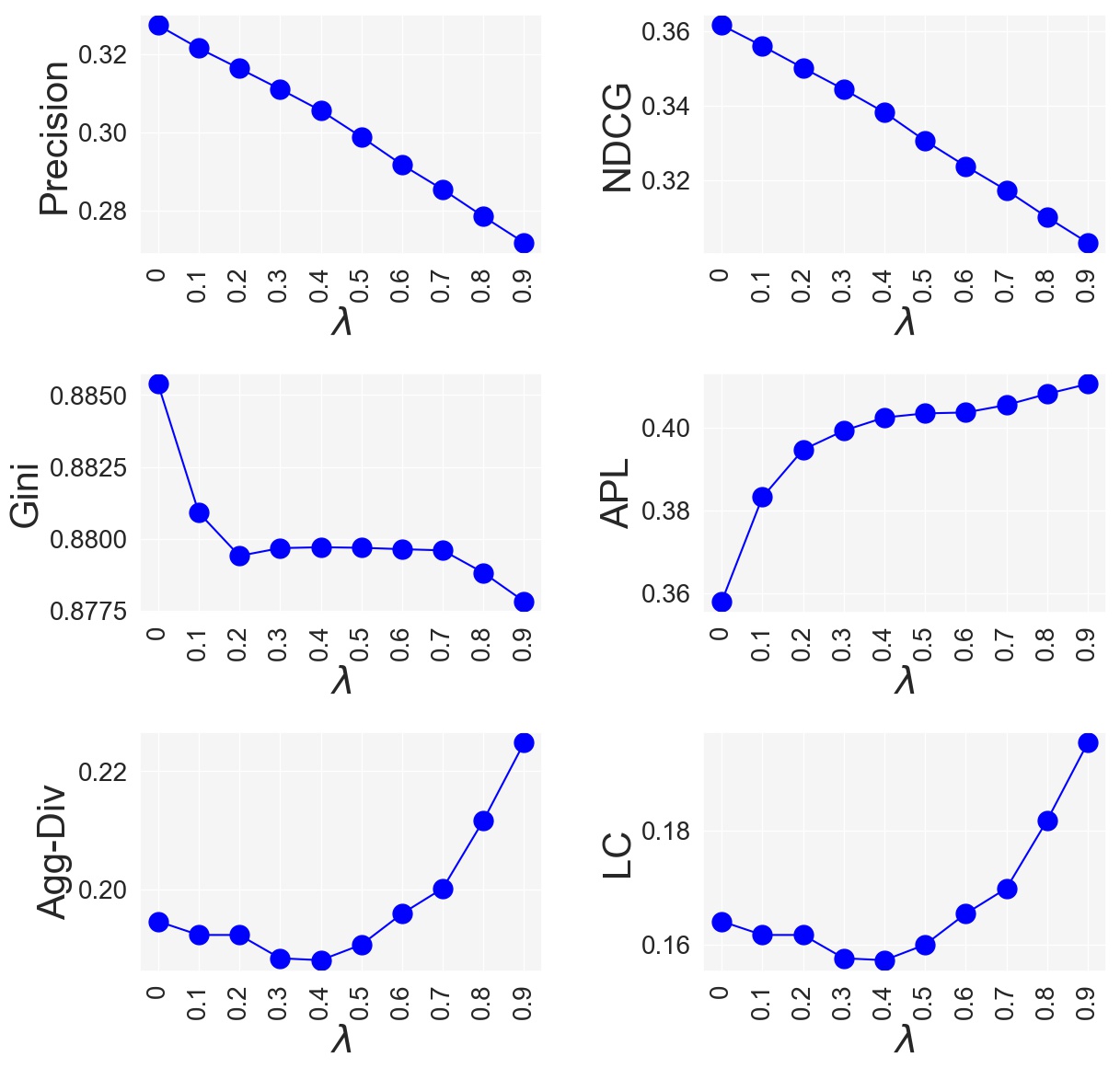}\label{fig:rg_results_movielens_1}}
 \subfigure[Equality of attention Supplier Fairness (ESF), Item Popularity Deviation (IPD), User Popularity Deviation (UPD) and Supplier Popularity Deviation (SPD) of our model-based technique on MovieLens data]{\includegraphics[width=4in]{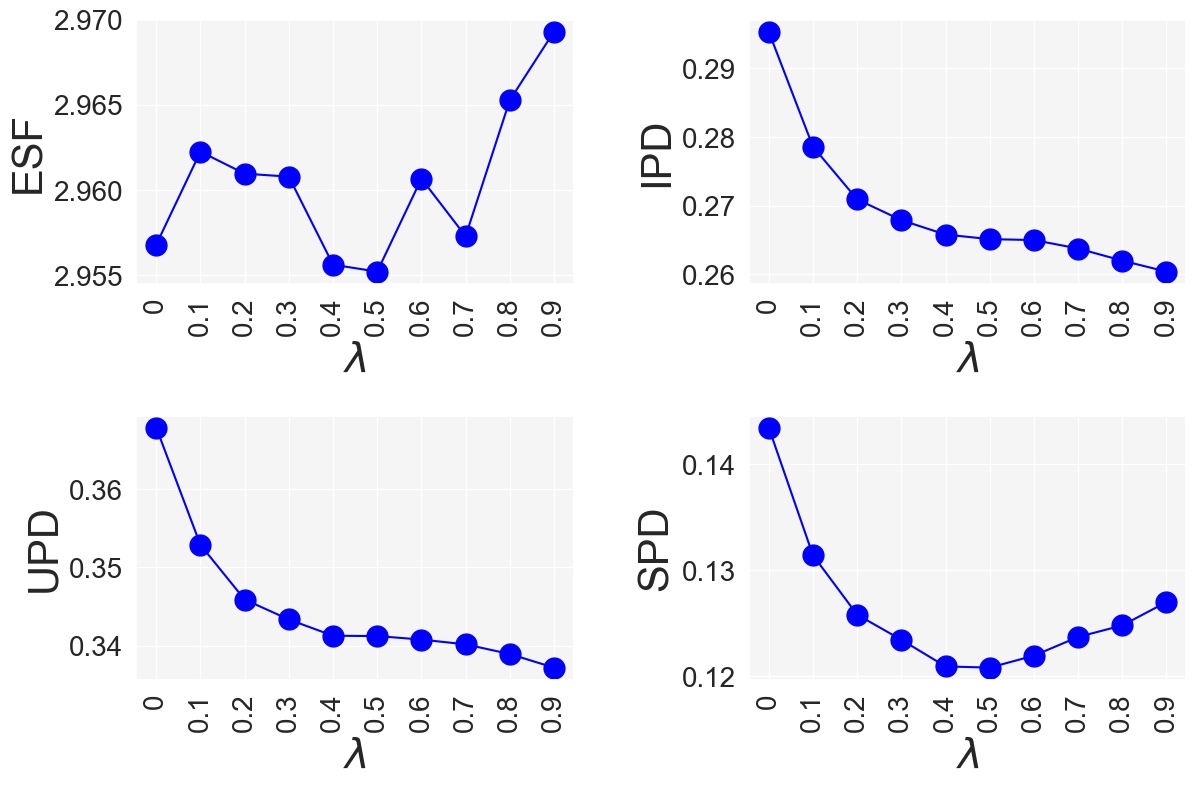}\label{fig:rg_results_movielens_2} }  
 \caption{Precision, NDCG, Gini, APL, Aggregate Diversity and Long-tail coverage, Equality of attention Supplier Fairness (ESF), Item Popularity Deviation (IPD), User Popularity Deviation (UPD) and Supplier Popularity Deviation (SPD) of our model-based technique on MovieLens data}

\end{figure}

\begin{figure}
    \centering
    \SetFigLayout{3}{2}
 \subfigure[Precision, NDCG,Gini, APL, Aggregate Diversity and Long-tail coverage for our model-based approach on Last.fm data]{\includegraphics[width=4in]{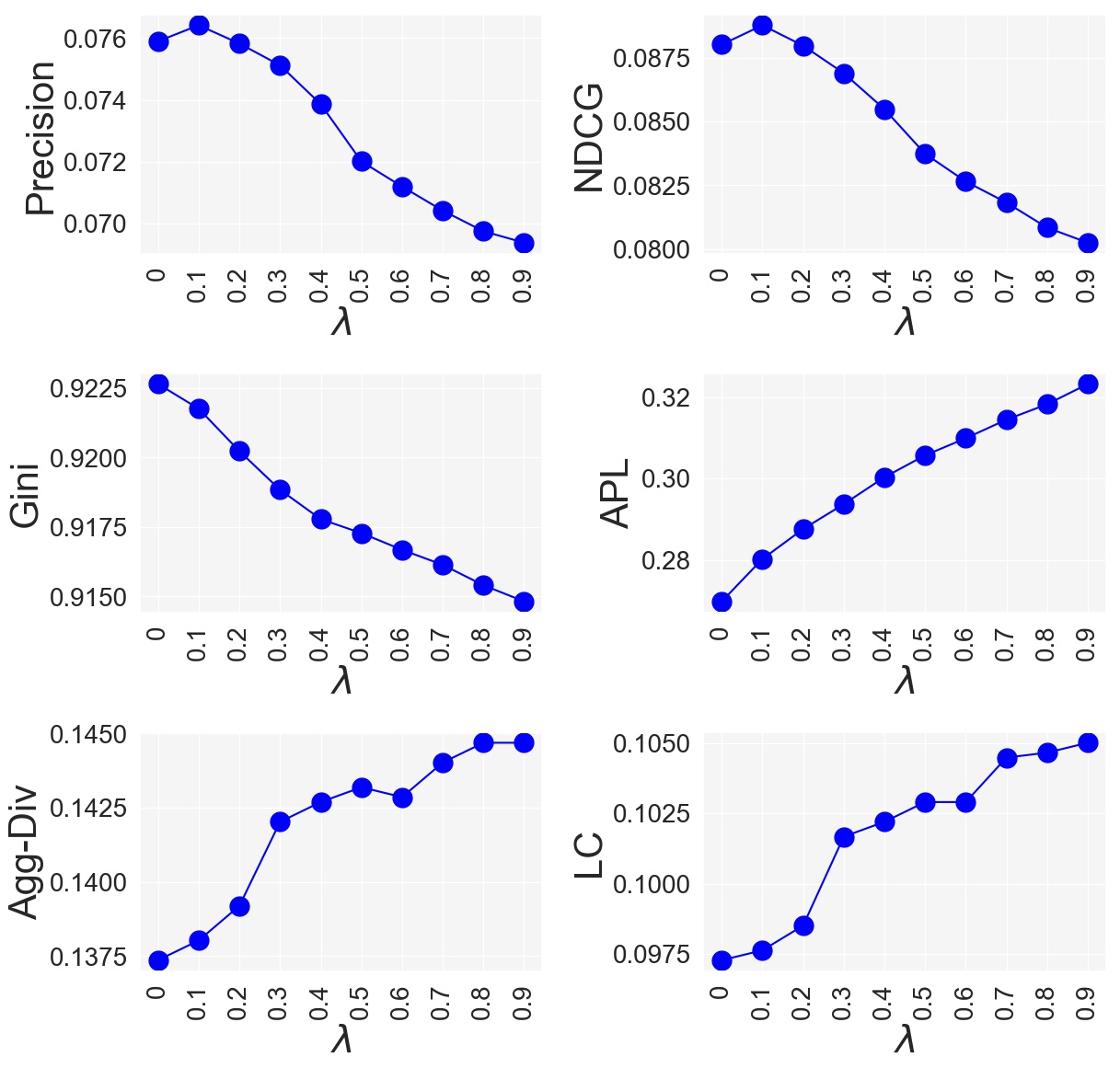}\label{fig:rg_results_lastfm_1}}
 \subfigure[Equality of attention Supplier Fairness (ESF), Item Popularity Deviation (IPD), User Popularity Deviation (UPD) and Supplier Popularity Deviation (SPD) of our model-based technique on Last.fm data]{\includegraphics[width=4in]{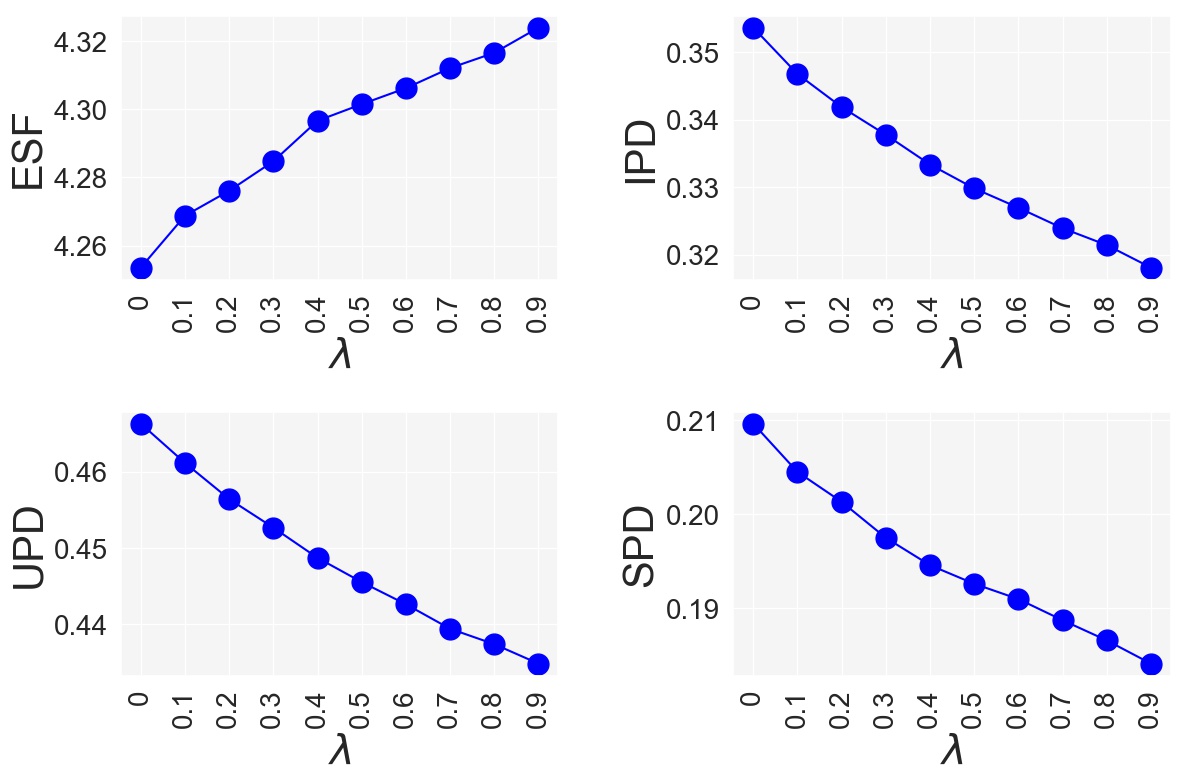}\label{fig:rg_results_lastfm_2}}   
 \caption{Precision, NDCG, Gini, APL, Aggregate Diversity and Long-tail coverage, Equality of attention Supplier Fairness (ESF), Item Popularity Deviation (IPD), User Popularity Deviation (UPD) and Supplier Popularity Deviation (SPD) of our model-based technique on Last.fm data }
 \label{fig:rg_results_lastfm}
\end{figure}

Figure \ref{fig:rg_results_movielens_1} shows the behavior of our model-based method when we vary the weight of the regularization: the $\lambda$ parameter. As we can see, the algorithm loses some accuracy (both precision and NDCG) when we increase the $\lambda$ and give more weight to the fairness criteria. The Gini slightly improves meaning the recommendations are more uniformly distributed compared to the base algorithm with no fairness component. 

Other metrics related to long-tail performance and fairness, as can be seen in Figure \ref{fig:rg_results_movielens_2} have all improved for the largest value of $\lambda$ but for some metrics such as $LC$, $ESF$, and $SPD$ the behavior of the algorithm is not completely monotonic. For instance, up until $\lambda=0.4$, the $LC$ actually slightly decreases and then it starts to increase for larger values of $\lambda$. However this anomaly is only seen on the MovieLens data. The trend for $ESF$ on MovieLens data seems to be unpredictable. However, looking at the scale for the vertical axis, we can see that the changes are not statistically significant($p$-value $>0.05$). $IPD$ and $UPD$, on the other hand, show a monotonic improvement over different values of $\lambda$. $SPD$ improves up until $\lambda=0.4$ and then it starts to go up again for larger values of $\lambda$. $APL$ on the other hand, which measures a more localized view of long-tail diversity in the individual lists and more in line with what I am optimizing for is improving when we increase the $\lambda$. I believe the reason for the behavior of $LC$ can be explained by the nature of our objective function. our $ILBU$ metric only looks at the individual lists in terms or the ratio of short-head versus long-tail items. Therefore, it is possible for an algorithm to make the individual lists more diverse in terms of item popularity (higher $APL$) but overall, the number of unique long-tail items in the recommendations across all users might decrease (low $LC$) as we also saw in our illustrative example in Figure \ref{fig:recs_example} in Chapter \ref{data_method}.  

Figure \ref{fig:rg_results_lastfm} shows the results on Last.fm data. On this dataset, unlike MovieLens, the metrics show a more monotonic behavior and we can see, generally, when we increase the $\lambda$, the algorithm loses accuracy (both Precision and NDCG) while other metrics all are improving.  


What we see in these results is similar to the findings of \cite{wasilewski2016incorporating}, namely that a regularization term incorporating list-wise properties of interest can be incorporated successfully into a learning-to-rank framework. In Wasilewski and Hurley's work, this was a similarity-based term that enabled them to enhance list-wise diversity; in our case, it is a group-membership-based term that enables us to enhance list-wise coverage of long-tail items.

Note in particular that because this loss function is local to individual recommendation lists, it does not aim to increase catalog coverage in long-tail directly. However, our results show that, for these datasets, the optimization of our local objective also increases overall coverage of less popular items which makes sense since when we increase the long-tail coverage of individual lists, the overall long-tail coverage may also increase. 

\section{Re-ranking Approach}\label{reranking_section}

\begin{figure*}
    \centering
    \includegraphics[width=4in]{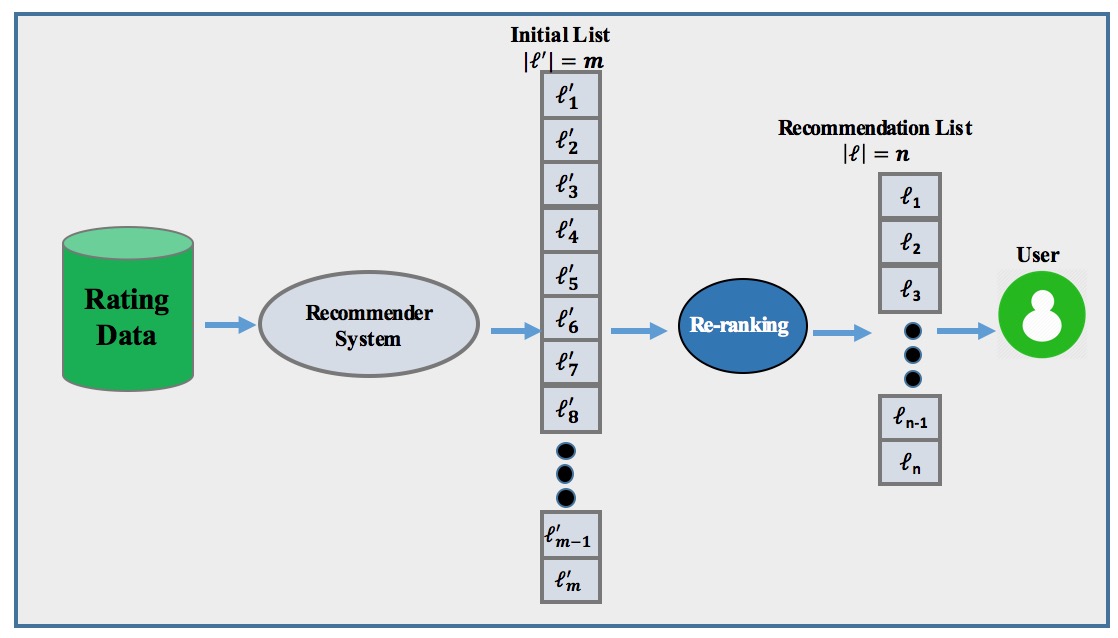}
    \caption{The schema for the re-ranking popularity bias mitigation}
    \label{fig:reranking}
\end{figure*}

The second class of approaches for mitigating popularity bias is based on the re-ranking of an original list generated by a given recommendation algorithm. The idea is, for each item in the original list, both the relevance of the item to the user and its popularity are taken into account so items with high relevance and also lower popularity could be pushed up to the top of the list \cite{adomavicius2009toward}. Figure \ref{fig:reranking} shows the general schema for this class of algorithms. As we can see, the recommendation component is not modified but, instead, the recommender system generates a larger list of recommendations $\ell^\prime$ of size $m$ and the re-ranking algorithm creates a shorter list $\ell$ of size $n$ ($m>>n$) that has better long-tail properties.

The advantage of the re-ranking techniques over the model-based ones is that they are not dependent on a particular recommendation algorithm. Unlike the model-based techniques, where popularity mitigation is done within the recommendation algorithm itself, the re-ranking methods do not modify the underlying recommendation algorithm but rather acts as a post-processing technique that could be applied on top of any existing recommendation algorithm.

There are many existing works for mitigating popularity bias that are based on the re-ranking approach. Miyamoto et al. \cite{miyamoto2018improving} proposed a post-processing re-ranking technique for reducing frequently recommended items. The algorithm, at first, predicts the ratings a user would give to a list of items. Then, it keeps those items whose predicted ratings is above a certain threshold and, for each of those items, the algorithm calculates how many times the item was recommended to other users. After that, the items are sorted based on this score and the items with the lowest score are put in the final recommendation list and the score for those items increases by one. This process repeats for every user. In another work \cite{adomavicius2011maximizing} propose a graph-based approach based on the maximum flow theory for improving the aggregate diversity of the recommendations. In this work, the max-flow value will be equal to the largest possible number of recommendations than can be made from among the available (highly predicted) items, where no user can be recommended more than $n$ items, and no item can be counted more than once, which is precisely the definition of the aggregate diversity. In a similar work which is also based on maximum flow theory \cite{antikacioglu2017post}, Antikacioglu et al. use minimum-cost network flow method to efficiently find recommendation sub-graphs that optimize diversity. Authors in this work define a target distribution of item exposure (i.e. the number of times each item should appear in the recommendations) as a constraint for their objective function. The goal is therefore to minimize the discrepancy of the recommendation frequency for each item and the target distribution. In addition, Jannach et al. in \cite{jannach2015recommenders} propose a personalized re-ranking technique that adjusts the relevance scores of the items so it can boost the predicted rating for less popular items to push them to the top of the list. The adjustment of relevance is done by taking into account the propensity of the users towards popular items and also the popularity of the recommended list and tries to minimize the norm-2 of this distance. In this dissertation, I also propose two re-ranking based techniques for controlling popularity bias that have the tolerance of the users towards item popularity incorporated in the algorithm.

\subsection{Re-ranking Technique 1: Fair Long-tail Promotion:}

I present a general and flexible approach for controlling the balance of item exposure in different portions of the item catalog as a post-processing phase for standard recommendation algorithms. Our work is inspired by \cite{santos2010exploiting} where authors introduced a novel probabilistic framework called \textit{xQuAD} for Web search result
diversification which aims to generate search results that explicitly account for various aspects associated with an under-specified query. One motivating example for the usefulness of such diversification in search results is the ambiguity of certain queries. For instance, if a user types the word \textit{Apple} in the search bar, s/he could be searcher for the fruit, or maybe the company. So, we can say, in this example, that the word apple has two aspects: 1) being a fruit and 2) being a company. Therefore, in order to make sure we are not limiting the choices for the user, the returned result is better to contain web pages related to both of those aspects. By doing so, we give the user the choice in selecting which of the search results s/he intended to find. The algorithm, re-ranks an initial list of search results so it contains a fair proportion of web pages related to each of these two different aspects. It is worth noting that a search query could be related to more than two aspects.

I adapt the xQuAD approach to the popularity bias problem. Here, the concept of diversification is slightly different as I no longer use aspects associated with each item. In my approach, each item is either popular and not. Therefore, I am dealing with a binary categorization and the goal of the re-ranking technique, is to return a list of recommendations that contains a balanced ratio of items belonging to either popular or less popular item categories. By doing that, we are helping the less popular items to have a better representation in the recommendation lists as, initially, they are under-represented. 

Given the nature of our proposed approach, it enables the system designer to tune the system to achieve the desired trade-off between accuracy and better coverage of long-tail, less popular items. 
 
\textit{xQuAD}

Result diversification has been studied in the context of information retrieval, especially for web search engines, which have a similar goal to find a ranking of documents that together provide a complete
coverage of the aspects underlying a query \cite{santos2015search}. EXplicit Query
Aspect Diversification (xQuAD) \cite{santos2010exploiting} explicitly accounts for the various aspects associated with an under-specified query. Items are selected iteratively by estimating how well a given document satisfies an uncovered aspect. 

In adapting this approach, I seek to recognize the difference among users in their interest in long-tail items. Uniformly-increasing diversity of items with different popularity levels in the recommendation lists may work poorly for some users. I propose a variant that adds a personalized bonus to the items that belong to the under-represented group (i.e. the long-tail items). The personalization factor is determined based on each user's historical interest in long-tail items.

\textit{Methodology}

I build on the \textit{xQuAD} model to control popularity bias in recommendation
algorithms. As described in section \ref{reranking_section}, I assume that for a given user $u$, a ranked recommendation list $\ell^\prime$ with size $m$ has already been generated by a base recommendation algorithm. The task of the modified xQuAD method is, therefore, to produce a new re-ranked list $\ell$ of size $n$ that manages popularity bias while still being accurate. 

The new list is built iteratively according to the following criterion:
\begin{equation}\label{eq:1}
    P(i|u)+\lambda P(i,u|i\notin \ell)
\end{equation}
where $P(i|u)$ is the likelihood of user $u \in U$ being interested in
item $i \in I$, independent of the items on the list so far as, predicted by the base recommender. The second term $P(i,u|i\notin \ell)$ denotes the likelihood of user u being interested in an item $i$ as an item not in the currently generated list $\ell$.

Intuitively, the first term
incorporates ranking accuracy while the second term promotes
diversity between two different categories of items (i.e. short head and long tail). The parameter $\lambda$ controls how strongly controlling popularity bias is weighted in general. The single item
that scores most highly under the equation \ref{eq:1} is added to the
output list $\ell$ and the process is repeated until $\ell$ has achieved the
desired length.

To achieve more diverse recommendation containing items from both short-head and long-tail items, the marginal likelihood $P(i,u|i\notin \ell)$ over both
item categories long-tail ($\gamma=M \cup T$) and short head ($H$) is computed by:

\begin{equation}\label{eq:2}
    P(i,u|i\notin \ell)=\sum_{c \in \{H , \gamma\}}P(i,u| c \notin \ell)P(c|u)
\end{equation}

Following the approach of \cite{santos2010exploiting}, I assume that the remaining
items are independent of the current contents of $\ell$ and that the items
are independent of each other given the short-head and long-tail categories. Under these assumptions, we can compute $P(i,u| c \notin \ell)$ in Eq.\ref{eq:2} as

\begin{equation} \label{eq:3}
    P(i,u|c \notin \ell)=P(i|c)P(c \notin \ell)=
    P(i|c)\prod_{j \in \ell} (1-P(j|c \in \ell))
\end{equation}

By substituting equation \ref{eq:3} into equation \ref{eq:2}, we can obtain
\begin{equation} \label{eq:4}
     score=(1-\lambda) P(i |u)+\lambda\sum_{c \in \{H, \gamma\}}P(c|u)P(i|c)\prod_{j \in \ell}(1-P(j|c \in \ell))
\end{equation}
where $P(i|c)$ is equal to 1 if $i \in c$ and 0 otherwise. 

I measure $P(j|c \in \ell)$ as the ratio of items ($j \in \ell$) that covers category $c$. 

The likelihood $P(c|u)$ is the measure of user preference over different item categories. In other words, it measures how much each user is interested in short head items versus long tail items. I calculate this likelihood by the ratio of items in the user profile which belong to category $c$. 

In order to select the next item to add to $\ell$, I compute a re-ranking score for each item in $\ell^\prime \backslash \ell$ according to Eq. \ref{eq:4}. For an item $i^\prime \in c$, if $\ell$ does not cover $c$, then an additional positive term will be added to the estimated user preference $P(i^\prime|u)$. Therefore, the chance that it will be selected is larger, balancing accuracy and popularity bias. 


\textit{Evaluation}

As with other re-ranking techniques, our proposed re-ranking method also needs a base algorithm to generate the initial list of recommendations for post-processing. Since our model-based algorithm I proposed in section \ref{model_based_RG} works on the RankALS algorithm, for comparison purposes, I chose this algorithm as the base and I call it \textit{Base}. 

Similar to \cite{kaya2019comparison} I set the size of the recommendations generated by the $Base$ algorithm to 100 ($m=100$), and the size of the final recommendation list is 10 ($n=10$).

\begin{figure}
    \centering
    \SetFigLayout{3}{2}
 \subfigure[Precision, NDCG, Gini, APL, Aggregate Diversity and Long-tail coverage for our fairness-aware re-ranking approach on MovieLens data ]{\includegraphics[width=4in]{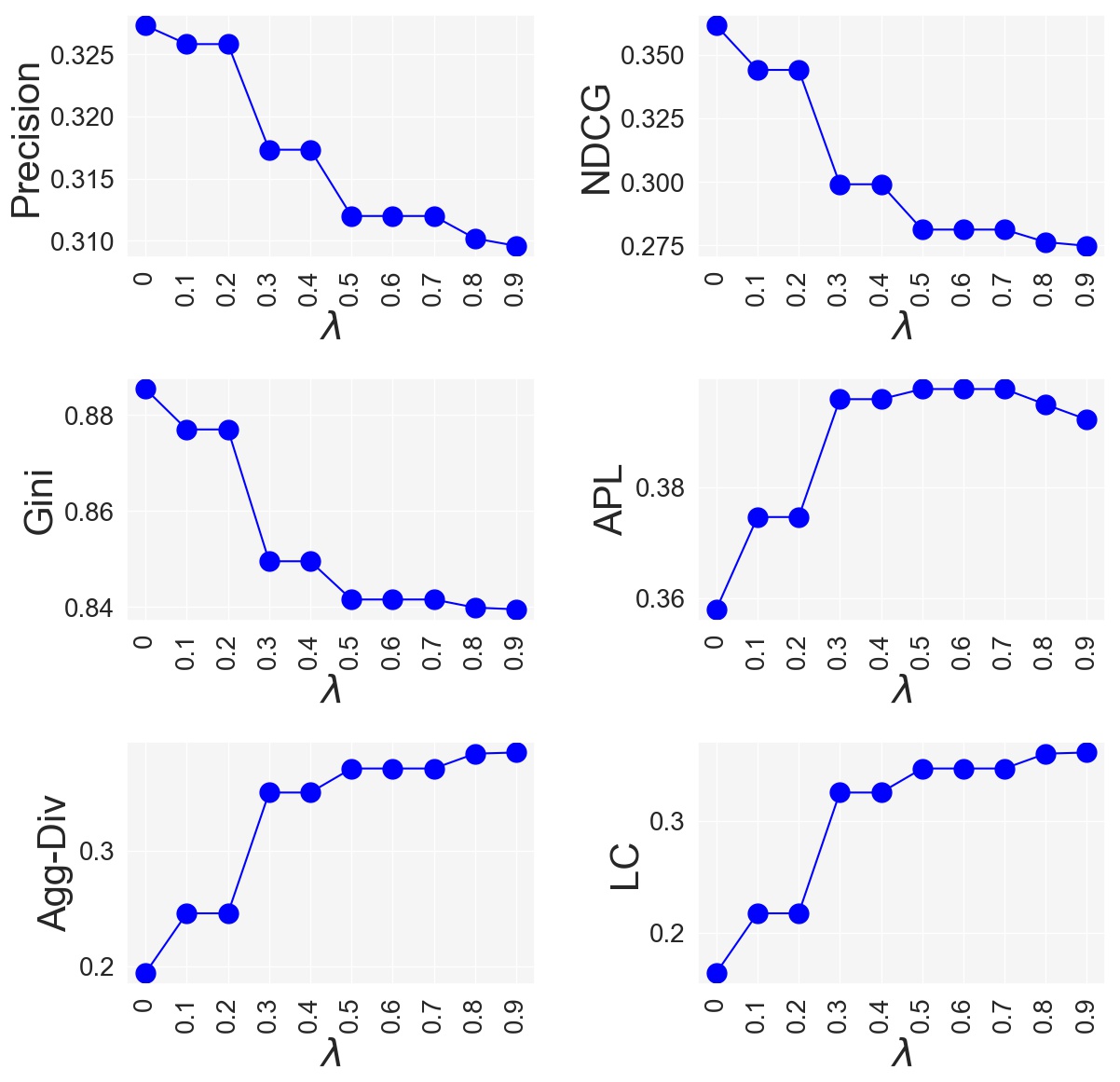}\label{fig:xq_results_movielens_1}}
 \subfigure[Equality of attention Supplier Fairness (ESF), Item Popularity Deviation (IPD), User Popularity Deviation (UPD) and Supplier Popularity Deviation (SPD) of our fairness-aware re-ranking approach on MovieLens data]{\includegraphics[width=4in]{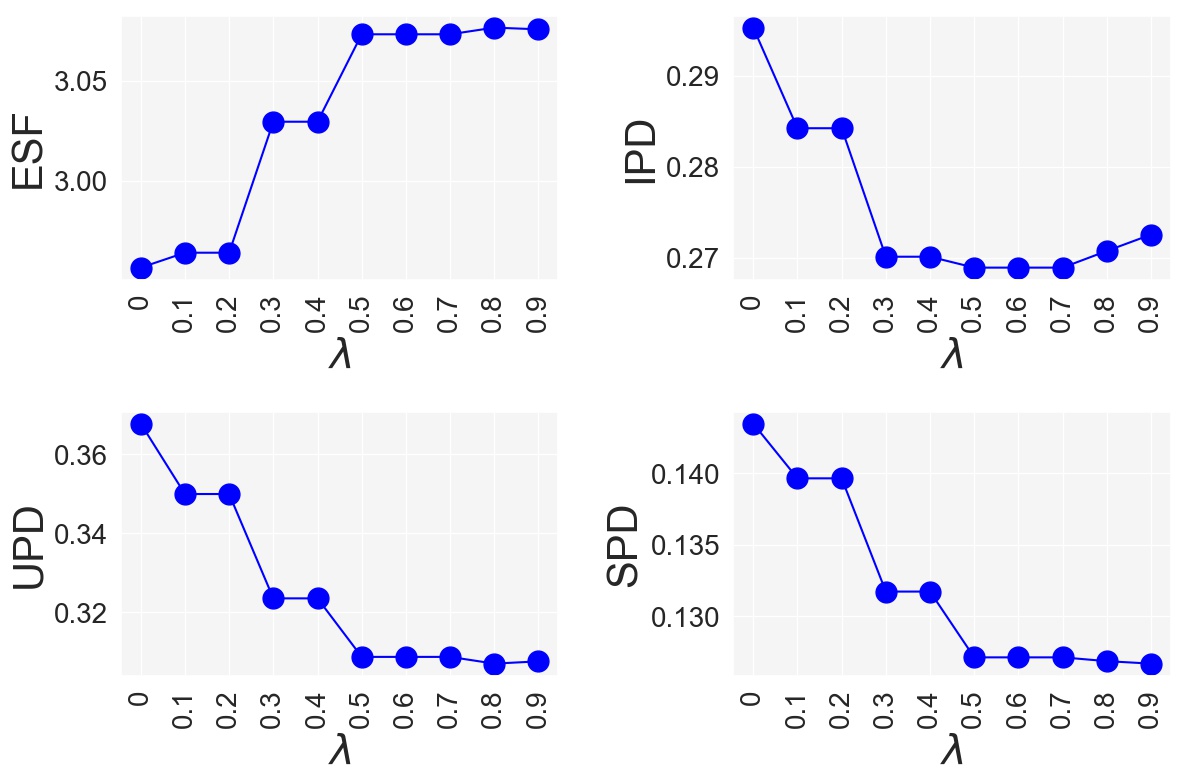}\label{fig:xq_results_movielens_2} }  
 \caption{Precision, NDCG, Gini, APL, Aggregate Diversity and Long-tail coverage, Equality of attention Supplier Fairness (ESF), Item Popularity Deviation (IPD), User Popularity Deviation (UPD) and Supplier Popularity Deviation (SPD) of our fairness-aware re-ranking on MovieLens data}

\end{figure}

\begin{figure}
    \centering
    \SetFigLayout{3}{2}
 \subfigure[Precision, NDCG, Gini, APL, Aggregate Diversity and Long-tail coverage for our fairness-aware re-ranking approach on Last.fm data]{\includegraphics[width=4in]{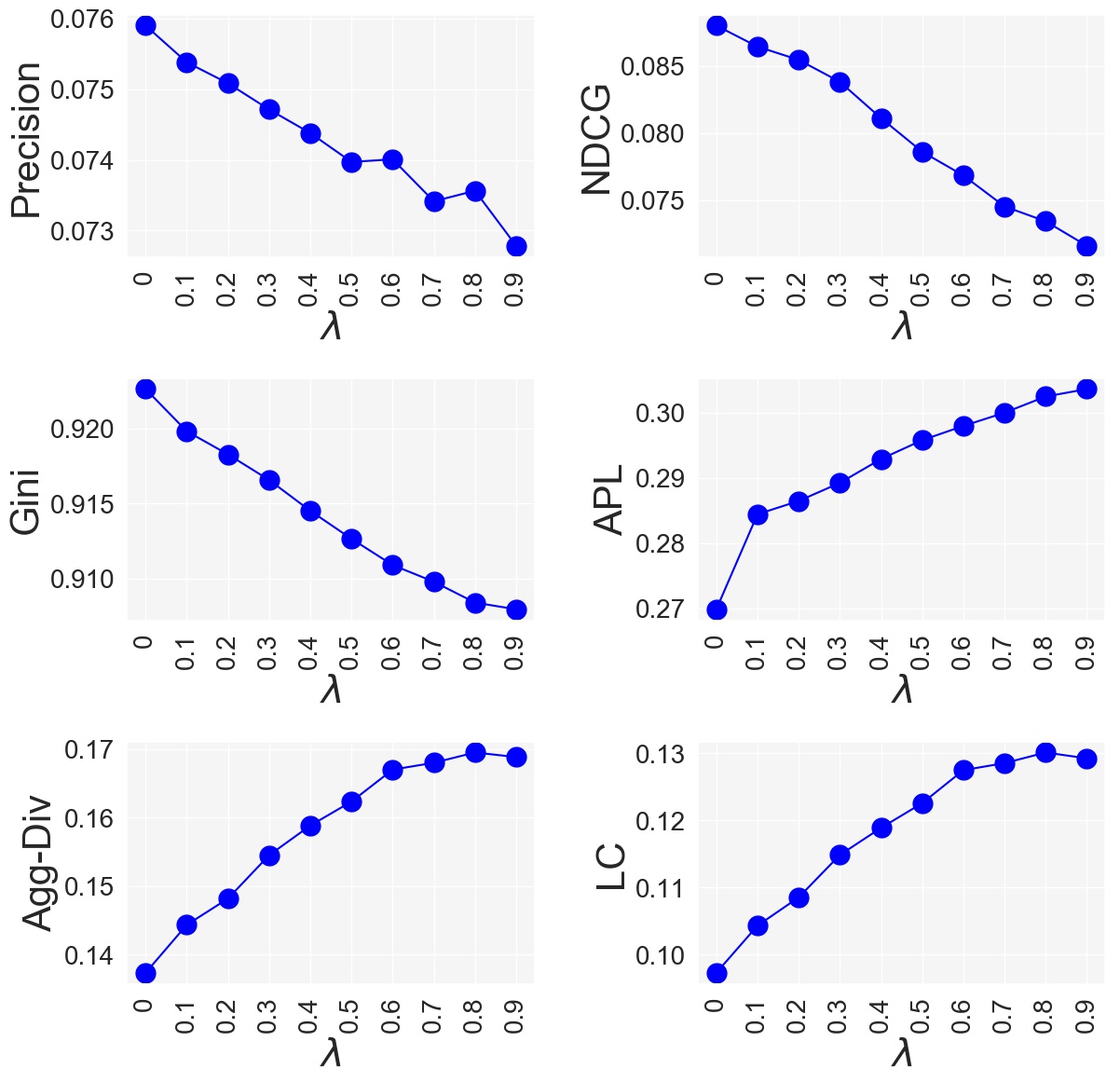}\label{fig:xq_results_lastfm_1}}
 \subfigure[Equality of attention Supplier Fairness (ESF), Item Popularity Deviation (IPD), User Popularity Deviation (UPD) and Supplier Popularity Deviation (SPD) of our fairness-aware re-ranking approach on Last.fm data]{\includegraphics[width=4in]{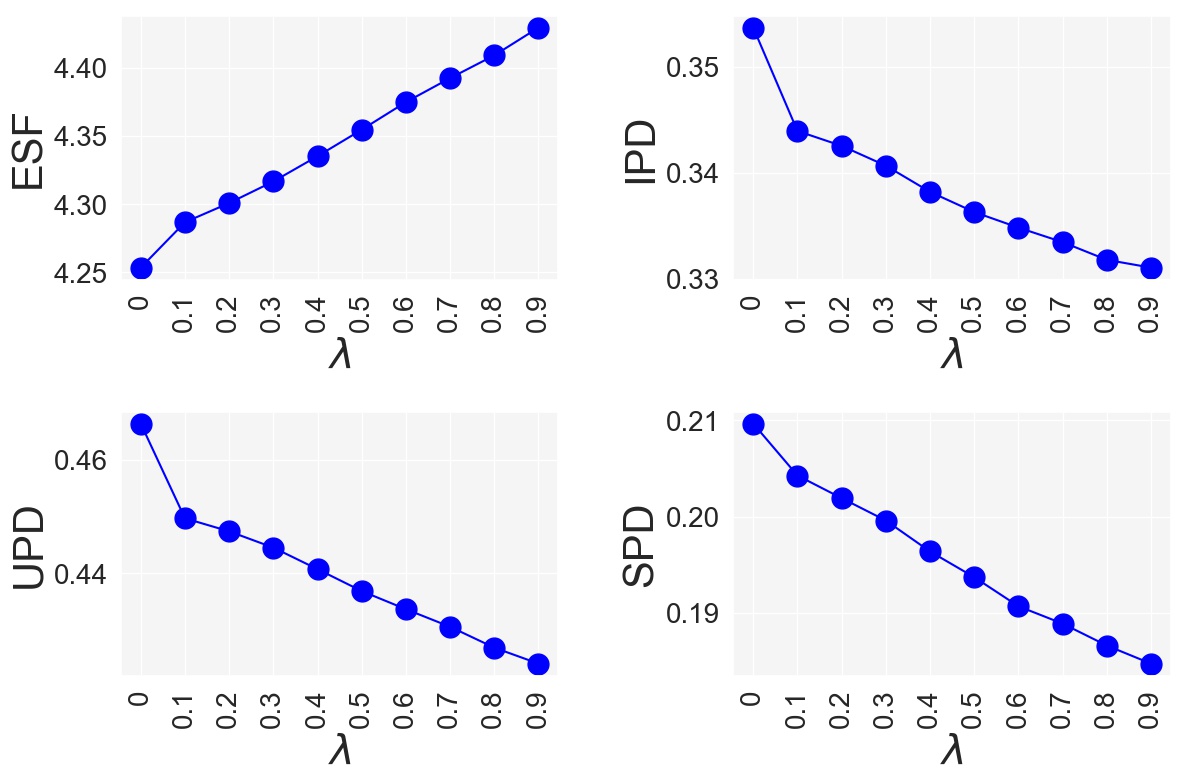}\label{fig:xq_results_lastfm_2}}   
 \caption{Precision, NDCG, Gini, APL, Aggregate Diversity and Long-tail coverage, Equality of attention Supplier Fairness (ESF), Item Popularity Deviation (IPD), User Popularity Deviation (UPD) and Supplier Popularity Deviation (SPD) of our fairness-aware re-ranking on Last.fm data}
 \label{fig:xq_results_lastfm}
\end{figure}

Figure \ref{fig:xq_results_movielens_1} shows the behavior of our fairness-aware re-ranking method for different values of $\lambda$ which controls the weight of the fairness(i.e. long-tail enhancement) compared to accuracy. Similar, to model-based method we saw in previous section, we can see that the re-ranking algorithm also loses accuracy (precision and NDCG) when we increase the $\lambda$. The Gini improves when we increase the $\lambda$ parameter indicating the algorithm achieves a more uniformly distributed recommendations across all users. 

The long-tail enhancement and fairness metrics, however, are all improving as we can be seen in Figure \ref{fig:xq_results_movielens_1} and \ref{fig:xq_results_movielens_2}. Although, almost for all metrics, we see a stepwise behavior meaning the curve goes up / down, stays constant at some points, and then goes up / down again. Looking at the scale of the y-axis, we can see this re-ranking approach performs better than the model-based technique when it comes to long-tail enhancement and fairness. I will compare the performance of all three algorithms proposed in this dissertation later in section \ref{calibration}.

\subsection{Re-ranking Technique 2:  Calibrated Popularity} \label{calibration}
The second re-ranking method I propose in this dissertation is \textit{Calibrated Popularity}.

One of the problems of algorithmic popularity bias is the tendency of over-recommending popular items even to those users who do not have much interest in popular items. However as we saw in Chapter ~\ref{pop_bias} there are many users who are interested in less popular items and hence exclusively recommending popular items to these users would hurt their experience in interacting with the recommender system. Therefore, I propose an algorithmic approach based on the novel notion of \textit{popularity calibration} that provides a comprehensive solution to the problem of popularity bias by addressing the needs of multiple stakeholders. A recommendation list is calibrated based on popularity when the range of items it covers matches the user's profile in terms of item popularity. For example, if a user has 20\% $H$ items, 40\% $M$ items and 40\% $T$ items in her profile, we call the recommendations to be calibrated when the ratio of each item category in the recommended list is consistent with the aforementioned ratios. I show that my approach, while yielding more calibration recommendations for users, also results in fairer recommendations from the suppliers' perspective. 

\begin{figure}
    \centering
    \includegraphics[width=6in]{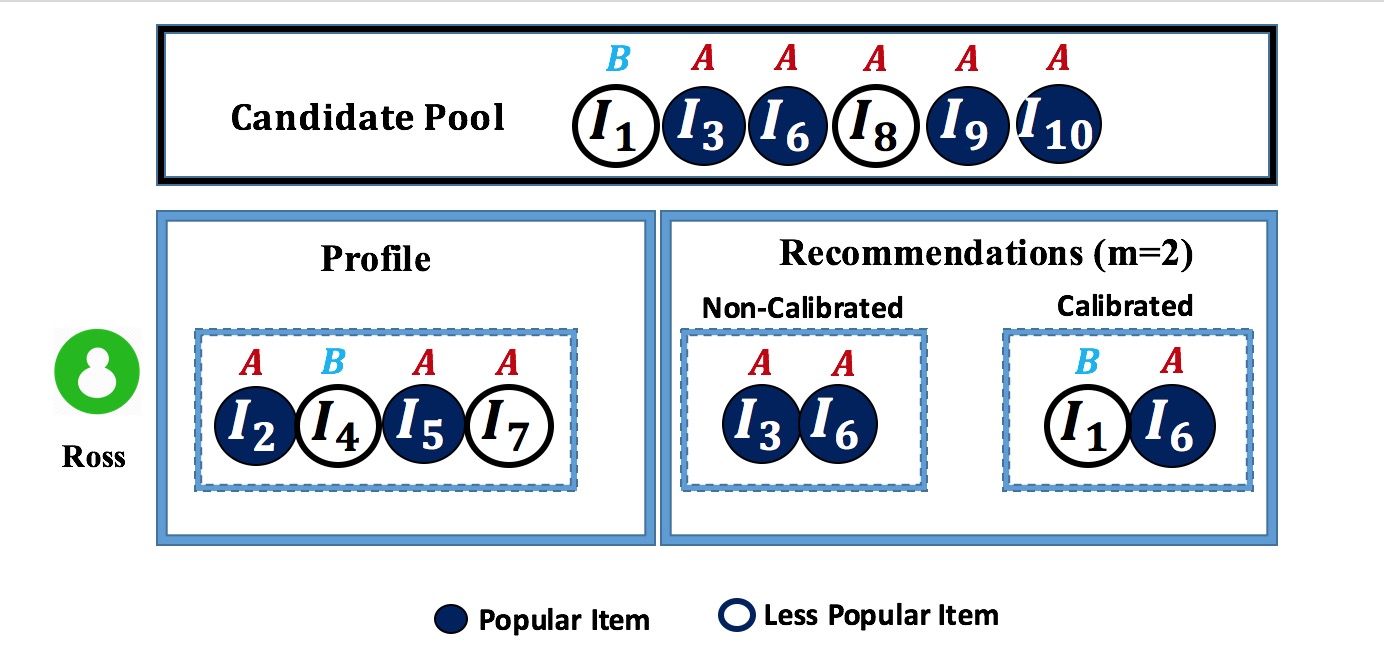}
    \caption{Popularity calibration example}
    \label{fig:rec_graph2}
\end{figure}

\subsubsection{Motivating Example}

Figure ~\ref{fig:rec_graph2} shows a user whose name is Ross\footnote{Ross is the name of the first author's favorite character in Friends.}. Out of all the items in a movie recommender's catalog, Ross has interacted with (or liked) four items $I_2$, $I_4$, $I_5$, and $I_7$. Suppose dark circles are popular items ($H$) and white ones are less popular items (either $M$ or $T$). That means his profile consists of an equal ratio of popular versus non-popular items (50\% each), and I assume that the contents of his profile represents the scope of his interest in movies. In addition, each item is provided by one of the suppliers $A$ or $B$. $A$ (red color) who is a popular supplier (based on the average popularity of their items) owns $I_2$, $I_3$, $I_5$, $I_6$, $I_7$, $I_8$, $I_9$ and $I_{10}$. $B$ who is a less popular supplier owns $I_1$ and $I_4$. Note that a popular supplier could also own some less popular items ($A$ owns $I_8$). Since Ross is equally interested in popular and less popular items based on his profile, a well-calibrated recommender should seek to deliver the same ratio in his recommendations, reflecting the diversity of interests that he has shared with the system. Assuming the size of the recommendation set is 2 (for illustration purposes), he should get one popular and one less popular recommendation. A recommendation algorithm influenced by popularity bias would be more likely to generate a recommendation list containing only popular items: ``Non-Calibrated'' in the figure. With this set of recommendations, the supplier $B$ is completely out of the picture and has received zero exposure even though Ross's profile does show interest in less popular movies. The calibrated recommendations, on the right side, shows preferred situation where the user has received recommendations that match his interests across the item popularity spectrum, while at the same time both suppliers have received exposure. This shows that the popularity calibration for the users would also benefit the suppliers.

\subsubsection{Calibrated Popularity}
\label{sec:calibration}

Our proposed technique, \textit{Calibrated Popularity} (CP), is a re-ranking method. I build on the work of Steck~\cite{steck2018calibrated} in using re-ranking to provide results that better match the distributional properties of user profiles. In Steck's case, the distribution of interest was the distribution of genres across recommended movies. In our case, it is the distribution of the item popularity I seek to control\footnote{Note that unlike the genre labels in \cite{steck2018calibrated} where it is possible for a movie to have multiple genres, each item only belongs to one item group.}. 

\begin{figure}
    \centering
    \includegraphics[width=4in]{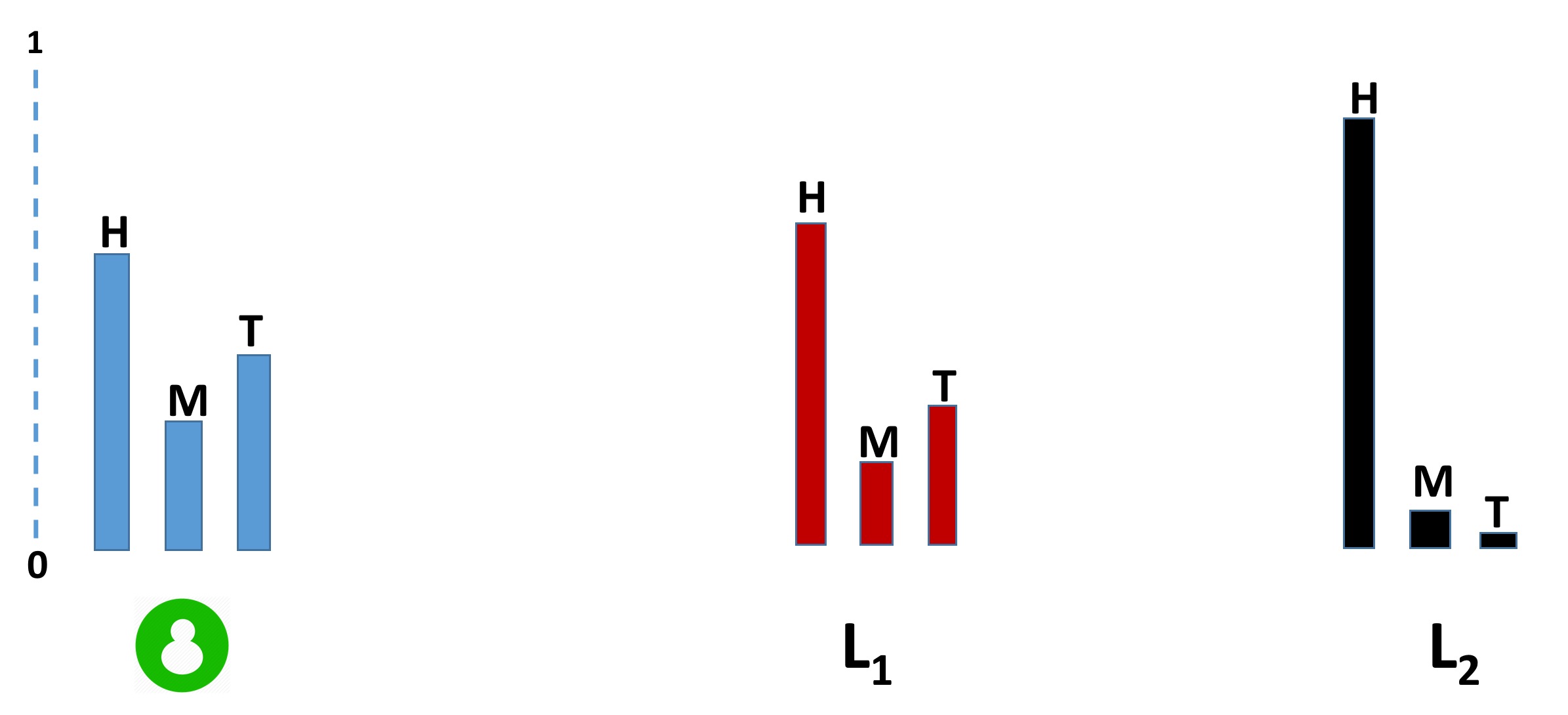}
    \caption{An illustration of the statistical distance between the distribution of item popularity in the user's profile and in the recommendations.}
    \label{fig:calibration}
\end{figure}

Figure \ref{fig:calibration} shows the distribution of three item groups $H$, $M$, and $T$ in a given user's profile. The distribution of the popularity item groups for two recommendation lists $L_1$ and $L_2$ are also shown. We can see that the ratios of different item groups in $L_1$ better match that of user's. The ratios of $H$, $M$, and $T$ in $L_2$, on the other hand, has no similarity with the those ratios in the user's profile. Therefore, we can say that the distribution of item popularity groups in $L_1$ has smaller distance to that of user's profile and, by definition, is more calibrated. 

CP algorithm operates on an initial recommendation list $\ell'$ of size $m$ generated by a base recommender to produce a final recommendation list $\ell$ of size $n$ ($m>>n$). Similar to \cite{steck2018calibrated}, I measure distributional differences in the categories (groups) to which items belong $C=\{c_1$,$c_2$,...,$c_k\}$. For our purposes, these are the three $H$, $M$ and $T$ item groups described above (i.e. $C=\{H,M,T\}$).

To do this comparison, I need to compute a discrete probability distribution $P$ for each user $u$, reflecting the popularity of the items found in their profile $\rho_u$ over each item group $c \in C$. I also need a corresponding distribution $Q$ over any given recommendation list $\ell$, indicating what item popularity groups are found among the listed items. For measuring the interest of each user towards each item popularity group, I use Vargas et al.'s \cite{vargas2013exploiting} measure of category propensity. Specifically, I calculate the propensity of each user $u$ towards each item group $c \in C$ in her profile $\rho_u$ ($p(c|u)$) and the ratio of such item group in her recommendation list $\ell_u$ ($q(c|u)$) as follows:  

\begin{equation}
p(c|u)=\frac{\sum_{i \in \rho_u}r(u,i)\mathbbm{1}(i \in c)}{\sum_{c_j \in C}\sum_{i \in \rho_u }r(u,i)\mathbbm{1}(i \in c_j)} \;, \;\;\;\;\; q(c|u)=\frac{\sum_{i \in \ell_u}\mathbbm{1}(i \in c)}{\sum_{c_j \in C}\sum_{i \in \ell_u }\mathbbm{1}(i \in c_j)}
\end{equation}   
$\mathbbm{1}(.)$ is the indicator function returning zero when its argument is False and 1 otherwise.

In order to determine if a recommendation list is calibrated to a given user, I need to measure the distance between the two probability distributions $P$ and $Q$. There are a number of metrics for measuring the statistical distance between two distributions~\cite{lin1991divergence}. Steck \cite{steck2018calibrated} used Kullbeck-Liebler (KL) Divergence. In this paper, I am using Jensen–Shannon divergence, which is a modification of KL Divergence that has two useful properties which KL divergence lacks: 1) it is symmetric: $\mathfrak{J}(P,Q)=\mathfrak{J}(Q,P)$ and 2) it has always a finite value even when there is a zero in $Q$. For our application, it is particularly important that the function be well-behaved at the zero point since it is possible for items from certain item groups to be missing from the recommendation list.

Given the $KL$ as the KL divergence function, the Jensen–Shannon divergence ($\mathfrak{J}$) between two probability distributions $P$ and $Q$ is defined as follows:
\begin{equation}
    \mathfrak{J}(P,Q)=\frac{1}{2}KL(P,M)+\frac{1}{2}KL(Q,M), \; \; M=\frac{1}{2}(P+Q)
\end{equation}

Similar to \cite{steck2018calibrated,wasilewski2018intent}, I use a weighted sum of relevance and calibration for creating our re-ranked recommendations. In order to determine the optimal set $\ell^*$ from the $m$ recommended items, I use the weighted combination of relevance and miscalibration as follows:
\begin{equation}\label{mmr}
    \ell^*= \argmax_{\ell, |\ell|= n} (1-\lambda)\cdot Rel(\ell)-\lambda \mathfrak{J}(P,Q(\ell))
\end{equation}
\noindent where $\lambda$ is the weight controlling the relevance versus the popularity calibration and $Rel(\ell)$ is the sum of the predicted scores for items in $\ell$. Since smaller values for $\mathfrak{J}$ are desirable, I used its negative for our score calculation. 

As noted in \cite{steck2018calibrated}, finding the optimal set $\ell^*$ is an NP-hard problem. However, similar to the first re-ranking method I proposed earlier, a simple greedy optimization approach is computationally effective. The greedy process starts with an empty set and iteratively adds one item from the larger list to the under-construction list until it reaches the desired length. At each step $j$, both $Rel(\ell)$ and $\mathfrak{J}(P,Q(\ell))$ are calculated using the union of the items that are already in the under-construction list and the item $i$ that is a candidate to be added to the list (i.e. $\ell@j \cup i$) and the item that gives the highest score will be added to the list. This greedy solution achieves a $(1-1/e)$ optimality to the best possible $\ell^*$ list with $e$ being Euler's number. 

\subsubsection{Evaluation}
Similar to the two other algorithms I discussed in previous sections, I evaluate the performance of our Calibrated Popularity algorithms with respect to different metrics explained in Chapter \ref{data_method}.

\begin{figure}
    \centering
    \SetFigLayout{3}{2}
 \subfigure[Precision, NDCG, Gini, APL, Aggregate Diversity and Long-tail coverage for our Popularity Calibration Re-ranking approach on MovieLens data ]{\includegraphics[width=4in]{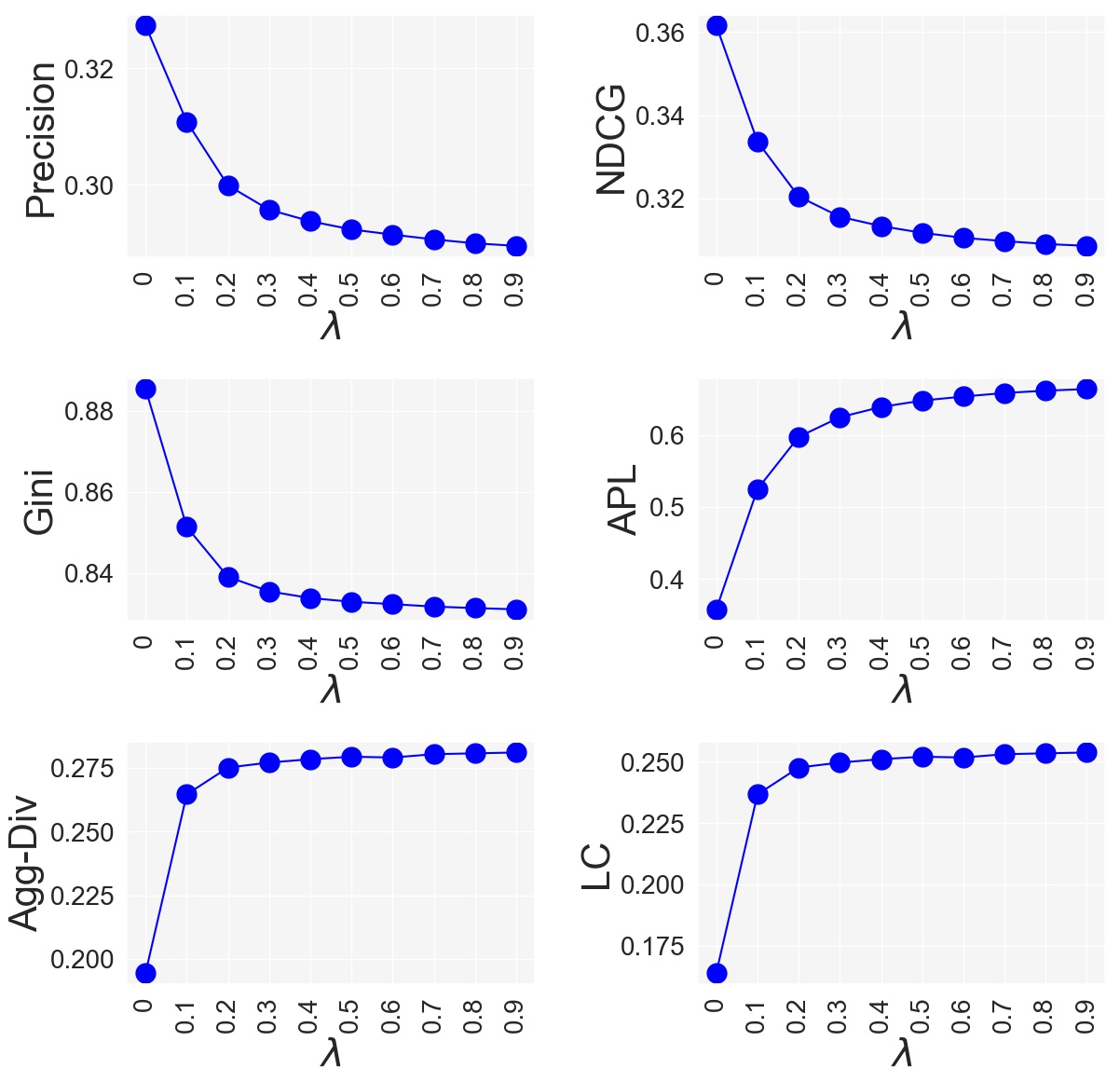}\label{fig:cp_results_movielens_1}}
 \subfigure[Equality of attention Supplier Fairness (ESF), Item Popularity Deviation (IPD), User Popularity Deviation (UPD) and Supplier Popularity Deviation (SPD) of our Popularity Calibration Re-ranking approach on MovieLens data]{\includegraphics[width=4in]{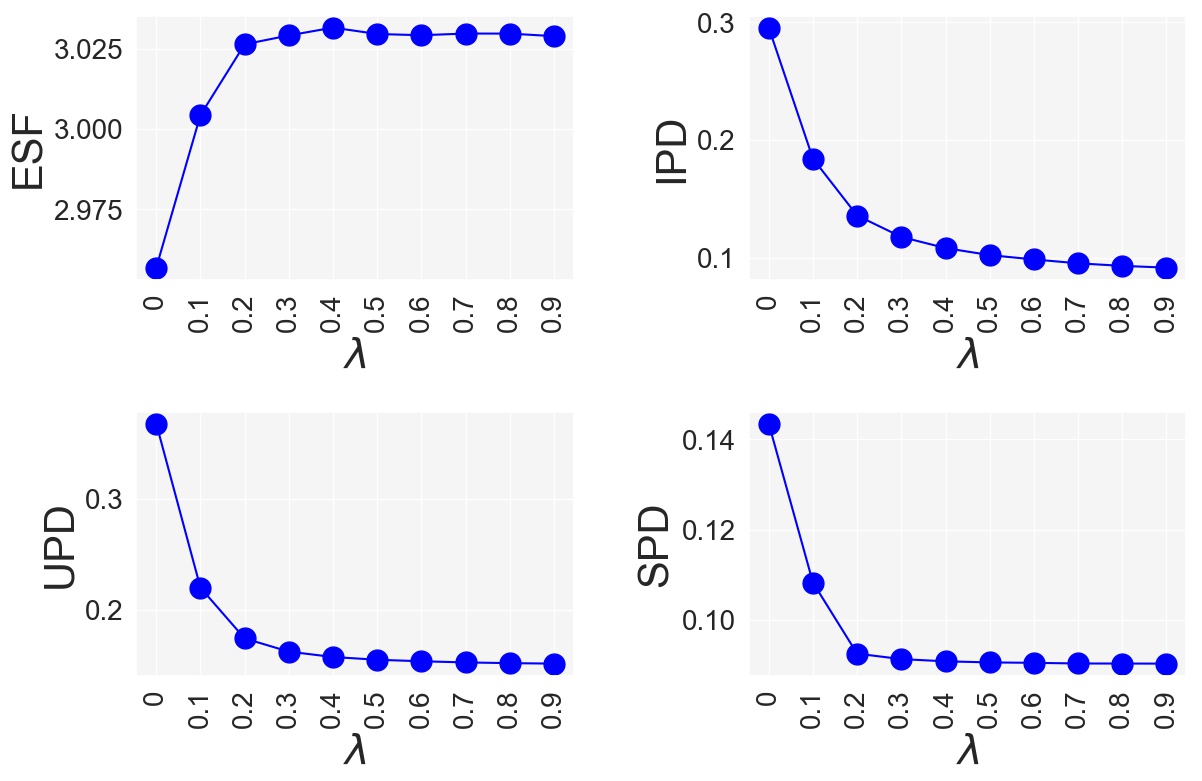}\label{fig:cp_results_movielens_2} }  
 \caption{Precision, NDCG, Gini, APL, Aggregate Diversity and Long-tail coverage, Equality of attention Supplier Fairness (ESF), Item Popularity Deviation (IPD), User Popularity Deviation (UPD) and Supplier Popularity Deviation (SPD) of our Popularity Calibration Re-ranking approach on MovieLens data}

\end{figure}

Figure \ref{fig:cp_results_movielens_1} shows the results for the calibrated popularity algorithm on precision, NDCG, Gini, average percentage of long-tail items in recommendation lists, aggregate diversity, and long-tail coverage. Similar to previous techniques, there is a loss in precision when we increase the weight for the calibration component ($\lambda$). Gini, aggregate diversity and long-tail coverage (both list-wise (APL) and overall (LC)) are all improving when we increase the values for $\lambda$ parameter meaning more unique items are recommended, recommendation are more uniformly distributed and also more long-tail items are appeared in the recommendations. It is interesting that aggregate diversity and long-tail coverage significantly improve up until a certain value for $\lambda$ ($\lambda=0.2$) and then it loses this sharp momentum for larger values of $\lambda$. The reason is, the calibrated popularity algorithm does not directly aim for increasing aggregate diversity or long-tail coverage but rather tries to give more exposure to different items in different item popularity groups. In other words, after a certain point, it does not add any new item to its list of covered items but tries to give these already covered items more exposure as can be seen from the monotonic improvement for Gini and also $IPD$ in Figure \ref{fig:cp_results_movielens_2} which measures how much the exposure for different items varies from their popularity in the original rating data.

\begin{figure}
    \centering
    \SetFigLayout{3}{2}
 \subfigure[Precision, NDCG, Gini, APL, Aggregate Diversity and Long-tail coverage for our Popularity Calibration Re-ranking approach on Last.fm data]{\includegraphics[width=4in]{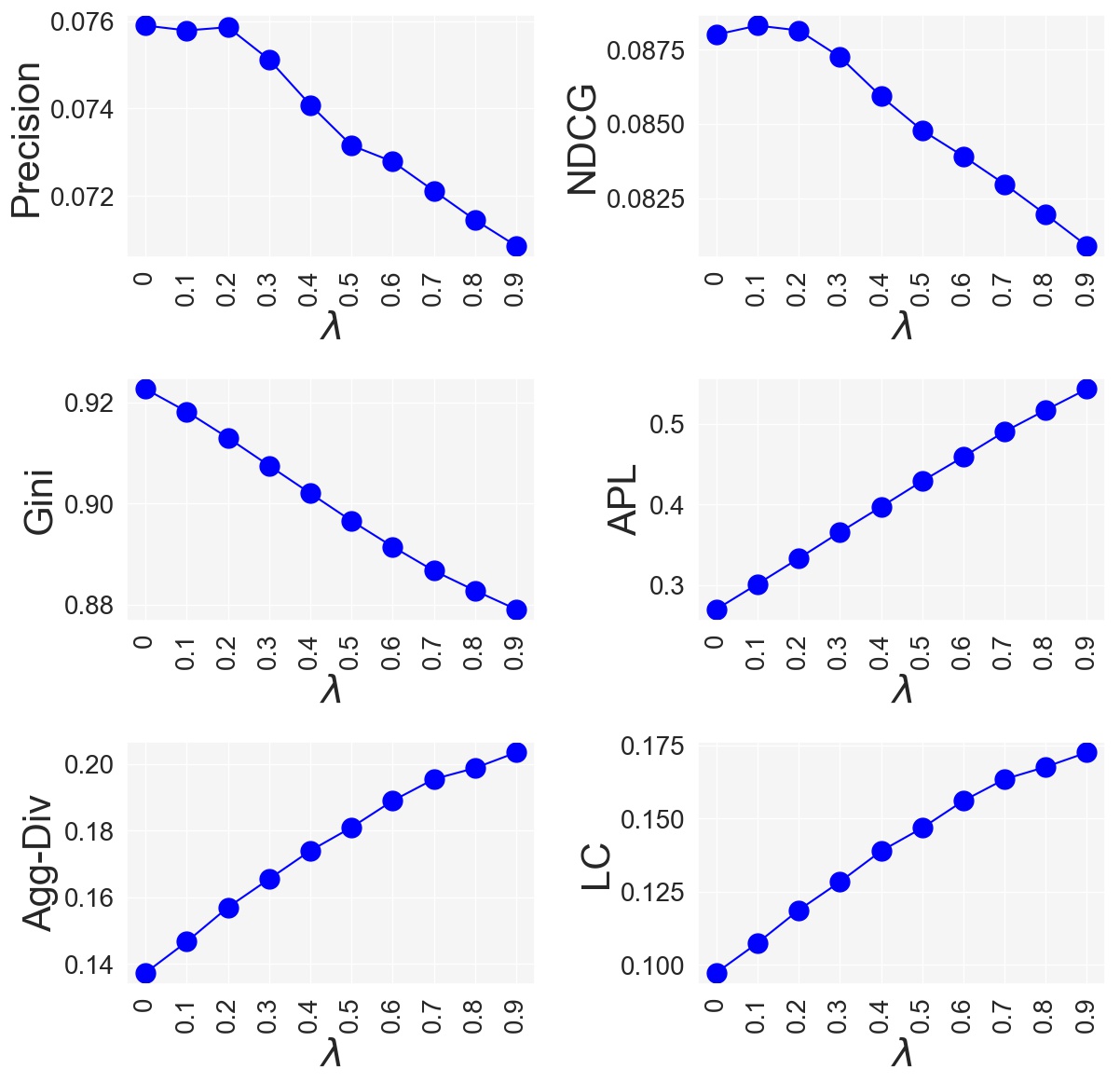}\label{fig:cp_results_lastfm_1}}
 \subfigure[Equality of attention Supplier Fairness (ESF), Item Popularity Deviation (IPD), User Popularity Deviation (UPD) and Supplier Popularity Deviation (SPD) of our Popularity Calibration Re-ranking approach on Last.fm data]{\includegraphics[width=4in]{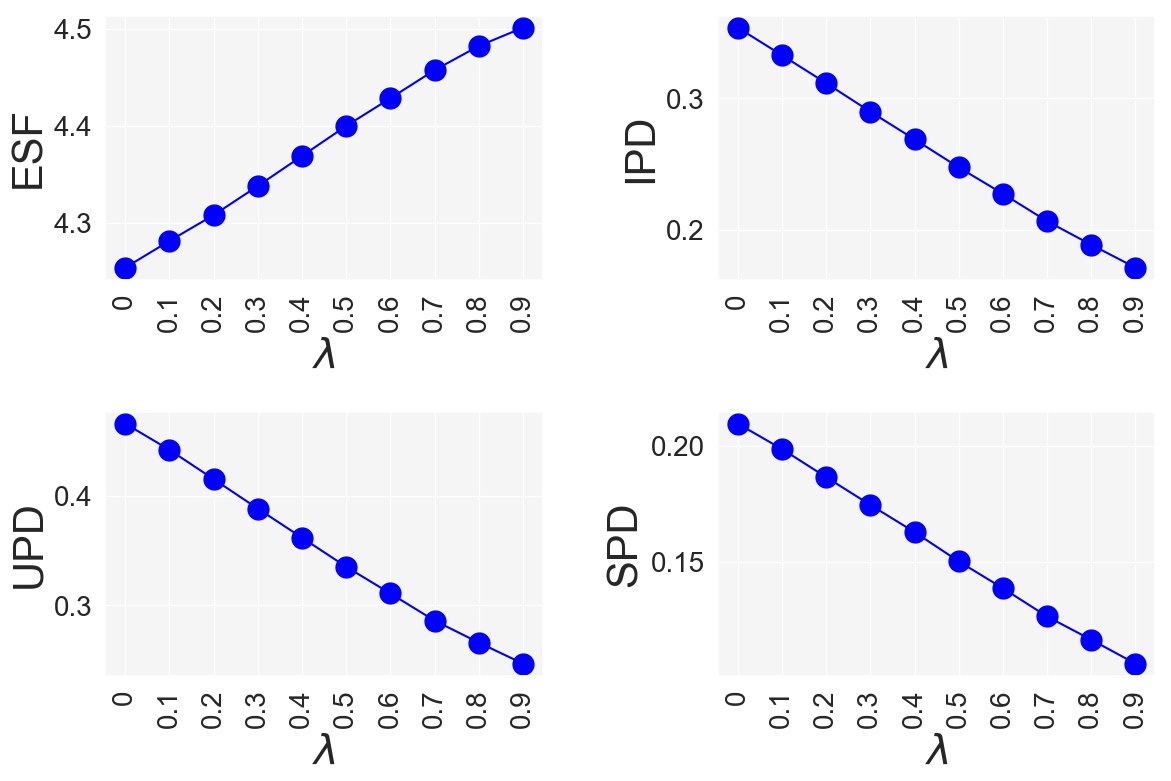}\label{fig:cp_results_lastfm_2}}   
 \caption{Precision, NDCG, Gini, APL, Aggregate Diversity and Long-tail coverage, Equality of attention Supplier Fairness (ESF), Item Popularity Deviation (IPD), User Popularity Deviation (UPD) and Supplier Popularity Deviation (SPD) of our Popularity Calibration Re-ranking approach on Last.fm data}
 \label{fig:cp_results_lastfm}
\end{figure}

The long-tail enhancement and fairness metrics are all improved as we can see from Figure \ref{fig:cp_results_movielens_2}. Similarly, Figure \ref{fig:cp_results_lastfm_1} and \ref{fig:cp_results_lastfm_2} show the same trends for the calibrated popularity on different metrics on Last.fm dataset.

\begin{figure}
    \centering
    \SetFigLayout{3}{2}
 \subfigure[]{\includegraphics[width=4in]{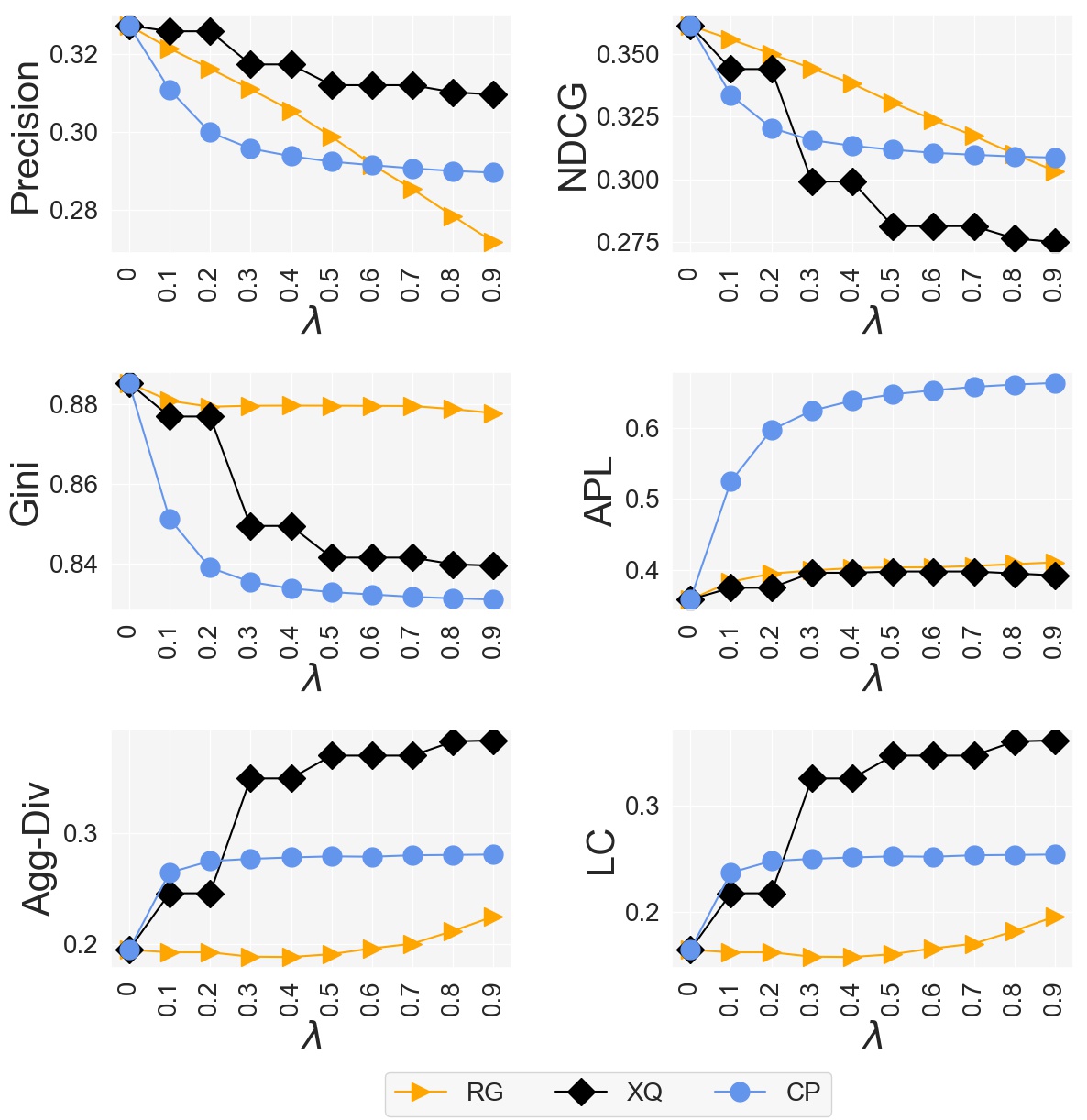}\label{fig:all_results_movielens1}}
 \subfigure[]{\includegraphics[width=4in]{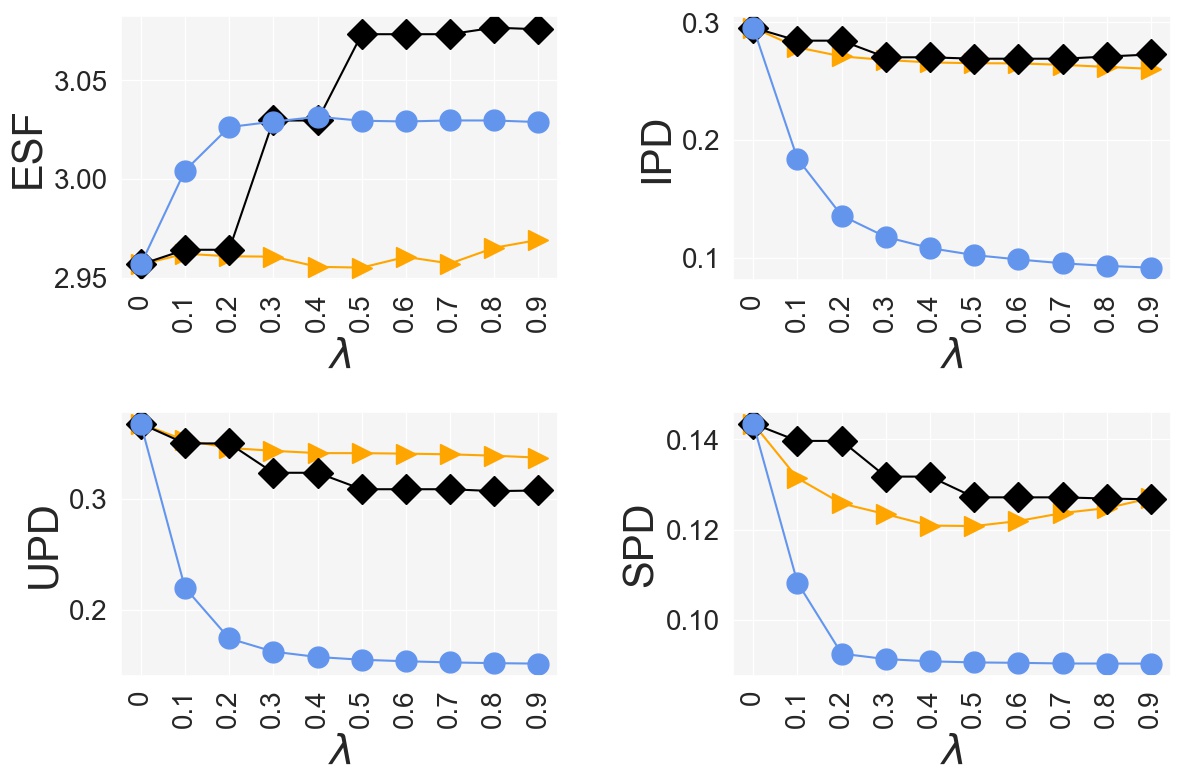}\label{fig:all_results_movielens2}} 
 \caption{Precision, Gini, Gini, APL, Aggregate Diversity and Long-tail coverage, Equality of attention Supplier Fairness (ESF), Item Popularity Deviation (IPD), User Popularity Deviation (UPD) and Supplier Popularity Deviation (SPD) for all three popularity mitigating approaches on MovieLens data}
    \label{fig:all_results_movielens}
\end{figure}

\begin{figure}
    \centering
    \SetFigLayout{3}{2}
 \subfigure[]{\includegraphics[width=4in]{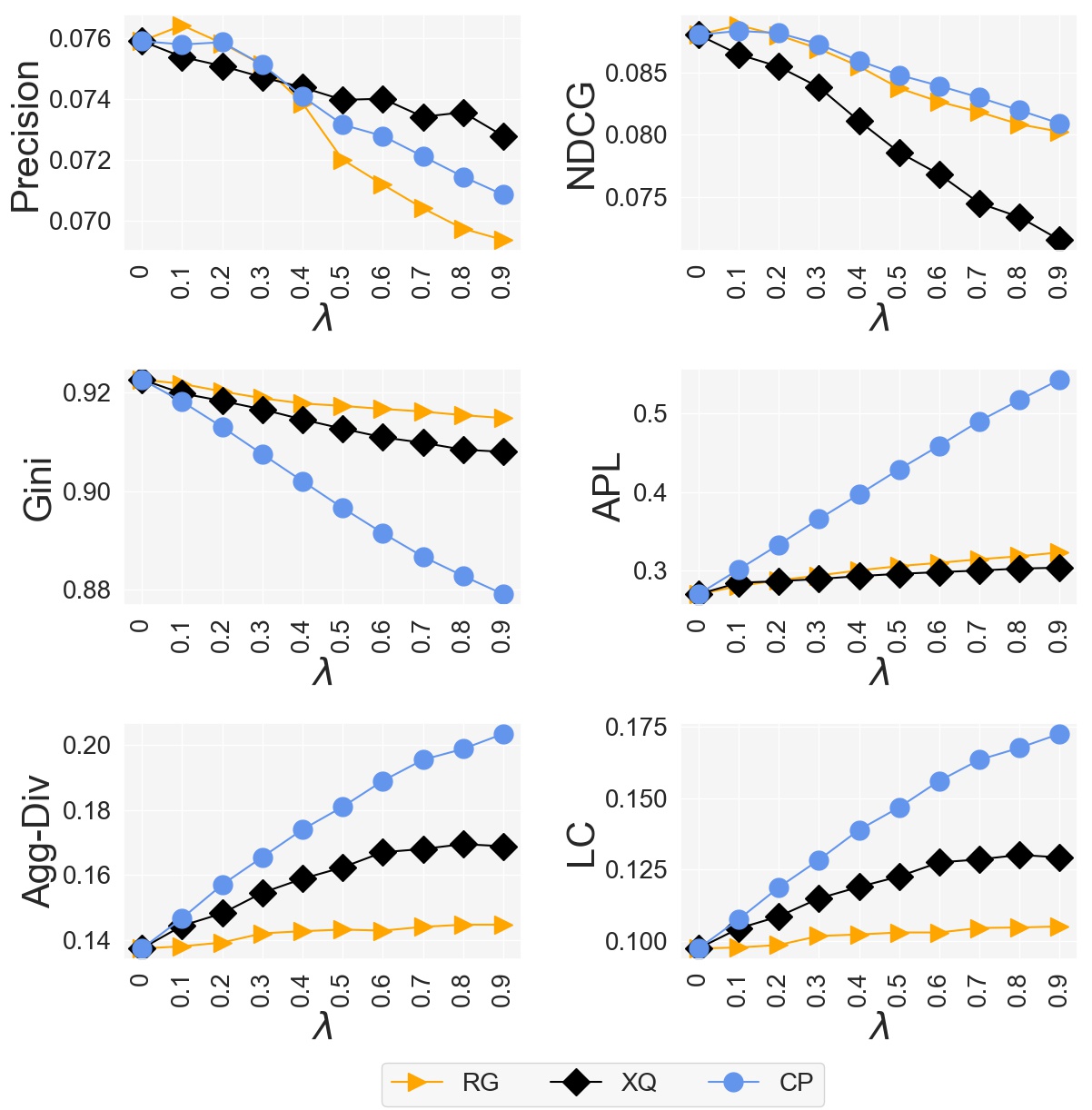}\label{fig:all_results_lastfm1}}
 \subfigure[]{\includegraphics[width=4in]{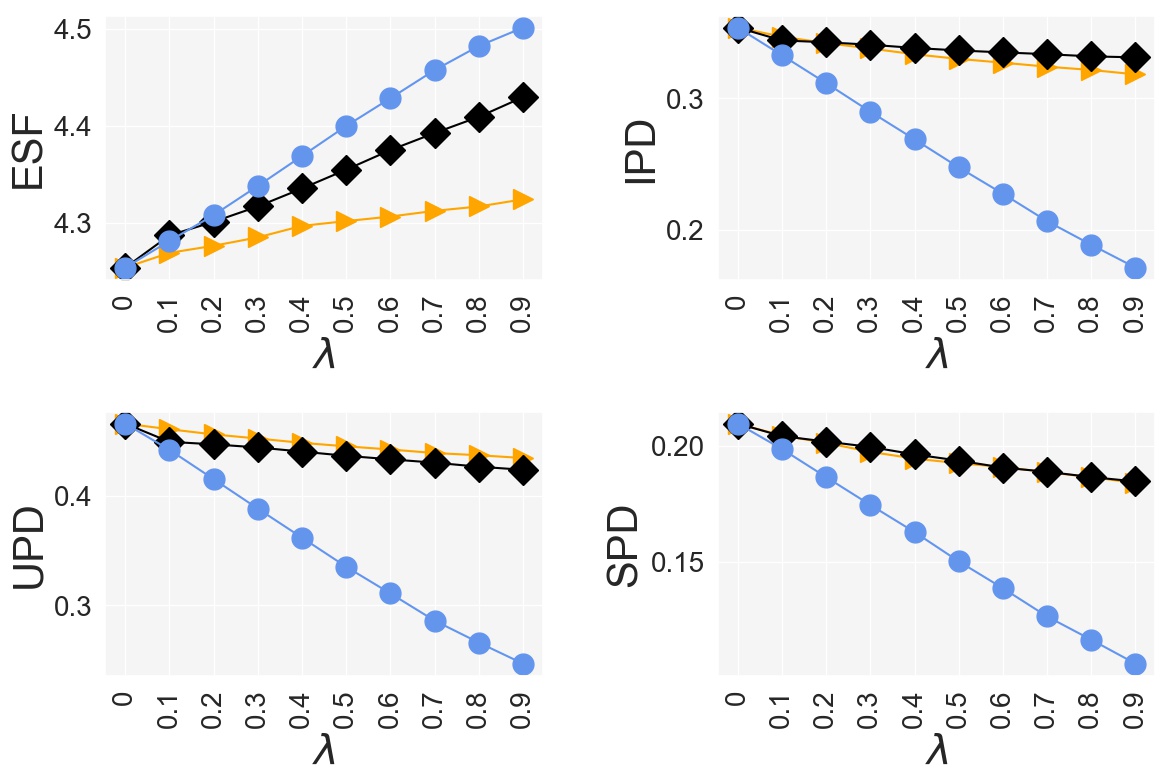}\label{fig:all_results_lastfm2}}    \caption{Precision, Gini, Gini, APL, Aggregate Diversity and Long-tail coverage, Equality of attention Supplier Fairness (ESF), Item Popularity Deviation (IPD), User Popularity Deviation (UPD) and Supplier Popularity Deviation (SPD) for all three popularity mitigating approaches on Last.fm data}
    \label{fig:all_results_lastfm}
\end{figure}

Figures \ref{fig:all_results_movielens} and \ref{fig:all_results_lastfm} compares the three algorithms I proposed in this dissertation all within the same plot. $RG$ is the model-based technique based on regularization, $XQ$ is the fairness-aware re-ranking technique based on $xQuAD$, and $CP$ is the popularity calibration approach I proposed in this section. 

First of all, the superiority of both re-ranking algorithms ($XQ$ and $CP$) over the model-based technique ($RG$) is quite obvious in many of the metrics. Overall, the precision for all three algorithms, as mentioned earlier, is decreasing over different values for $\lambda$. The precision for $XQ$, overall, has experienced a lower drop compared to other two indicating it keeps its accuracy better. Also, for smaller values of $\lambda$, the precision of $RG$ is higher than that of $CP$ but for larger values it experiences a larger loss. 

Looking at Gini for three algorithms we can see that the Gini of $RG$ looks almost like a straight line when it is contrasted with the other two re-ranking techniques. That means, $RG$ does not improve the uniformity of the recommendation distribution compared to $CP$ or $XQ$. $CP$ is clearly outperforming $XQ$ here. 

When it comes to aggregate diversity and long-tail coverage, $XQ$ outperforms the other two. $ESF$ which also has a strong connection to aggregate diversity is higher for $XQ$ compared to $CP$ and $RG$ when we increase $\lambda$. However, all the metrics for measuring the proportionality of popularity ($IPD$, $UPD$, and $SPD$) indicate that $CP$ outperforms the other two by a large margin. Another interesting pattern we can see in the results for MovieLens dataset, is that $RG$ performs worse than $XQ$ on $UPD$ but it outperforms $XQ$ on $SPD$. In other words, $XQ$ better keeps the ratio of different item popularity groups in the users' profile when generating the recommendations but the popularity of different suppliers is not as well-kept as in $RG$.

I also want to compare all the three proposed methods I proposed in this dissertation with some of the state-of-the-art techniques for mitigating popularity bias. I used two recent state-of-the-art re-ranking techniques for mitigating popularity bias in the literature. These two methods are as follows:

\begin{itemize}
    \item \textbf{Discrepancy Minimization (DM)} \cite{antikacioglu2017}:
    In this method, the goal is to improve the total number of unique recommended items, also referred to as aggregate diversity (see Equation \ref{agg_diver} in Chapter \ref{background_chapter}) of recommendations using minimum-cost network flow method to efficiently find recommendation sub-graphs that optimize diversity. Authors in this work define a target distribution of item exposure (i.e. the number of times each item should appear in the recommendations) as a constraint for their objective function. The goal is therefore to minimize the discrepancy of the recommendation frequency for each item and the target distribution. 
    
     \item \textbf{FA*IR (FS) \cite{zehlike2017}:} This method was originally used for improving group fairness in job recommendation and was adapted here to improve the fairness of recommendations in terms of head ($H$) versus long-tail ($M \cup T$) items in recommendations. The algorithm creates queues of protected and unprotected items and merges them using normalized scoring such that protected items get more exposure. I define protected and unprotected groups as long-tail and head items, respectively. I performed grid search over the two hyperparameters, proportion of protected candidates in the top $n$ items \footnote{Based on suggestion from the released code, the range should be in $[0.02,0.98]$} and significance level\footnote{Based on suggestion from the released code, the range should be in $[0.01,0.15]$}, using values of $\{0.25,0.5,0.75,0.95\}$ and $\{0.05,0.1,0.15\}$, respectively. 
     
\end{itemize}

The two re-ranking baselines ($DM$ and $FS$) and the proposed techniques in this dissertation ($RG$, $XQ$ and $CP$) have hyperparameters that control the trade-off between relevance and a second criterion: long-tail regulariation in $RG$, diversification in $XQ$, fairness in $FS$, aggregate diversity in $DM$, and popularity calibration in $CP$. To establish a point of comparison across the algorithms, I varied these trade-off hyperparameters for each algorithm, and chose, for each, a hyperparameter that yields a recommendation precision that is within a small margin (Difference in precision is less than 0.01 for MovieLens and less than 0.001 for Last.fm) compared to that off other algorithms since getting exactly an equal precision is not possible for certain algorithms.



\begin{table*}[t]

\centering
\captionof{table}{Results of different algorithms on two datasets. The setting that yields the same precision for all three reranking methods XQ, DM and CP is chosen and the results is reported. Bold values are statistically significant compared to the second best value in each column with significance level of 0.05 ($\alpha$=0.05) } 
\label{tables/movielens_rankals}
\begin{tabular}{lccccccl}
\toprule
           & Agg-Div$\uparrow$ & LC$\uparrow$ &$Gini$$\downarrow$ &$ESF$$\uparrow$&$IPD$$\downarrow$&$UPD$$\downarrow$&$SPD$$\downarrow$
           \\\midrule

Base    & 0.194  & 0.164 &0.885& 2.956&0.295& 0.368 &0.143\\
RG    & 0.196 & 0.167 &0.872&2.960 &0.265& 0.341 &0.127\\
XQ & 0.384  & 0.3610 &0.839&3.055&0.273&0.308 &0.127\\
DM & \textbf{0.519}  & \textbf{0.501}&\textbf{0.623} &3.057&0.213&0.302 &0.109\\
FS &0.198 & 0.168&0.863&2.965&0.170  &0.268&0.109\\
CP   & 0.281  & 0.253& 0.831&3.056& \textbf{0.092}&\textbf{0.152}&\textbf{0.090}\\
\bottomrule
\end{tabular}
\end{table*}


\begin{table*}[t]

\centering
\captionof{table}{Results of different algorithms on two datasets. The setting that yields the same precision for all three reranking methods XQ, DM and CP is chosen and the results is reported. Bold values are statistically significant compared to the second best value in each column with significance level of 0.05 ($\alpha$=0.05) } 
\label{tables/lastfm_rankals}
\begin{tabular}{lccccccl}
\toprule
          & Agg-Div$\uparrow$ & LC$\uparrow$ &$Gini$$\downarrow$ &$ESF$$\uparrow$&$IPD$$\downarrow$&$UPD$$\downarrow$&$SPD$$\downarrow$
          \\\midrule

Base    & 0.137  & 0.0970 &0.922& 4.253& 0.354 &0.466&0.21\\
RG    & 0.145  & 0.105 &0.915& 4.324& 0.318 &0.435&0.184\\
XQ & 0.169  & 0.129 &0.908&4.429&0.331 &0.424&0.185\\
DM & 0.175  & 0.134&0.905 & 4.274&0.339 &0.453&0.2\\
FS &0.176 & 0.142&0.891 &4.440&0.186  &0.328&0.129\\
CP   & \textbf{0.203}  & \textbf{0.172}& \textbf{0.879}& \textbf{4.501}&\textbf{0.171}&\textbf{0.246}&\textbf{0.106}\\
\bottomrule
\end{tabular}
\end{table*}

First I look at the overall performance of these algorithms. The overall results for different algorithms on both MovieLens and Last.fm datasets can be seen in Table ~\ref{tables/movielens_rankals} and ~\ref{tables/lastfm_rankals}. The arrow next to the metric name indicates the direction of preference for a better outcome. 

On MovieLens, with respect to aggregate diversity (\textit{Agg-Div}), long-tail coverage ($LC$), and Gini, On MovieLens, the $DM$ algorithm has the best performance on these metrics. $CP$ outperforms $XQ$ on aggregate diversity and long-tail coverage but they have similar performance with respect to Gini. Even though $RG$ and $FS$ have both improved these metrics over the \textit{Base} algorithm they still stand behind the other algorithms. Not only is greater coverage for $DM$ achieved for the whole catalog, but a better distribution as shown by its low Gini index. The results for Last.fm dataset in Table \ref{tables/lastfm_rankals} shows that $CP$ outperforms the other algorithms with respect to all metrics followed by $FS$ and $XQ$. $RG$ still performs the worst among all. 


\begin{figure*}
\centering
\SetFigLayout{3}{2}
 \subfigure[MovieLens]{\includegraphics[width=5.9in]{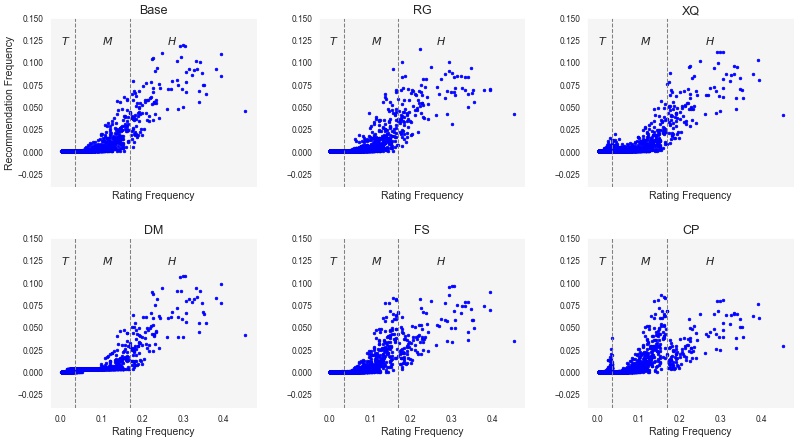}\label{expoure_scatter_movielens}}
\subfigure[Last.fm]{\includegraphics[width=5.9in]{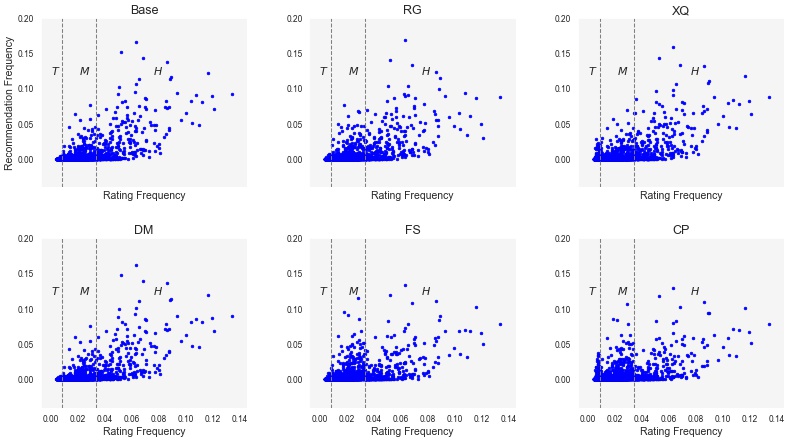}\label{expoure_scatter_lastfm}}
\caption{Exposure rate vs popularity across different re-ranking algorithms (MovieLens)} \label{expoure_scatter}
\end{figure*}

A more detailed picture of the way that item popularity is transformed in recommendation frequency is available in Figure~\ref{expoure_scatter_movielens} and \ref{expoure_scatter_lastfm}. These scatter plots have rating frequency and popularity segments (i.e. item groups) on the x-axis with the \textit{Tail} items ($T$) having low rating frequency on the left, moving to the popular \textit{Head} items ($H$) on the right. Also, y-axis shows the frequency of the recommendations for each item (i.e. exposure rate). Looking first at part (a) showing the MovieLens results, we see that, in the \textit{Base} algorithm, the $T$ items are rarely recommended, as previously noted. The $XQ$, and $CP$ boost the $T$ and $M$ items significantly, while a more limited effect is seen for $DM$, $RG$ and $FS$. The division of the items into groups has an obvious effect as the promotion of lower-popularity items is concentrated at the top end of the $T$ and $M$ items, which could be expected as these are the items in this set with the most number of ratings. Note also that $CP$ and $FS$ algorithms reduce the number of items with the maximum rating frequency (contained in more than 12.5\% of recommendation lists). 

In Figure \ref{expoure_scatter_lastfm}, we see the results for Last.fm. We can clearly see that the popularity bias of the \textit{Base} recommender is lower compared to that of for MovieLens in its neglect of the lower parts of the popularity distribution here and that gives all of the re-rankers more to work with. $CP$ and $XQ$ boost the frequency of the recommendations for $T$ items somewhat and reduce the extremes of recommendation frequency. Also, in addition to $CP$, $FS$ improves the exposure for the items in the $M$ category. Overall, the $CP$ algorithm has performed the best with respect to giving items from different item groups the exposure that they deserve. 

There is another interesting pattern that helps explain the performance of the $DM$ algorithm on the \textit{Agg-Div}, \textit{LC}, and \textit{Gini} metrics on Movielens dataset. On this dataset, the $DM$ algorithm promotes some of the $T$ and $M$ items as can be seen in the small step increase in the plot in these regions. Thus, it increases aggregate and long-tail coverage since in both of these two metrics, the recommendation frequency does not matter meaning recommending an item even once would be counted the same as recommending it to many users. By giving very small but consistent exposure to this range of items, it improves the Gini index as well.

\begin{figure*}
\centering
\SetFigLayout{3}{2}
 \subfigure[MovieLens]{\includegraphics[width=6.2in]{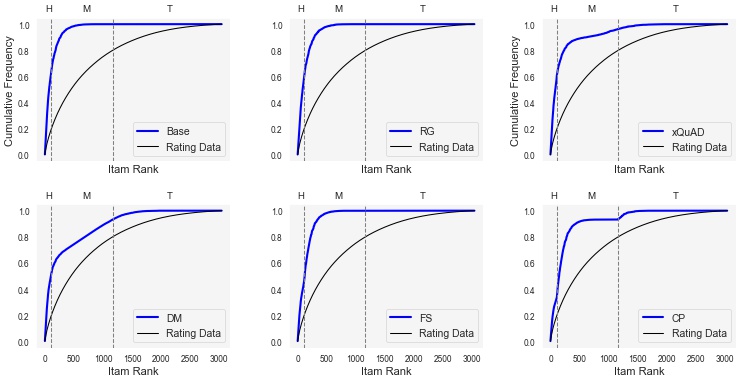}}
\subfigure[Last.fm]{\includegraphics[width=6.2in]{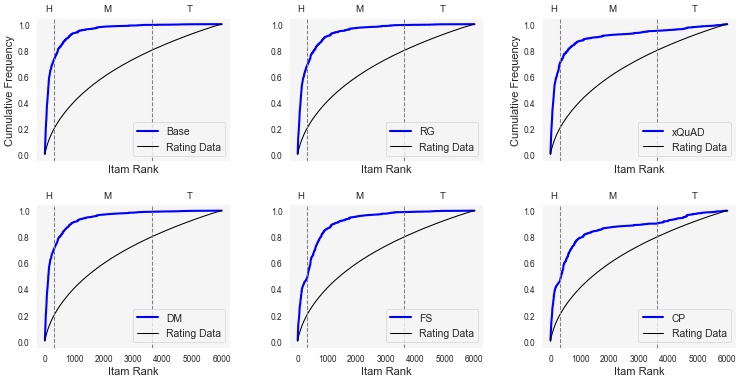}}
\caption{Cumulative frequency of the ratings for different items at a given rank and their frequency of appearance in the recommendations} \label{reranking_item_groups}
\end{figure*}

Figure \ref{reranking_item_groups} shows a similar plot to Figure 5.7 in Chapter \ref{pop_bias} but for the re-ranking algorithms. The black curve shows the cumulative frequency of ratings for different items at a given rank in MovieLens and Last.fm datasets. The x-axis indicates the rank of each item when sorted from most popular to least popular. The blue curves in each figure show the cumulative frequency of the recommendations for each item at a given rank. Three item groups $H$, $M$, and $T$ are depicted on the top of each plot. The point that the blue curve intersects with each of the dotted vertical lines shows the cumulative percentage of the recommendations that come from items up that point. For example, on MovieLens, in the \textit{Base} algorithm, approximately 60\% of the recommendation come from items in the $H$ which is significantly higher that what they originally had in rating data ($H$ items take up 20\% of the ratings). The lower this number is for the re-ranking algorithms, the better they have controlled this bias. $RG$ does not seem to perform well as this intersection is still at around 60\%. $XQ$ has slightly lowered this bias to around 50\%. $CP$ seems to have the best performance across all re-ranking algorithms and it has lowered the frequency of $H$ items to around 30\% which is very close to their original popularity. $CP$ has also given the highest exposure to the $T$ items. The same pattern repeats for the Last.fm data where $CP$ performs well on different item groups compared to other algorithms.  

Without a multi-stakeholder analysis, we might conclude from the overall metrics shown in Table~\ref{tables/movielens_rankals} that $DM$ is a superior approach for remediating popularity bias, on MovieLens data. However, evaluating the recommendations from the perspective of the different stakeholders reveal a different story. Because $CP$ aims to improve popularity calibration, we might expect that \textit{UPD}, which more or less measures the inverse of average calibration, would be improved and this is indeed a strong effect across both datasets. $CP$ is applying its long-tail enhancement where it matters for users and therefore is able to have a big impact on calibration. 
Thus, we might say that \textit{UPD} is measuring \textit{useful long-tail diversity} in recommendation results. Interestingly, this emphasis on users also has a beneficial effect on the metrics related to other recommendation stakeholders as shown in the superior performance on \textit{IPD}, and \textit{SPD}. All algorithms have a comparable $ESF$.

\begin{figure*}
\centering
\SetFigLayout{3}{2}
 \subfigure[MovieLens]{\includegraphics[width=5.9in]{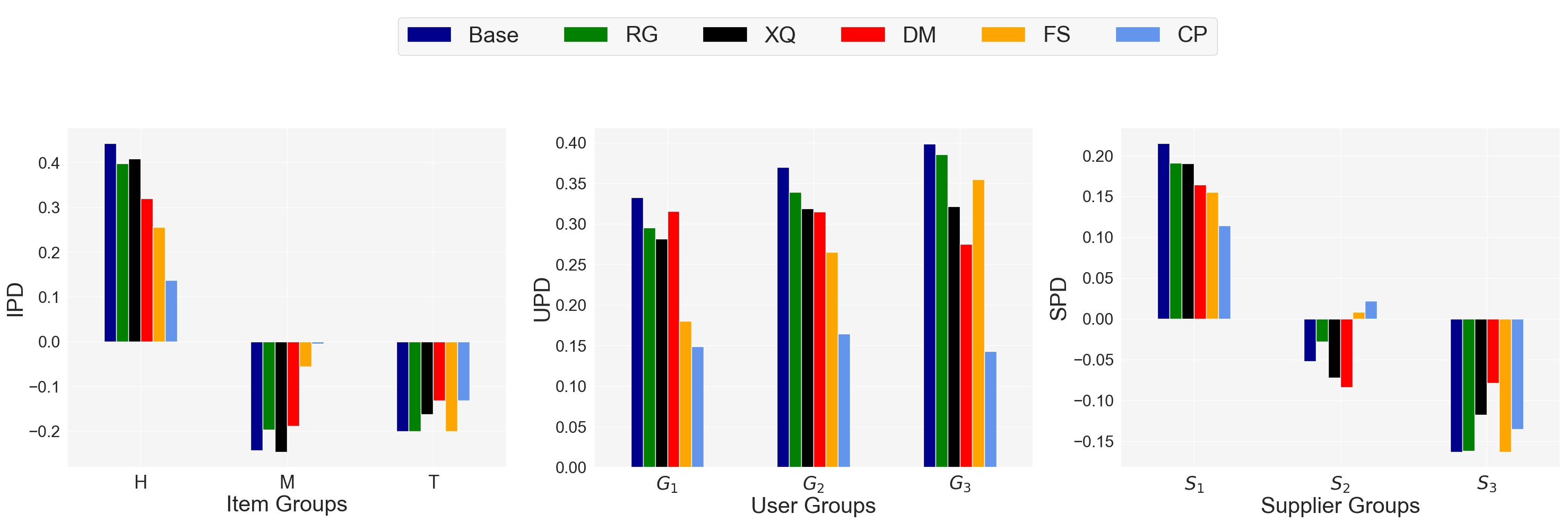}}
\subfigure[Last.fm]{\includegraphics[width=5.9in]{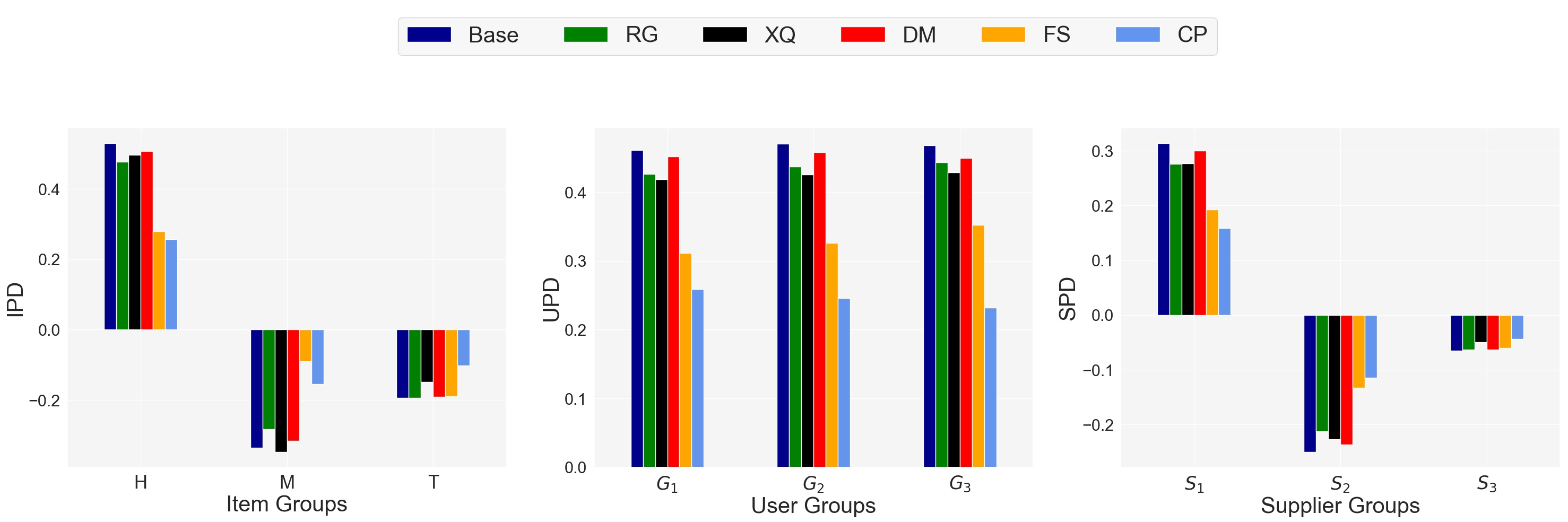}}

\caption{Popularity deviation metrics for different stakeholder groups} \label{groups_analysis_deviation}
\end{figure*}

Additional detail on the comparative performance for different stakeholders and sub-groups can be seen in Figure~\ref{groups_analysis_deviation}. These results compare the popularity deviation metrics (\textit{IPD}, \textit{UPD}, \textit{SPD}) for different sub-groups. 

\subsubsection{Item Groups}
First, we can see that all algorithms have positive \textit{IPD} values for items in \textit{H} meaning they all over-recommend popular items. This is not surprising since the \textit{Base} algorithm exhibits a stronger bias towards these items and although the re-ranking techniques have reduced this over-concentration they still have not removed it completely. The $CP$ algorithm comes the closest to doing so. Also, on Movielens, in all algorithms except for $CP$, items from the \textit{M} group have negative deviation meaning they are under-represented in the recommendations. But again, $CP$ performs significantly better in the \textit{M} group giving these items a slight boost, as we also saw in Figure~\ref{expoure_scatter}. 

The result for the \textit{T} items show that $CP$ and $XQ$ have lower deviation for this item group on MovieLens compared to other algorithms, but all algorithms including $CP$ have performed poorly. On Last.fm, $CP$ and $XQ$ performs better than other algorithms on this item group. So overall, no algorithm has outperformed $CP$ in any of these item groups.  

\subsubsection{User Groups}
Looking at the charts for user groups, again, we see the superiority of $CP$. Firstly, on MovieLens, $G_1$, the group with the highest interest towards popular items, has the lowest \textit{UPD} (i.e. lowest miscalibration) using all algorithms and $G_3$ has the highest miscalibration. That shows users with high interest towards popular items are served much better as expected. Regardless, all user groups have experienced the lowest miscalibration using $CP$. $FS$ also performs well here with lower \textit{UPD} compared to $RG$, $DM$, and $XQ$. What is interesting about $CP$ and $FS$ with regard to \textit{UPD} is that they both have comparable \textit{UPD} for $G_1$ but the \textit{UPD} of $CP$ is significantly lower for $G_2$ and $G_3$ indicating $CP$ also performs well for users with lesser interest in popular items as it calibrates its recommendations better to different users' profiles. The same for $DM$ and $FS$ where $FS$ works better for users in $G_1$ (lower \textit{UPD}) while $DM$ is superior for $G_3$. These are distinctions that we would not have observed without a multi-stakeholder approach to evaluation.

\begin{figure*}
\centering
\SetFigLayout{3}{2}
 \subfigure[MovieLens]{\includegraphics[width=5.9in]{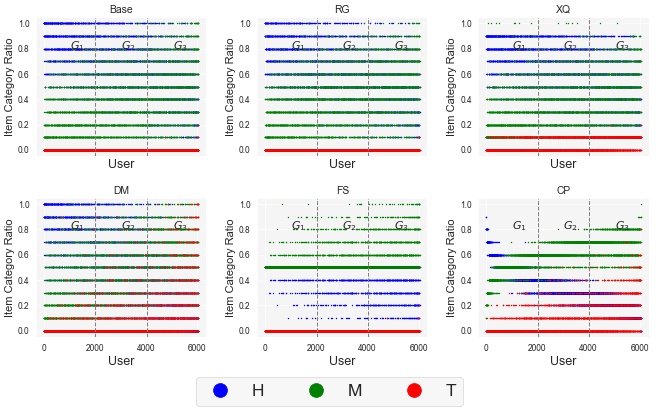}}
\subfigure[Last.fm]{\includegraphics[width=5.9in]{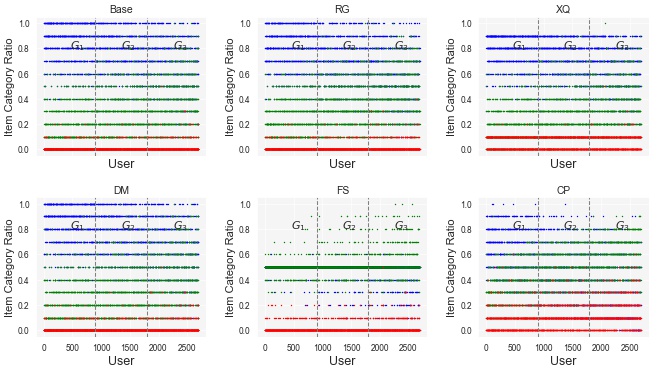}}
\caption{The ratio of different item groups in  the recommendations of different users using the popularity bias mitigation techniques} \label{reranking_recs_user_groups}
\end{figure*}

\begin{figure*}
\centering
\SetFigLayout{3}{2}
 \subfigure[MovieLens]{\includegraphics[width=5in]{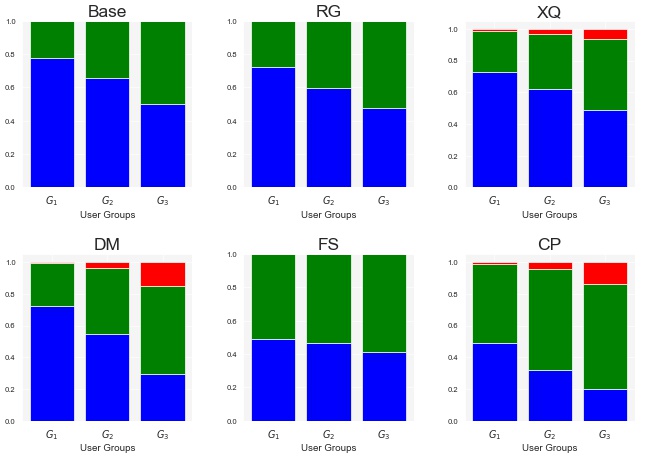}}
\subfigure[Last.fm]{\includegraphics[width=5in]{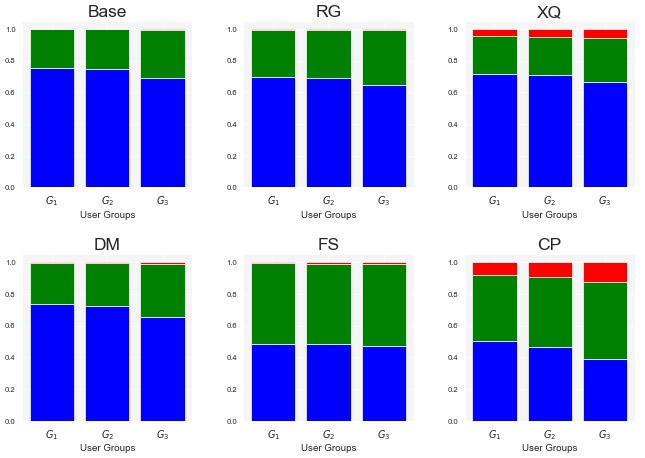}}
\caption{The ratio of different item groups in  the recommendations of different user groups using the popularity bias mitigation techniques} \label{reranking_recs_user_groups_stacked}
\end{figure*}

Another view of the performance of different algorithm for different user groups is shown in Figure \ref{reranking_recs_user_groups} which is a similar plot to what we saw in Figure \ref{user_propensity} in Chapter \ref{pop_bias}. Users are ranked based on their interests towards popular items from left to right and the three user groups are shown in each plot. We can see that different recommendation algorithms produced different density of recommended items from different item groups. Some still have recommended many popular items even to the users who have lesser interest items in these items and some have done a better job in giving items from different popularity bins to different users according to their interest. $RG$ still does not look good from the users' perspective as it almost has the shape as the original $RankALS$. $DM$ seems to the most aggressive one in recommending $T$ items. $CP$ shows a better overall performance as the ratio of different recommended item groups seem to better match the original ratio of these items in the users' profiles. 

The difference of performance for different algorithms in terms of recommending the right ratio of different item groups for users in different user groups can be seen in Figure \ref{reranking_recs_user_groups_stacked}. The ratio of tail items recommended to all three user groups using $RG$ and $FS$ is essentially zero. $DM$, $XQ$, and $CP$ have all recommended tail items but the proportion of different item categories recommended using $CP$ is closer to the historical users' interaction with these different item categories as we saw in Figure \ref{user_group_propensity} in Chapter \ref{pop_bias}.

\subsubsection{Supplier Groups}\label{supplier_groups}

Finally, the chart for supplier groups similarly confirms the superiority of $CP$ for $S_1$ and $S_2$ in this multi-stakeholder analysis. On both datasets, $CP$ has the lowest \textit{SPD} for the $S_1$ (group with high popularity). On Movielens, this metric for $S_2$ has a positive value meaning $CP$ has actually given a boost to suppliers from this group in terms of exposure.

One other interesting finding that can be seen in this chart for MovieLens dataset is the relationship between item popularity and supplier popularity. One might expect that when an algorithm performs better than another on items with lower popularity (e.g. $T$) it should also perform better on the suppliers with lowest popularity as these two seem to be correlated. However, as we can see on MovieLens dataset, $CP$ and $DM$ have comparable performance on $T$ items (both have roughly equal $IPD$ ) but $DM$ outperforms $CP$ on $S_3$ (lower $SPD$). The reason is, we observed not every item from a popular / non-popular supplier is necessarily popular / non-popular and therefore it is possible for an algorithm to perform differently on item groups and supplier groups as it is the case on MovieLens dataset for $CP$ and $DM$.

\begin{figure*}
\centering
\SetFigLayout{3}{2}
 \subfigure[MovieLens]{\includegraphics[width=6.2in]{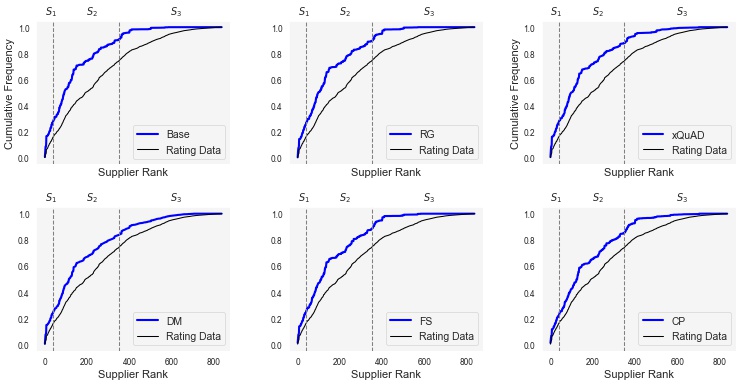}}
\subfigure[Last.fm]{\includegraphics[width=6.2in]{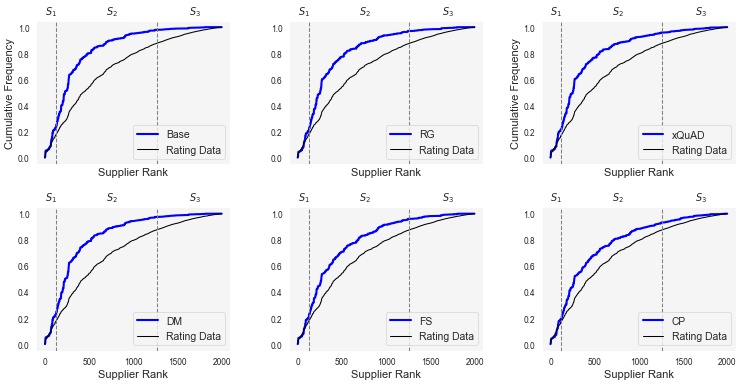}}
\caption{Cumulative Frequency of recommendations from different supplier groups versus their popularity in rating data} \label{reranking_supplier_groups}
\end{figure*}

Figure \ref{reranking_supplier_groups} shows the cumulative Frequency of the recommendations from different supplier groups versus their popularity in rating data using different popularity bias mitigation techniques. All algorithms (even $RG$ to some extent) have reduced the over-concentration of the recommendations on the items from popular supplier groups. Especially, $CP$ and $FS$ seem to have narrowed down the gap between the blue and black curve which shows their superior performance in giving proportionate exposure to different suppliers with a varying degree of popularity.

Overall, we learned that, among all the popularity bias mitigation approaches we studied in this chapter, $CP$ can be considered the superior approach since it performs better than the others from the perspective of different stakeholders. This is an interesting finding which means matching the recommendations with the users' range of interests in terms of item popularity not only does it calibrate the recommendations to the users' profiles, it also gives a fairer exposure to items from different popularity groups and also suppliers with different levels of popularity.

\chapter{Conclusion and future work}

\section{Summary of Contributions}

\subsection{Multi-stakeholder recommendation}
In Chapter \ref{ms_rs} I described a new paradigm for recommender systems' design, implementation, and evaluation. I defined what the characteristics of multi-stakeholder recommendations are and I discussed its connection to some of the existing work in the literature such as group recommendation, people recommendation and matching, fairness-aware recommendation and value-aware recommendation. 

Generally speaking, multi-stakeholder recommendation refers to new practice of taking into account the needs and interests of several stakeholders in the recommendation process. The reason this is an important consideration is that, in many real-world recommendation platforms a simple focus on the users may not address all the complexities of the recommender systems. For instance, in an online dating application, both parties (e.g. men vs women) should be satisfied with a given recommendations. In addition, the platform owner may also have other objectives that is beyond the preferences of users such as fairness concerns, premium users vs free subscriptions. As another example, we can think of a music recommendation system as a multi-sided platform where on one side we have the listeners and on the side we have the artists. It is clear that only focusing on the listeners may not be ideal for the overall long-term health of the business as it may lead to the unsatisfaction of the artists and, hence, losing the artists on the platform altering the balance of the marketplace. 

I stated that typically three main stakeholders can be observed in many recommendation platform: 1) Users or consumers, 2) Suppliers of the recommendations and 3) the system owner. The ideal situation would be to have a recommender system where every stakeholder is satisfied with the outcome of the recommendations. However, this might not always happen in reality. Depending on the situation, time, and other considerations, the system owner may decide to tune the recommender system so it improves the outcome for a particular stakeholder or group of stakeholders. 

\subsection{Multi-stakeholder Impact of Popularity Bias}

In Chapter \ref{pop_bias}, I have taken a multi-stakeholder view of the problem of popularity bias. I showed that the popularity bias is not just an overall system-level problem which alters the balance of the distribution of the recommendations towards popular items but rather one that impacts the entire eco-system including the users and the suppliers of the recommendations.

Regarding the users, I showed that large segments of the user population in a typical recommender system have a strong interest in items outside of the ``short head'' of the distribution. Consistent with prior work, I showed that because these long-tail items are recommended less frequently, these groups of users are not well-served and their recommendations are heavily concentrated on popular items despite their interest towards less popular ones.

I also showed that the popularity bias has a ripple-out effect on the other recommendation stakeholders. For example, the less popular suppliers are extremely being impacted by this bias and are not recommended to the users at which they have shown interest at. For example a less popular artist such as a local artist may have significant number of people interested in her/his music but because the recommendations are dominated by the songs from popular artists, the algorithmic popularity bias does not let her/his songs to be recommended to the desired audience. 

This multi-stakeholder evaluation of popularity bias has revealed certain aspects of the algorithms' behavior that cannot be captured using standard evaluation methods and metrics that are focused on overall outcomes, including differences across results returned to different user, supplier, and item groups. For example, in Chapter \ref{tackle_pop_bias} we saw that some algorithms perform better than others for certain user /item /supplier groups and worse on some other groups. That indicates that, when we take an average perspective over all users or items, we cannot see the differences of different algorithms in their performance for different stakeholders.

\subsection{Algorithmic solutions to tackle popularity bias}

In Chapter \ref{tackle_pop_bias}, I presented three different recommendation algorithms to control popularity bias. As I discuss in that chapter, generally two types of solutions can be found to address the popularity bias in recommendation: 1) model-based and 2) post-processing re-ranking. I proposed one model-based approach and two re-ranking methods.

The first approach I presented in this dissertation was a model-based technique that works on top of a particular learning-to-rank algorithm ($RankALS$) and changes its objective function so it incorporates the popularity of the recommendations in each given recommendation list. The algorithm aims to diversify the recommendations in each list according to whether they belong to popular items or the long-tail ones. In other words, it penalizes the lists that only contain popular items or only long-tail items. Experimental results showed that our proposed method was able to improve the long-tail properties of the recommendations such as catalog coverage and long-tail coverage. 

The second method I proposed to control popularity bias was in the post-processing re-ranking category. Generally speaking, in a re-ranking technique, the algorithm first generates a larger list of recommendations and the re-ranking acts as a post-processing step to extract the best top recommendations that have both high relevance to the user and also desired characteristics of long-tail. The algorithm was inspired by a well-known diversification method in information retrieval called xQuAD which aims at covering different aspects related to a given query in the returned search results. In my case, I did not have aspects associated with items but rather a category to which each item belongs. I defined two item categories short-head and long-tail and, therefore, the diversification task was to make sure items from different item categories are fairly represented in the recommendations. The experimental results indicated a significant improvement in long-tail properties of the recommendations with a small loss in accuracy. 

The third algorithm I proposed in this dissertation was also a post-processing technique but with a different objective. In this method which I called \textit{Calibrated Popularity} the goal was to make sure the recommendation lists for each user are calibrated in terms of the range of popularity of the items in the list compared to the range of item popularity the user had originally shown interest at. For instance, if a user has 20\% popular items, 50\% items with medium popularity and 30\% less popular items in his profile, the recommender system should try to keep this ratio as much as possible while keeping the accuracy loss as small as possible. Experimental results showed that, when evaluated from the multi-stakeholder paradigm discussed in Chapter \ref{data_method}, this method outperforms our two other proposed methods and also two other state-of-the-art methods proposed by other researchers. Compared to the other algorithms, the Calibrated Popularity algorithm matches the distribution of the recommendations in terms of item popularity with that of rating data much better and hence gives a more calibrated lists to the users. I also showed that this approach improves supplier fairness, as measured by the exposure of items from different supplies groups in recommendation lists.  

\section{Future Work}
The contributions presented in this thesis extend the state-of-the-art of long-tail recommendation, however they are not exhaustive and a number of questions remain for future work. In this section, I outline
some of the envisioned lines of work that could follow the contributions of this
thesis.

\subsection{User Studies}

Throughout the thesis I have been performing evaluation of the proposed algorithms for long-tail recommendation. This has been done in an offline fashion, where we use the historical interaction data of the users with recommendations as a ground truth to measure the effectiveness of the proposed methods. Such evaluation is based on a number of assumptions
about how users interact with recommendations and how they perceive these. Although the adapted evaluation protocol follows some of the common approaches taken in recommender systems, conducting user studies may be particularly important when evaluating the long-tail promoting algorithms. The reason for that is the complexity of user needs and interests that are commonly not well reflected in the recorded datasets. Long-tail promotion and diversification
is meant to bring (sometimes) content that is not well explored by a user, and its
evaluation is not very conclusive in an offline setting since only items that are already in the user's profile are considered good recommendations. In other words, in offline evaluation we have no knowledge about whether a recommended item that was not in the user's profile was a good recommendation or not and, by the nature of this type of evaluation, we consider them as bad recommendations.

Metrics such as long-tail promotion, diversity, and novelty are frequently placed together with accuracy
— as I did — and a trade-off of these is analysed in order to develop a better
method, a method which offers smaller loss in accuracy and more prevalent diversity or long-tail
improvements. However prior research such as the one by Ekastrand et al. \cite{ekstrand2014user} has suggested that non-accuracy metrics actually could lead to accuracy improvements. In fact, this is the ultimate goal of making recommendations more diverse and calibrating the recommendations with the range of users' interests — improving user’s satisfaction, which means
producing recommendations that users feel address their needs. We can say that, all the algorithms I proposed in this dissertation are evaluated in a rather pessimistic way meaning the real trade-off between accuracy and long-tail diversification would probably be better if they were evaluated on a live system and the users had interacted with those recommendations. 

\subsection{Directly Optimizing for Supplier Fairness}

The beneficial side-effect of supplier fairness was not directly optimized as our algorithm focused on the user calibration and no information about the suppliers was used in the calibration algorithm. However, because of the positive correlation between item popularity and the average popularity of the corresponding suppliers, improving popularity calibration has this indirect effect of improving supplier fairness. An interesting extension of this work would be to incorporate the tendency of each user towards different groups of suppliers as well as different popularity categories in the calibration algorithm. I will leave this for future work.

For each user we can define a vector of popularity interest which can encompass both item categories and supplier groups. For instance, sticking with the triple grouping I presented in this thesis, one could define a vector of size 6 that has the interest of the user towards different item categories and supplier groups and the recommendation algorithm should try to calibrate the recommendation list according this vector. How this extended method would perform for different stakeholders and also how it matches the distributional properties of the item popularity of the recommendations with that of rating data will remain as an open question.  

\subsection{Connection Between Popularity Bias and other Types of Fairness}

In this dissertation, our main concentration was on measuring the fairness of the recommendations with respect to popularity. In other words, I intended to evaluate the unfairness of a given recommender system for users with different levels of interest in item popularity and also for suppliers with different degrees of popularity. For example, our grouping strategy I described in Chapter \ref{data_method} was solely based on the information about the popularity of the items, suppliers, or the users' interest towards popular items. However,in many domains, certain aspects associated with users or suppliers might be important for the system designer to consider them as a group. These could be gender, age, race, nationality, or any other types of pre-defined labels for determining a group of people. The connection between popularity bias and the unfairness in those scenarios is an interesting open question that needs further attention. Can controlling popularity bias help making the recommendation to become fairer with respect to those pre-defined labels? What is the connection of what is referred to "protected group" or under-represented group and the group with lesser popularity? In what situations can popularity bias mitigation can be also helpful in alleviating other types of unfairness? Our preliminary results show that popularity bias is indeed connected to certain types of unfairness in recommendation such as miscalibration of the recommendations for certain groups of users \cite{abdollahpouri2019impact}.

\section{Limitations}
This dissertation, like many other scientific works, is not perfect and there are questions that are not addressed. Two main limitations are as follows:

\begin{itemize}
    \item \textbf{Not Investigating Pre-processing Techniques}: In Chapter \ref{tackle_pop_bias} three approaches for mitigating popularity bias were mentioned: pre-processing, in-processing, and post-processing. My model-based technique ($RG$) is an in-processing method since it modifies the existing recommendation algorithm. Tow other methods $XQ$ and $CP$ are both post-processing methods since they work on top of an existing recommendation algorithm. I have not proposed any pre-processing method for two reasons: these methods require sampling the data and this usually leads to increased data sparsity. In addition, the focus of this dissertation was not on comparing different processing methods for mitigating popularity bias but rather to study the impact of popularity bias on different stakeholders and the extent to which different algorithms can mitigate this bias for these stakeholders. Nevertheless, trying different pre-processing techniques to remove the bias in data would be an interesting thread of research. 
    \item \textbf{Not taking into account the dynamics of item popularity:} In real world, items become popular over time and might lose popularity when new items arrive. In fact, if this was not the case, music in 50's would have been still the most popular music now. My analysis does not take into account this dynamic as I was mainly interested in how popularity bias impacts the performance of the algorithms for different stakeholders and my analysis has been under the fact that certain items are popular and others are not. In other words, my analysis sees a snapshot of the current situation of a recommendation algorithm. However, the dynamics of item popularity could be incorporated as an interesting future work to analyze the popularity bias of different algorithms on different stakeholders over time. 
\end{itemize}


\bibliographystyle{plain}	
\bibliography{refs}		


\end{document}